\documentclass[a4paper,twoside,12pt]{report}
\usepackage{setspace}
\usepackage[dvips]{graphicx}
\usepackage{times}
\usepackage{amsmath}
\usepackage{amssymb}
\usepackage[sort]{natbib} 
\usepackage{aas_macros}
\usepackage{index}
\usepackage{color}
\usepackage{multirow}
\usepackage{rotating}
\usepackage[nottoc,numbib]{tocbibind} 
\usepackage{fixltx2e}
\usepackage{textcomp}
\usepackage{titlesec}
\usepackage[textwidth=16cm, bindingoffset=1.2cm, voffset=0pt, bottom=2.5cm, top=3.2cm]{geometry} 
\usepackage{fancyhdr} 
\titleformat{\chapter}{\normalfont\huge\bf}{\thechapter.}{20pt}{\huge\bf} 

\newcommand{\blankpage}{
\newpage
\thispagestyle{empty}
\mbox{}
\newpage
}

\begin{document}

\onehalfspacing

\begin{titlepage}
\null\vfill
\begin{center}
\Large \textbf{Habitability of Exoplanetary Systems}\par
PhD Dissertation
\vskip0.5cm
\textbf{Vera Dobos}\par
\normalsize
Konkoly Observatory, Research Centre for Astronomy and Earth Sciences, Hungarian Academy of Sciences\\
\vskip0.5cm
R\'eszecskefizika \'es csillag\'aszat doktori program\par
Head of program: L\'aszl\'o Palla
\vskip0.3cm
PhD School of Physics, E\"otv\"os University\par
Head of School: Tam\'as T\'el

\begin{figure}[h]
\center
\includegraphics[width=40mm]{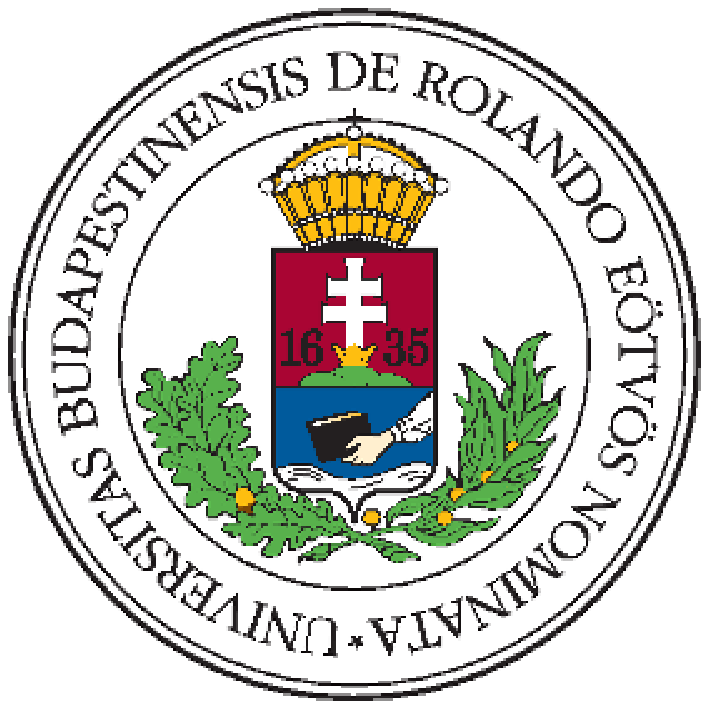}
\end{figure}

\textbf{Supervisors:}\\
Imre Nagy, College Associate Professor at the Nat. Univ. of Public Service, PhD\\
Emese Forg\'acs-Dajka, Senior Lecturer at E\"otv\"os University, PhD\\
\vskip0.3cm
\textbf{Associate Supervisors:}\\
L\'aszl\'o L. Kiss, Director of the Konkoly Observatory, Corr. Member of MTA\\
Edwin L. Turner, Professor of Astrophysical Sciences at Princeton University, PhD
\vskip0.8cm
Department of Astronomy, E\"otv\"os University
\vskip0.8cm
2016
\end{center}\vfill
\end{titlepage}

\blankpage

\tableofcontents

\newpage

\chapter{Preface}

The aim of my dissertation is to investigate habitability in extra-Solar Systems. Most of the time, only planets are considered as possible places where extraterrestrial life can emerge and evolve, however, their moons could be inhabited, too. I present a comprehensive study, which considers habitability not only on planets, but on satellites, as well.

My research focuses on three closely related topics. The first one is the circumstellar habitable zone, which is usually used as a first proxy for determining the habitability of a planet around the host star. The word \textit{habitability} is used in the sense that liquid water, which is essential for life as we know it, may be present on the planetary surface.

Whether the planet is habitable or not, its moon might have a suitable surface temperature for holding water reservoirs, providing that tidal heating is in action. Tidal heating is generated inside the satellite and its source is the strong gravitational force of the nearby planet. The second topic of my research explores tidal heating and the habitability of extra-solar moons with and without stellar radiation and other related energy sources.

Life is possible to form even on icy planetary bodies, inside tidally heated subsurface oceans. The third topic probes the possibility of identifying an ice-covered satellite from photometric observations. A strong indication of surface ice is the high reflectance of the body, which may be measured when the moon disappears behind the host star, so its reflected light is blocked out by the star.

Before submerging in these topics, I give an overview in Chapter \ref{intro} about the current state of research regarding the habitability of planets and their moons outside the Solar System. I present the used methods for my investigations for all three major topics in Chapter \ref{methods}. My work regarding the habitable zone is described in Chapter \ref{calcHZ}, the tidal heating calculations are shown in Chapter \ref{viscoelastictidal} and the proposed method for selecting icy moons by observations is presented in Chapter \ref{possibalbedoest}. Finally, I summarise the results in Chapter \ref{bigsummary}.

\chapter{Introduction} \label{intro}

 \section{Habitability of exoplanets}
 
More than three thousand extra-solar planets (exoplanets) are known as of today (source: \textit{exoplanet.eu}, July 2016) and even more planet candidates wait for confirmation. The first planets discovered were hot Jupiters that are of super-Jupiter size and orbit the central star at a very close distance. We do not expect that these planets could harbour life as we know it, but with current technology we are capable of detecting Earth-size planets, as well.
  
  \subsection{Habitable zones}
   
    The usual method for searching life is to find those planets which are located
    in their star's habitable zone, hence they may be suitable for supporting life.
    Depending on the inspected conditions, several habitable zones can be
    distinguished. These are the \textit{Galactic Habitable Zone} \citep[for details see][]{lineweaver04}, the \textit{Liquid Water Habitable Zone} \citep{kasting93}, the \textit{Ultraviolet Habitable Zone} \citep{buccino06, dobos10}, the
    \textit{Tidal Habitable Zone} \citep{barnes09}, the
    \textit{Photosynthesis-Sustaining Habitable Zone} \citep{franck00, bloh09} and the \textit{Life Supporting Zone} \citep{neubauer12}.
    
    The emergence of life needs long time, given that on Earth it took about a
    billion years after the end of the heavy bombardment \citep[][Chapter 3]{jones04}. The
    orbit of the planet needs to be stable during this long time interval to stay in
    the central star's habitable zone.
    
    The so-called \textit{Continuous Habitable Zones} describe whether a planet is
    located continuously inside the habitable zone for a long time. As the star
    ages, its radiation increases, hence the habitable zone moves outward from the
    star \citep[see for example][]{kasting93, underwood03, buccino06, guo09}. This means that
    a planet which was initially inside the habitable zone gets stronger stellar
    radiation with time, which can be harmful for living organisms. On the
    other hand a farther planet, that was too far from the star and did not receive
    sufficient energy may become habitable.

In the followings, the Liquid Water Habitable Zone will be discussed in details.

\subsection{Liquid water habitable zone}
    
\begin{figure}
 	\centering   
 	\includegraphics[width=25pc]{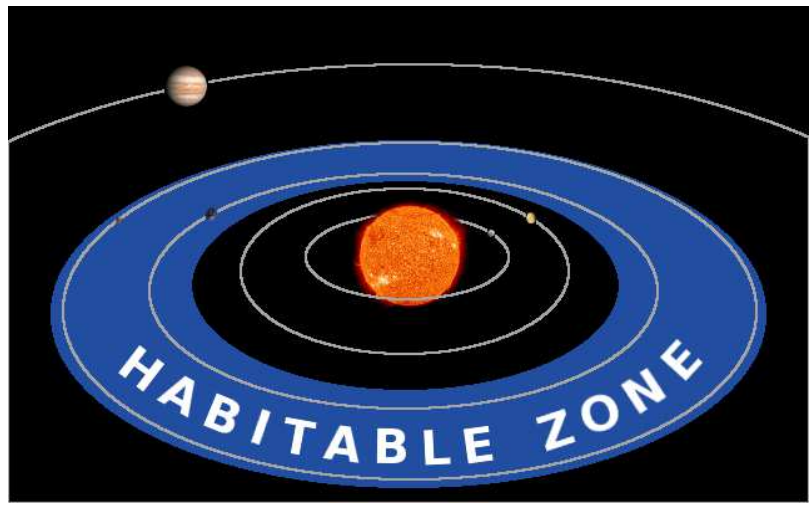}     
 	\caption{\label{HZ-SS}Schematic figure of the liquid water habitable zone in the Solar System.}
 \end{figure}

  The most studied habitable zone is the liquid water habitable zone (HZ) which is
  usually described as a region around a star where a hypothetical Earth-like
  planet could support liquid water on its surface \citep{kasting93}. A schematic figure of the HZ in the Solar System can be seen in Fig. \ref{HZ-SS}.
  In order to calculate the boundaries of the HZ, 
  the stellar parameters are needed. Other environmental elements, like the presence of an
  atmosphere, its thickness, composition, atmospheric pressure, or cloud formation
  can influence the surface temperature, i.e. the state of the water, as well.
  \citet{kasting93} used a climate model based on
  carbonate-silicate cycle, atmospheric composition and stellar radiation. This
  model is widely used and is considered to be the best tool to estimate the
  boundaries of the liquid water habitable zone.
   
On the surface of the Earth, the water reservoirs wouldn't boil at 100~$^\circ$C, because if the atmosphere were at this temperature, than the atmospheric pressure would be 2 bars \citep[Chapter 4]{sullivan07}. However, water molecules can move to the atmosphere approximately at 60~$^\circ$C, enhancing the greenhouse effect, which leads to a warmer temperature. Because of the higher temperature even more water molecules evaporate, resulting in a positive feedback process. In the upper atmosphere dissociation occurs, meaning that ultraviolet photons break up the molecules to atoms, and finally hydrogen atoms escape to space, preventing new water molecules to form in the future. This process is called the runaway greenhouse effect \citep{kasting93}. Usually, this runaway greenhouse limit is used as the inner boundary of the liquid water habitable zone.

\begin{figure}
 	\centering   
 	\includegraphics[width=25pc]{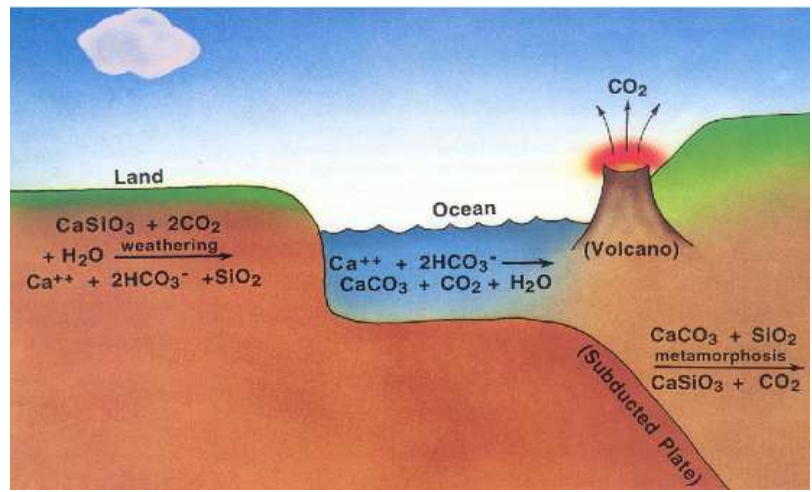}     
 	\caption{\label{CO2}Schematic figure of the carbonate-silicate cycle. The term ``metamorphosis" should read ``metamorphism". The reference for the figure is \citet{kasting95}.}
 \end{figure}

Water vapour is responsible for about the two-third of the total greenhouse effect on Earth. The other major greenhouse gas is carbon-dioxide, which is essential in the carbonate-silicate cycle, too \citep[Chapter 4]{sullivan07}. The importance of this cycle is that the circulation of CO\textsubscript{2} stabilizes the climate on Earth over a long timescale. Atmospheric carbon-dioxide dissolves in rainwater and forms carbonic acid (H\textsubscript{2}CO\textsubscript{3}, see Fig. \ref{CO2}). Silicate rocks are dissolved by this weak acid and the products of this silicate weathering (e.g. bicarbonate and dissolved silica) are transported by rivers to oceans. In the ocean these are built into shells in the form of calcium carbonate (CaCO\textsubscript{3}). When these organisms die, they fall to the bottom of the ocean, where the shells either redissolve, or become buried in sediments on the seafloor. At plate boundaries, when the oceanic plate slides beneath the continental plate, the carbonate sediments sink to high temperature and high pressure depths. In the process of volcanic eruption, CO\textsubscript{2} is released into the atmosphere, which is called carbonate metamorphism. This is a full cycle of carbone-dioxide, which has a timescale of approximately 200 million years \citep{kasting93, sullivan07}.
   
The climate regulation is due to the negative feedback mechanism of carbonate-silicate cycle. If the temperature is higher, then precipitation rate will be higher, too, removing more CO\textsubscript{2} from the atmosphere. Consequently, the greenhouse effect will lessen, resulting in a lower surface temperature. On the other hand, decreasing temperature leads to slower weathering rates, leading to the accumulation of CO\textsubscript{2} in the atmosphere, which will strengthen the greenhouse effect. This process can even bring a planet back from a snowball state, providing that the temperature is not low enough to permit the formation of CO\textsubscript{2} ice clouds, which would further cool the surface because of the release of latent heat, and the increased albedo \citep{caldeira92}.

Since carbonate-silicate cycle can reverse the frozen state of a planet, it can push the outer boundary of the habitable zone relatively far out from the star. For this reason, the carbonate-silicate cycle is usually taken into account when calculating the boundaries of the HZ. Mostly, calculations are made for Earth-like bodies, since the climate models are very complex, and different planetary sizes, atmospheric compositions, cloud coverage, or continent-to-ocean ratios would fundamentally change the required models. The HZ in the Solar System is located between 0.97 and 1.67 AU distances from the Sun \citep[these are the runaway greenhouse and the maximum greenhouse limits, respectively,][]{kopparapu13, kopparapu13err}.

\begin{figure}
 	\centering   
 	\includegraphics[width=27pc]{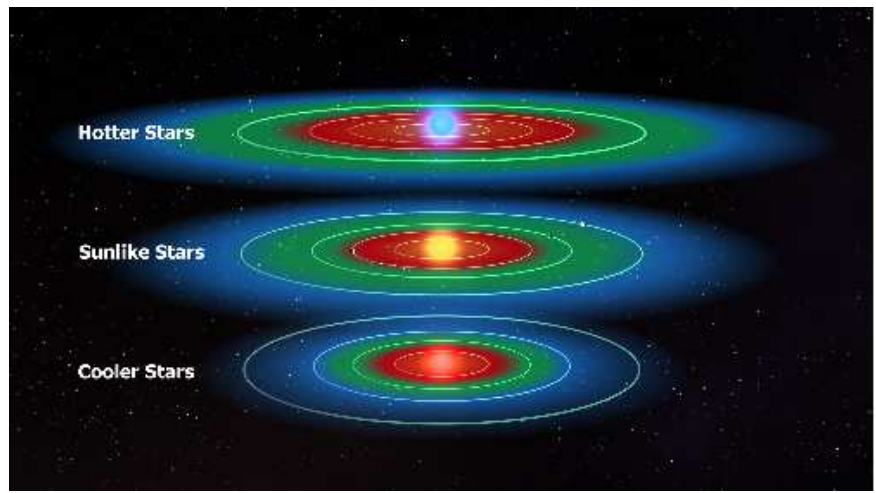}     
 	\caption{\label{diffstars}The location of the liquid water habitable zone depends on the central star's radiation. Red and blues colours indicate that the surface temperature of an Earth-like planet would be too hot and too cold for the liquid state of water, respectively. The green area is the habitable zone. \textit{Credit: NASA/Kepler Mission/Dana Berry.}}
 \end{figure}

The HZ of hotter stars is located farther than for cooler stars, because their stellar radiation is stronger (see Fig. \ref{diffstars}). Also, the HZ is wider for hotter stars, for example it is approximately 18 AU wide for a B spectral type main sequence star, but only 0.5 AU wide for an M class main sequence dwarf.

The habitable zone is a useful tool for selecting those planets where life could potentially emerge, however, even if a planet can be called `habitable' based on liquid water, it does not necessary mean that it meets every requirements for life. A habitable planet, as described above, is one that can maintain liquid water reservoirs on the surface for a long period of time.

In general, F, G and K class main sequence stars are in the main focus in connection with
   habitability, since they are similar to our Sun, and spend sufficiently long time on
   the main sequence for the emergence of life. The habitable zone of the less
   massive M class stars is much closer to the star. M type red dwarfs are among
   the most active stars, with erosive flares, which can be harmful to life on close-in
   planets. A strong magnetosphere can protect the planet from cosmic rays, however, planets in the HZ of M dwarfs are probably tidally locked, which may result in a small magnetic moment \citep{greissmeier05}.
   Even for a magnetosphereless body, flares may not induce a direct hazard on the
   planetary surface, if the planet has a protective, oxygen-rich atmosphere \citep{segura10}. In addition, the abundance and long
   lifetime on the main sequence make M stars preferable targets in searching for
   life.

The climate model developed by \citet{kasting93} is complex, since it takes into account the carbonate-silicate cycle, the atmospheric composition of the planet and the stellar properties. For faster estimations of the location of the HZ, \citet{underwood03} fitted parabolic equations to the results of \citet{kasting93}
and used these equations for calculating the boundaries of
the liquid water habitable zone. Later one of the fitted equations was modified \citep{jones06, jones10}. \citet{selsis07} proposed similar equations by extrapolating the results of
\citet{kasting93}. \citet{catling16} also made
a parabolic fit to the same results. \citet{kopparapu13} proposed a quartic function for calculating the stellar flux from the effective temperature.
Note, that the correct coefficients can be found in \citet{kopparapu13err}. To calculate the boundaries of the HZ with any of these equations, only the stellar temperature and luminosity are needed.
\citet{dobos13} gave empirical formulae to calculate these parameters as functions of the stellar mass for F, G, K and M class main sequence stars.

It is also possible, that a planet is habitable, while it is outside the HZ. For example, the planet's atmospheric composition and thickness can strongly influence its surface temperature, as well as large amount of clouds, atmospheric pressure, salinity of water, high eccentricity of the orbit or tidal heating. These all influence habitability, although these parameters are not considered in the traditionally called habitable zone.

 \section{Habitability of exomoons}

 Beside exoplanets, their moons (exomoons) can be habitable, too. No exomoons have been discovered yet, but these measurements are expected in the next decade. \cite{bennett14} present a candidate, which has been detected via the MOA-2011-BLG-262 microlensing event. The best-fit solution for the data implies the presence of an exoplanet hosting a sub-Earth mass moon. This measurement however needs confirmation, since an alternate solution is also presented. Nevertheless, this measurement indicates that the era of exomoon detection is about to begin.
 
 The most favourable method for exomoon discoveries is photometry. An exoplanetary transit may reveal the presence of a moon in the light curve. Details of this method are thoroughly discussed in the literature \citep{simon07, kipping09a, kipping09b, simon10, kipping12, simon12}.
 
 In addition, habitability of exomoons is under examination as well \citep[see e.g.][]{kaltenegger10, heller13, helleretal14}. \cite{hinkel13} investigated the influence of eccentric planetary orbits on moons, and concluded that a moon with sufficient atmospheric heat redistribution may sustain suitable temperature for life on the surface even if the moon orbits a planet that moves temporarily outside of the circumstellar HZ at each orbital period.
  
Although moons in the Solar System are small compared to planets (the largest moon, Ganymede has a size of about 0.4 Earth-radius and a mass of 0.025 Earth-mass), it is widely believed that much larger moons can exist in other planetary systems, as well. One of the reasons is that much larger planets (hot Jupiters) exist, too, indicating that at the planet formation period they could have had much more massive protoplanetary discs, from which more massive moons might have been able to form. According to \citet{crida12}, the mass of a moon is limited by the mass of its host planet. Also, the mass of satellite systems in the Solar System is proportional to the mass of their host planet. \citet{canup06} showed that this might be the case for extra-solar satellite systems as well, giving an upper limit for the mass ratio at around 10\textsuperscript{$-4$}. Thus 10 Jupiter-mass planets may have 0.3 Earth-mass satellites. However, if the moon is not originated from the circumplanetary disk, but from collision, like in the case of the Earth's Moon, then even larger satellites might exist. When studying the habitability of exomoons, it is very common to consider Earth-size moons, as they can have significant atmosphere, and they could be similar to Earth.

M dwarfs could have specific characteristics regarding the exomoon identification and analysis. Although most search projects focus on solar-like stars, M dwarfs as targets should also be considered, because more than 70\% of the stars are of M spectral type. In theory it is possible that no large exomoons are present around M dwarfs, as their smaller stellar mass allow the formation of only smaller exoplanets and exomoons than in the Solar System. But based on the discovered exoplanets, 68 M dwarfs are known to have 96 exoplanets, from which the heaviest planet's mass is 62 $M_\mathrm{J}$, 20 other exoplanets have at least 10 Jupiter-masses, and there are 13 other planets more massive than 1 $M_\mathrm{J}$ (\textit{exoplanet.eu}, January 2016). The average mass of the exoplanets of the 68 M dwarfs is 4.8 $M_\mathrm{J}$.

  \subsection{Circumplanetary habitability}
   
\begin{figure}
 	\centering   
 	\includegraphics[width=25pc]{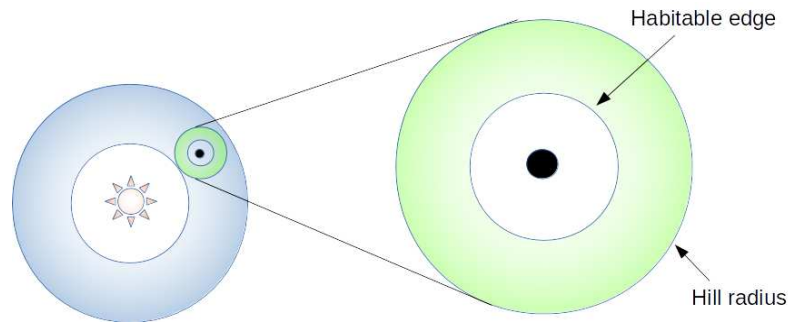}     
 	\caption{\label{HZmoon}Range of habitable orbits around a planet in the IHZ.}
 \end{figure}

When considering the habitability of exomoons, it is usually assumed that the host planet orbits the central star inside the circumstellar habitable zone (the blue region in Fig. \ref{HZmoon}), and the moon's orbit is between the so-called \textit{habitable edge} and the Hill radius, or half of the Hill radius of the planet (this is the green circumplanetary region in Fig. \ref{HZmoon}). If the satellite is too close to the planet, then the reflected stellar light from the planet, the thermal radiation and tidal heating will be strong enough to make the moon's surface too warm to be habitable. The habitable edge is defined by the runaway greenhouse limit, hence in orbits closer to the planet the moon will loose its water reservoirs, providing that the atmospheric conditions are similar to that of the Earth \citep{heller13}.  The Hill radius surrounds the gravitational sphere of influence for a body. This is the outermost distance where a moon can orbit a planet. However, according to \citet{domingos06} and \citet{donnison10}, the orbits of direct orbiting satellites can stay stable only inside the inner half of the Hill sphere.

Another outer limit can be defined by climate models. \citet{forgan13} developed a model for exomoons in order to investigate the energy balance of satellites. This climate model combined with eclipses and the ice-albedo feedback is capable of predicting the orbital parameters of a moon that will result in a snowball state, defining an outer limit for the circumplanetary habitable zone \citep{forgan14, forgan16}. The ice-albedo feedback is a positive loop where cooling results in larger ice coverage, which leads to a higher albedo that means that more stellar light will be reflected back, and less energy will be absorbed, so the cooling will be stronger.

A comprehensive model should include all the energy sources and energy sinks for the exomoon. The energy sources for an exomoon are: stellar insolation, atmospheric retention, radiogenic heat, planetary illumination, thermal heat from the planet and tidal heating. The possible energy sink sources are: infrared cooling, planetary eclipses and high albedo caused by oceans, ice or clouds. Eclipses can be frequent (in case of low orbital inclination relative to the planetary orbit) and prolonged on a satellite of a giant planet, hence they can reduce the incident flux by as much as 6\% \citep{heller13}. In calculations the albedo of exomoons are usually just estimated, or the albedo of Earth is used.

  \subsection{The importance of tidal heating}

A moon can be habitable outside the circumplanetary, or even outside the circumstellar habitable zones, as well. Since tidal heating of the moon is caused by the presence of the planet, this energy source is independent from the star. Tidal heating is an inner source of energy that is caused by the gravitational forces of the nearby planet. Since the planet is much more massive than the moon, the arising forces deform the moon and cause friction inside the body that leads to heat dissipation. Hence, opposed to stellar radiation, tidal heating comes from the inside. For this reason, the surface temperature of a moon can be suitable for liquid water for several billion years even in the case when the planet-moon pair does not have a central star.

There are a few factors that favour the detection of tidally heated exomoons: 1. they can be far more luminous than their host planet, 2. their luminosities are independent of their separations from the star, and 3. they are visible around nearby (thus bright) stars \citep{peters13}.

The tidal heat rate of a moon is usually calculated by the following expression \citep[e.g.][]{reynolds87, meyer07}:
 
 \begin{equation}
 \label{fixQ}
 \dot E_\mathrm{tidal} = \frac {21} {2} \frac {k_2} {Q} \frac {G M_\mathrm{p}^2 R_\mathrm{m}^5 n e^2} {a^6} \, ,
 \end{equation}
 
 \noindent where $G$ is the gravitational constant, $M_p$ is the mass of the planet, $R_\mathrm{m}$, $n$, $e$ and $a$ are the radius, mean motion, eccentricity and semi-major axis of the moon, respectively. $Q$ is the tidal dissipation factor and $k_2$ is the second order Love number:
 
 \begin{equation}
 \label{k2}
 k_2 = \frac {3/2} {1 + \frac { 19 \mu } { 2 \rho g R_\mathrm{m} } } \, ,
 \end{equation}
 
 \noindent where $\mu$ is the rigidity, $\rho$ is the density and $g$ is the surface gravity of the satellite. This calculation method is called the fixed $Q$ model, because $Q$, $\mu$ and $k_2$ are considered to be constants.
 
 \begin{figure}
 	\centering   
 	\includegraphics[width=25pc]{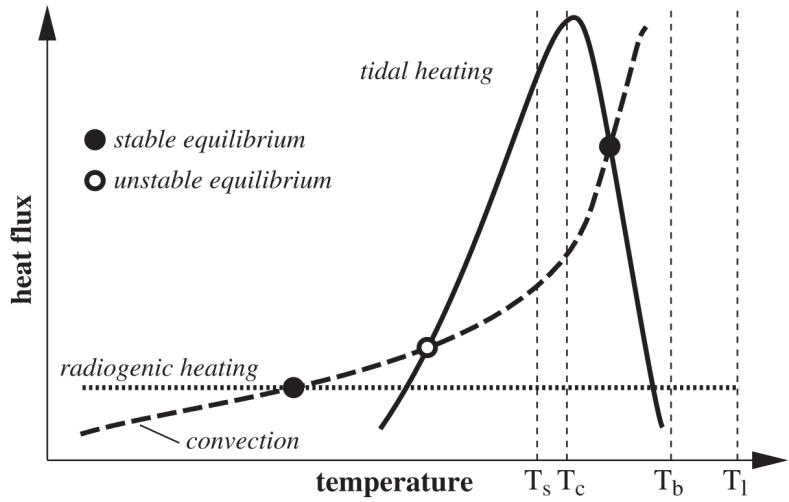}     
 	\caption{\label{Moore}Schematic figure of the temperature dependency of tidal heat flux and convective heat loss. The solid, dashed and dotted curves indicate the tidal heat flux, the convective cooling flux and the radiogenic heating flux, respectively. $T_\mathrm{s}$, $T_\mathrm{c}$, $T_\mathrm{b}$ and $T_\mathrm{l}$ indicate the solidus, critical, breakdown and liquidus temperatures, respectively \citep{moore03}.}
 \end{figure}
 
It can be seen from Eq. \ref{fixQ} that in case the moon orbits the planet in a circular orbit, then the tidal heat rate will be zero. Since tidal forces tend to circularize the orbit of satellites, tidal heating attenuates quickly, unless some other force continuously maintains the eccentricity. If other moons exist which orbit the same planet, then it is possible that their orbits will be in mean motion resonance, which means that the ratio of their orbital periods can be expressed with small integers. Due to this resonance, they will experience regular, periodic gravitational pull that excites their eccentricity. Therefore mean motion resonance between the orbits of two or more satellites maintains their eccentricities in long timescales, which continuously sustain tidal heating.
 
 The fixed $Q$ model is broadly used in tidal calculations, but highly underestimates the tidal heat of the body \citep{ross88, meyer07}. Moreover, both $Q$ and $\mu$ are very difficult to determine, and vary on a large scale for different bodies: from a few to hundreds for rocky planets, and tens or hundreds of thousands for giants \citep[see e.g.][]{goldreich66}. In addition, these parameters should not be constants, since they strongly depend on the temperature \citep{fischer90, moore03, henning09, shoji14}. As a consequence, tidal heat flux has a temperature dependency, as well: it reaches a maximum at a critical temperature ($T_\mathrm{c}$) as can be seen in Fig. \ref{Moore}. Between the solidus and the liquidus temperatures ($T_\mathrm{s}$ and $T_\mathrm{l}$, respectively) the material partially melts. Above the breakdown temperature ($T_\mathrm{b}$) the mixture behaves as a suspension of particles. The dashed curve represents the convective heat loss of the body. Circles indicate equilibria, for example, the solid circle between $T_\mathrm{c}$ and $T_\mathrm{b}$ is a stable equilibrium point. If the temperature increases, convective cooling will be stronger than the heat flux, resulting in a cooler temperature. In the case of decreasing temperature, the tidal heat flux will be the stronger, hence the temperature increases, returning the system to the stable point. The stable equilibrium between the tidal heat and convection is not necessarily located between $T_\mathrm{c}$ and $T_\mathrm{b}$, in fact, there are cases, when the two curves do not have intersection at all \citep[see][Fig. 6]{henning09}. In these cases tidal heating is not strong enough to induce convection inside the body.
 
 In contrast to the fixed $Q$ model, viscoelastic models take into account the temperature dependency of the body, hence are more realistic. The first application of a viscoelastic tidal heating model to exomoons was introduced by \citet{dobos15}. This model did not include any other energy sources than tidal heating. For this reason it can only be used in cases when stellar radiation and other energy sources are negligible.

As mentioned above, comprehensive models exist which take into account all of the energy sources affecting moons \citep[see for example][]{heller13, forgan14}. These models however, use fixed $Q$ models for calculating the tidal heating of the moon. \citet{forgan16} merged a comprehensive climate model with viscoelastic tidal heating calculation and compared the results with fixed $Q$ calculations.

  \subsection{Icy moons}
   
    Among the satellites in the Solar System, Europa (moon of Jupiter) and Enceladus (moon of Saturn) are the best candidates for supporting subsurface liquid water and related potential habitability and hence these are the most actively studied satellites. Both moons have icy crusts, that are at least a few tens of kilometres thick. Underneath the ice layer, a huge ocean resides, which is believed to be global in the case of Europa, and probably on Enceladus, too \citep{kivelson00, thomas16}. Both bodies are tidally heated which maintains the liquid phase state of the oceans. At the boundary of the ocean and the silicate layers, where the temperature is warm enough, the environment can be suitable for the appearance of prebiotic processes thus potentially further development toward simple forms of life.

Ice is very reflective in the visual wavelengths and for this reason, the albedo of an icy celestial body is usually very high. The \textit{Bond albedo} (which is often called as \textit{albedo} for short) is defined as the ratio of the reflected and the incident flux at all phase angles. The \textit{phase angle} is the angle subtended at the observer between the star and a celestial body. In other words, the Bond albedo is the reflected radiation combined in all directions. The albedo of a fully reflective body, that reflects all incident radiation is one, and the lowest possible value for fully absorbing bodies is zero.

 \begin{figure}
 	\centering   
 	\includegraphics[width=13cm]{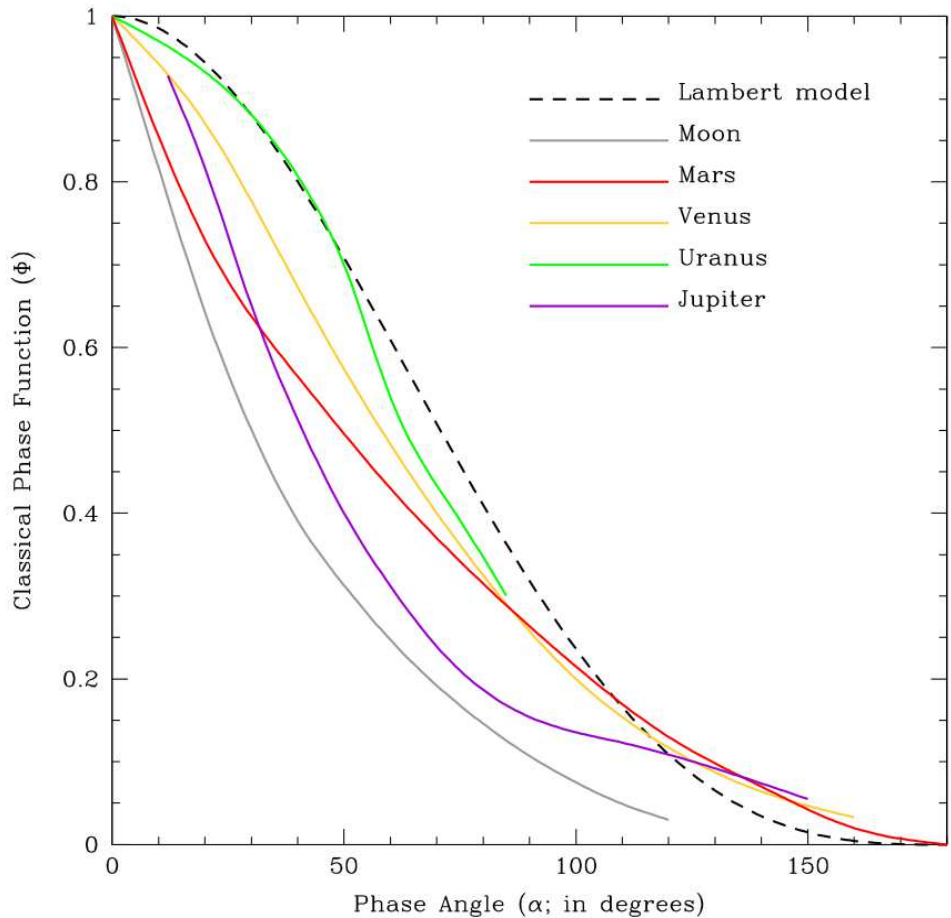}     
 	\caption{\label{phase_angle}Reflectivity of different Solar System bodies as function of the phase angle. The phase function on the ordinate expresses reflectivity in different phase angles and is always normalised to one at zero phase angle \citep{burrows10}.}
 \end{figure}

The \textit{geometric albedo} however, can exceed one in some cases. The geometric albedo is the ratio of brightness at zero phase angle compared to an idealized, flat, fully reflecting, diffusely scattering disk (which is called \textit{Lambertian reflectance}). Zero phase angle means that the observer measures the brightness of the body from the direction of the stellar light, so the visible hemisphere is illuminated completely. In most cases it is impossible to do such measurements in practice, since the Earth or the detector blocks out some part of the light. Usually the geometric albedo is measured in a phase angle very close to zero. This difference in the angle can be very significant, because in some cases the reflected light increases drastically when the phase angle approaches zero. This is due to the so-called \textit{opposition effect}. Fig. \ref{phase_angle} shows that Solar System planetary bodies reflect much more light when the phase angle is low. The decrease for higher phase angles is partly caused by the decreased area that is lit, but note that the Moon and Mars have the steepest phase function curve at low phase angles in Fig. \ref{phase_angle} which is caused by the opposition effect.

 \begin{figure}
 	\centering   
 	\includegraphics[width=16cm]{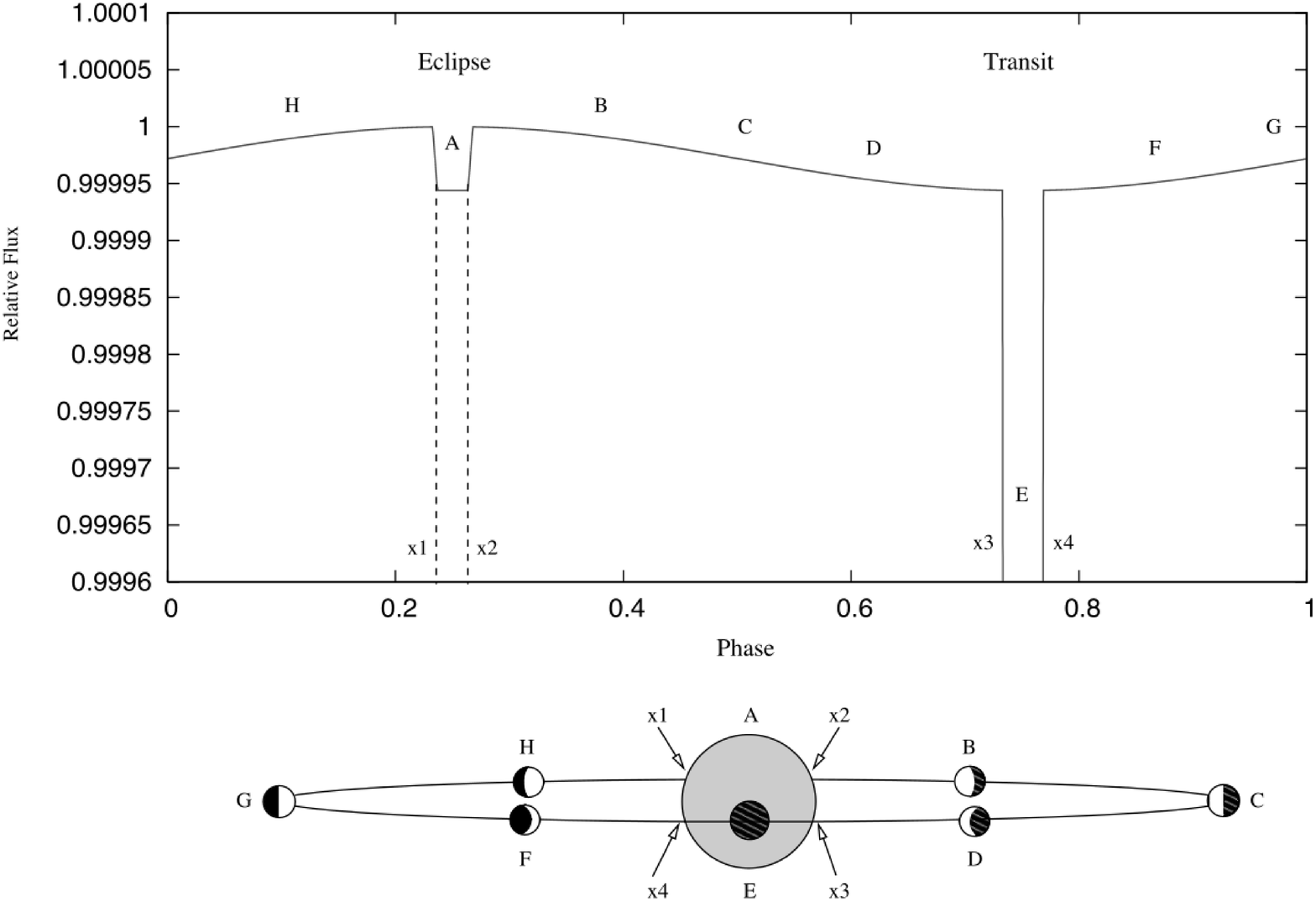}     
 	\caption{\label{Rowe}Schematic figure of the the stellar light curve with a planet during one orbital phase. At phase E the planet transits the star, and at phase A the planet occults. The stellar light alone is seen at phase A, the additional flux is caused by the reflected light from the planet \citep{rowe08}.}
 \end{figure}

There are two explanations for the opposition effect, one is called the \textit{shadow-hiding opposition effect} (SHOE) and the other is the \textit{coherent backscatter opposition effect} (CBOE). The SHOE is caused by the disappearing shadow at zero phase angle which is particularly important in fine powders with grain sizes less than 20 \textmu m, because the electrostatic and van der Waals forces exceed the gravitational force of the planetary body \citep{hapke12}. Consequently, the grain particles form towers and bridges which cast long shadows except at zero phase angle. The Moon is known to have such fine, porous powders on the surface that cause strong SHOE.

The CBOE is basically an interference phenomenon: an incident wave front on a particulate medium is scattered twice or more before exiting the medium, and from the same wave front light will travel in the same path but in the opposite direction, as well. At zero phase angle the phase difference of the emerging wavelets will be zero and the amplitudes add coherently as they interfere with each other, resulting in higher intensity up to a factor of two \citep{hapke12, verbiscer13}. Europa and Enceladus show significant CBOE \citep{verbiscer13}.

The high albedo of planetary bodies favours occultation measurements, because higher reflectivity means higher flux difference in the light curve. As can be seen in Fig. \ref{Rowe}, the stellar flux can be measured alone when the planet disappears behind the star (\textit{occultation}), and just before and after this occultation, the total flux is higher because of the reflected stellar light arriving from the planet. Conversely, if we can measure the flux difference in occultation, the albedo of the planet can be obtained. \citet{rowe08} estimated the geometric albedo of a planet using this method. Besides planets, this method can also be used to moons in theory. \citet{dobos16} investigated the possibilities of such measurements for exomoons with next generation telescopes.

  \subsection{Solar System analogues}

When considering the habitability of exomoons, Solar System satellites can serve as useful analogues, since similar moons could exist in other planetary systems, as well. Tidal heating is present in several moons in the Solar System, and lots of ice satellites orbit the giant planets, hence it may be useful to study them in more details before moving forward to the investigation of extra-solar moons.
   
\begin{figure}
 	\centering   
 	\includegraphics[width=12cm]{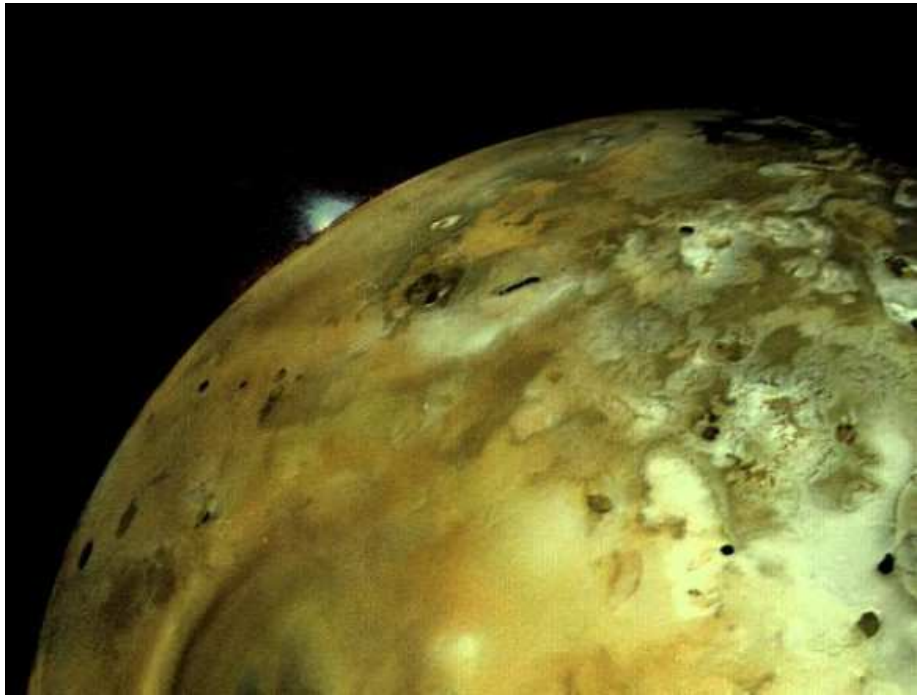}     
 	\caption{\label{Io}Volcanic eruption on Io as seen by Voyager-1 in 1979. \textit{Credit: NASA/JPL.}}
 \end{figure}

Io, one of the Galilean moons of Jupiter, is the most volcanically active satellite in the Solar System. Voyager-1 took the first photo of a volcanic eruption on another planetary body than the Earth (see Fig. \ref{Io}). Today several volcanoes are known on Io. According to \citet{spencer00}, the mean surface temperature of the satellite is approximately 90-95 K, but at volcanoes it can be around 200-400 K. The main source of this internal energy is tidal heating. Io, Europa and Ganymede are in a 1:2:4 mean motion resonance, respectively, which maintains their orbital eccentricities. This eccentricity excites continuous tidal heating inside the bodies.

\begin{figure}
 	\centering   
 	\includegraphics[width=9cm]{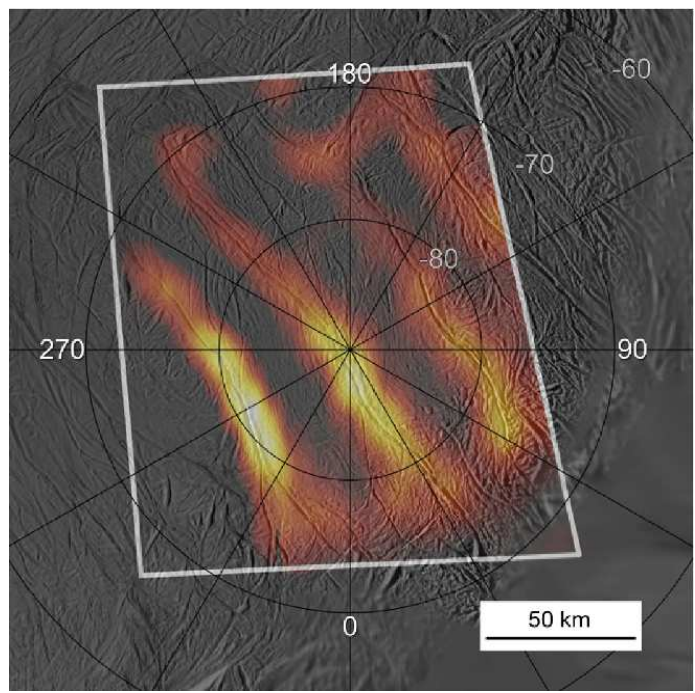}     
 	\caption{\label{tiger}Temperature map of the southern pole of Enceladus. Yellow colour indicates warmer areas which are called tiger stripes. \textit{Credit: NASA/JPL/Space Science Institute.}}
 \end{figure}

\begin{figure}
 	\centering   
 	\includegraphics[width=7cm]{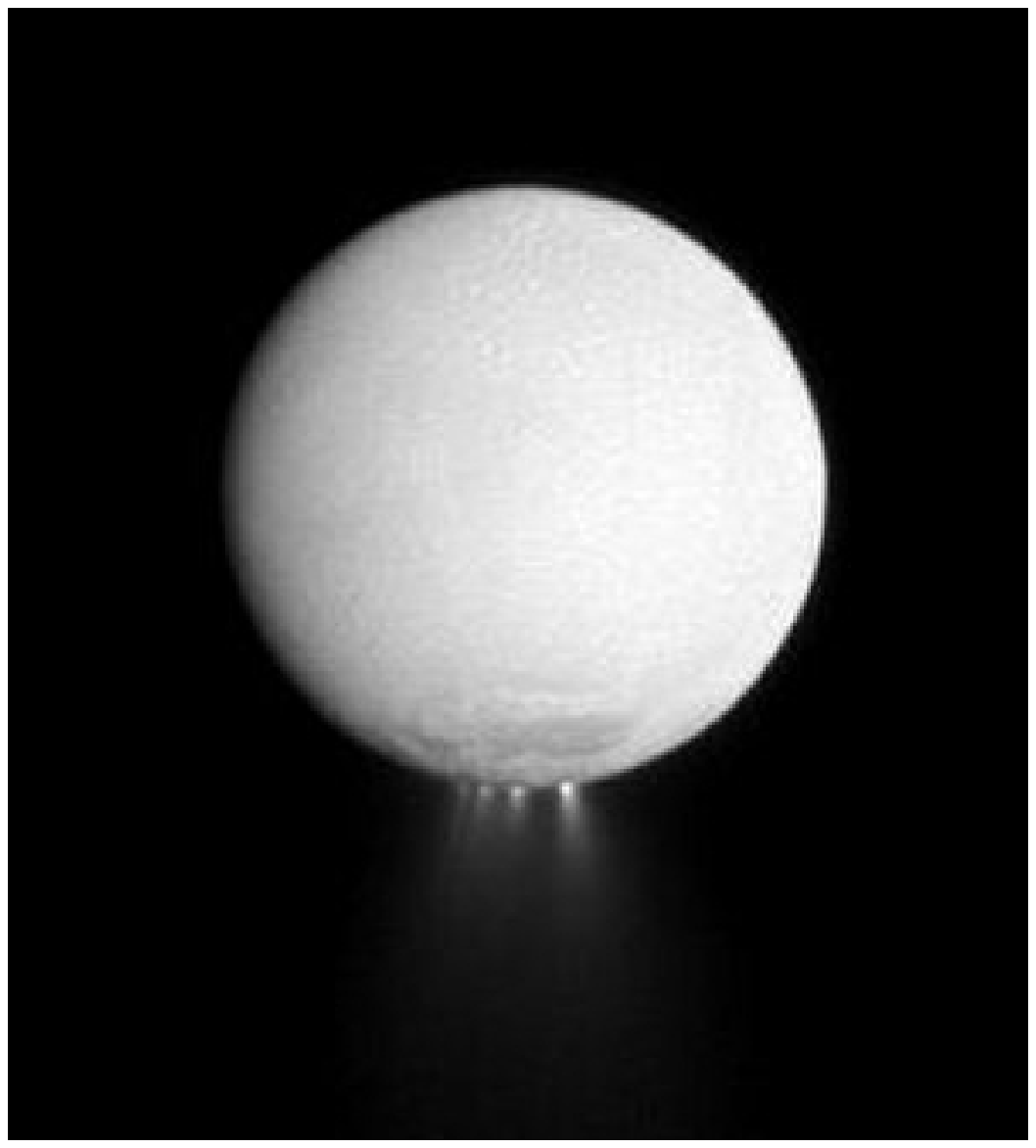}     
 	\caption{\label{plumes}Plumes at the south pole of Enceladus. \textit{Credit: NASA/JPL/Space Science Institute.}}
 \end{figure}

The temperature map of Enceladus shows warmer areas, the so-called \textit{tiger stripes} (see Fig. \ref{tiger}) at the south pole \citep{porco06}. The stripes coincide with the location of plumes that are geyser-like jets (see Fig. \ref{plumes}). The erupting water originates from the subsurface ocean and is ejected to such heights that it basically feeds the E ring of Saturn \citep{juhasz07}. \citet{kite16} showed that the eruptions are sustained by tidal heating on million year timescales.

On Europa, tidal friction leads to heat dissipation mainly in the silicate layer that is the underlying layer of the ocean. Part of the heat is dissipated probably through hydrothermal vents on the seafloor that heats the ocean locally. The warmer water moves upward to the ice layer, melting it and creating the so-called \textit{chaos terrains}. At these regions the ice sheet of Europa becomes approximately 2 km thin \citep{kereszturi13}. It is possible, that similarly to Enceladus, Europa has plume activity, too \citep{roth14}.

The satellites of the giant planets hold substantial mass of water ice that is often located on the surface \citep{brown97, verbiscer90}. Although there could be differences between satellite systems, the icy surface, including water ice, is abundant on them \citep{stevenson85, sasaki10}. Analysing the optical properties of the Solar System satellites, it can be seen that the high albedo in general indicates water ice. Non-ice covered satellites usually show low geometric albedo values between 0.04 and 0.15, while ice covered satellites show higher values, although different ingredients (mostly silicate grains) could decrease the albedo if they are embedded in the ice, or the ice is old and strongly irradiated/solar gardened \citep{grundy00, sheppard05}. These satellites show moderate geometric albedo between 0.21 for Umbriel and 1.00 for Dione \citep{veverka91, verbiscer13}. Beside water ice, nitrogen ice is also a strong reflector, producing albedo around 0.76 on Triton \citep{hicks04}. On this satellite the bright surface can partly be explained by its relatively young, 6-10 Myr old age. \citep{schenk07}. The most important analogues for high albedo in the Solar System could be Europa with 1.02 and Enceladus with 1.38 geometric albedo \citep{verbiscer13}. In the case of Europa the relatively young (about 50 million year old) surface accompanies with clean and bright ice, while on Enceladus freshly fallen ice crystals increase the albedo significantly \citep{pappalardo99, porco06}.

 \begin{figure}
 	\centering   
 	\includegraphics[width=9cm]{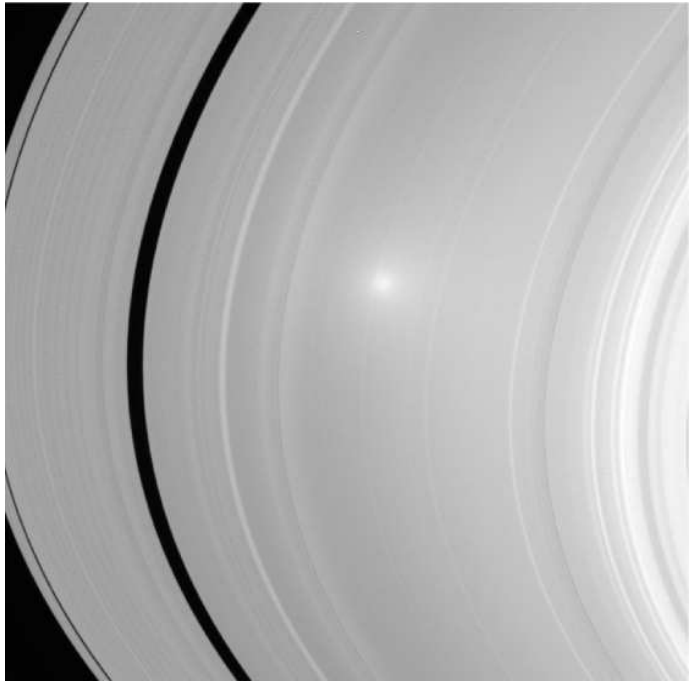}     
 	\caption{\label{rings}Opposition effect on the A ring of Saturn imaged by the Cassini spacecraft in 2006. \textit{Credit: NASA/JPL/Space Science Institute.}}
 \end{figure}

The opposition effect (CBOE) is prominent on the rings (see Fig. \ref{rings}) and nearby satellites of Saturn which results in high geometric albedos that in some cases exceed unity. \citet{verbiscer07} showed that there is a correlation between the reflectance profile of the E ring and the geometric albedo of nearby satellites. The plumes of Enceladus give material not only to the E ring, but eventually, the particles fall to the surface of the satellites, too. The clean ice coats them, enhancing their reflectance in opposition. The affected satellites are Mimas, Tethys, Dione and Rhea, and they have 0.96, 1.23, 1.00 and 0.95 geometric albedos, respectively.

\chapter{Methods} \label{methods}

The used calculation methods for the circumstellar and circumplanetary habitability, and also for the albedo estimation are presented in this section. The description of the habitable zone calculation (section \ref{habzone}) is based on the work of \citet{dobos13}, the tidal heating and the comprehensive climate models (section \ref{tidalheat}) are based on \citet{dobos15} and \citet{forgan16} and the method for albedo estimation (section \ref{albedoest}) is based on the work of \citet{dobos16}.

 \section{Habitable zone} \label{habzone}

   To calculate the boundaries of the liquid water habitable zone as a function of
   the stellar mass, the values of the stellar luminosity and
   effective temperature are needed. Many of the equations for calculating these parameters from the stellar mass have different terms for different stellar masses. The terms for F, G, K and M
   class main sequence stars are chosen and the equations are arranged to 
   similar forms to provide an easier comparison.
   
   The distance of the inner and the outer boundaries from the central star is
   given by
   \begin{subequations}
     \begin{align}
       \label{r_inner}
         r_{\mathrm{inner}}  = \sqrt { \frac { L } { S_{\mathrm{inner}} } } \, ,\\
       \label{r_outer}
         r_{\mathrm{outer}}  = \sqrt { \frac { L } { S_{\mathrm{outer}} } } \, ,
     \end{align}
   \end{subequations}
   \noindent where $r$ is the distance from the star in astronomical units, $L$ is
   the luminosity of the star in solar units and $S$ is the stellar flux at the
   boundaries in units of the solar constant \citep{kasting93}.
   
   The value of $S_{\mathrm{inner}}$ and $S_{\mathrm{outer}}$ can be calculated by
   the fitted equations of \citet{jones06}:
   \begin{subequations}
     \begin{align}
       \label{S_J_inner}
         S_{\mathrm{inner}} = 1.296 - 2.139 \cdot 10^{-4} T + 4.19 \cdot
   10^{-8} T^2 \, ,\\
       \label{S_J_outer}
         S_{\mathrm{outer}} = 0.234 - 1.319 \cdot 10^{-5} T + 6.19 \cdot
   10^{-9} T^2
     \end{align}
   \end{subequations}
   \noindent where $T$ is the effective temperature of the star. Similar 
   equations were given by \citet{catling16}. These equations are equivalent.
   
   The stellar temperature and luminosity can be calculated for main sequence stars
   from their mass as given by \citet{zaninetti08}:
   \begin{equation}
      \label{T_Zaninetti}
          T = 5760 \cdot M^{0.48}  \; \; \mathrm{ if } \; \; 0.3 M_\odot < M < 18.5
   M_\odot \, ,
   \end{equation}
   \begin{equation}
      \label{L_Zaninetti}
          L = 1.15 \cdot M^{3.43}\, .
   \end{equation}
   \noindent Here $M$ is the mass of the star in solar units and $M_\odot$
   represents one solar mass.
   
   Other possible solutions are presented by the simple power law (SPL):
   \begin{equation}
    \label{T_SPL}
       T = 5700 \cdot M^{0.5} \; \; \mathrm{ if } \; \; M \leq 10 M_\odot \, ,
   \end{equation}
   \begin{equation}
    \label{L_SPL}
       L = M^{3.6} \, ,
   \end{equation}
   \noindent or by the broken power law (BPL) of \citet{razzaque09}:
   \begin{equation}
    \label{T_BPL}
       T = 5700 \cdot M^{0.8} \; \; \mathrm{ if } \; \; 0.1 M_\odot \leq M < 2
   M_\odot \, ,
   \end{equation}
   \begin{equation}
    \label{L_BPL}
        L = M^{4.8} \; \; \mathrm{ if } \; \; M < 2 M_\odot\, .
   \end{equation}
   
   With these formulae the location of the liquid water habitable zone can be
   determined easily for different stellar masses between 0.3 and 2 $M_\odot$ which is
   the mass interval covered by all equations. Any of the equations above
   can be used for these stellar masses.
   
   For numeric comparison of the different
   calculation methods the reduced chi-square ($\chi^2$) values were calculated for all
   cases. The following equation was used:
   \begin{equation}
    \label{khi2}
        \chi^2 = \frac {1} {N} \sum_{i=1}^{N} \frac { \left[ y_i - f(x_i) \right]^2} {\sigma_i^2} ,
   \end{equation}
\noindent where $N$ is the number of data points, $y_i$ is the measured value, $f(x_i)$ is the calculated value and $\sigma_i$ is the uncertainty of the measurement.

 \section{Tidal heating} \label{tidalheat}

 \subsection{Viscoelastic model}
 
 In viscoelastic tidal heating models the term $k_2/Q$ in Eq. \ref{fixQ} is replaced by the imaginary part of the complex Love number Im($k_2$), which describes structure and rheology in the satellite \citep{segatz88}:
 
 \begin{equation}
 \label{viscel}
 \dot E_\mathrm{tidal} = - \frac {21} {2} \mathrm{Im}(k_2) \frac {R_\mathrm{m}^5 n^5 e^2} {G} \, .
 \end{equation}
 
 \noindent Note that in this expression the mass of the planet and the semi-major axis of the moon are eliminated by the mean motion ($n = \sqrt{G M_p / a^3}$).
 
 \citet{henning09} give the value of Im($k_2$) for four different models (see Table 1 in their paper). In this work we use the Maxwell model:
 
 \begin{equation}
 \label{Imk2}
 - \mathrm{Im}(k_2) = \dfrac {57 \eta \omega} { 4 \rho g R_\mathrm{m} \left[ 1 + \left( 1 + \dfrac { 19 \mu } { 2 \rho g R_\mathrm{m} } \right)^2 \dfrac { \eta^2 \omega^2 } { \mu^2 } \right] } \, ,
 \end{equation}
 
 \noindent where $\eta$ is the viscosity, $\omega$ is the orbital frequency and $\mu$ is the shear modulus of the satellite.
 
 The viscosity and the shear modulus of the body strongly depend on the temperature. Below the $T_\mathrm{s}$ the shear modulus is constant: $\mu = 50 \, \mathrm{GPa}$ and the viscosity follows an exponential function:
 
 \begin{equation}
 \label{viTs}
 \eta = \eta_0 \, \mathrm{exp} \left( \frac {E} {\mathcal{R} T} \right) \, ,
 \end{equation}
 
 \noindent where $\eta_0 = 1.6 \cdot 10^5 \mathrm{Pa \, s}$ (reference viscosity), $E$ is the activation energy, $\mathcal{R}$ is the universal gas constant and $T$ is the temperature of the material \citep{fischer90}.
 
 Between $T_\mathrm{s}$ and $T_\mathrm{b}$ the body starts to melt. The shear modulus changes by
 
 \begin{equation}
 \label{shTb}
 \mu = 10^{ \left( \frac {\mu_1} {T} + \mu_2 \right) } \mathrm{Pa} \, ,
 \end{equation}
 
 \noindent where $\mu_1 = 8.2 \cdot 10^4 \mathrm{K}$ and $\mu_2 = -40.6$ \citep{fischer90}. The viscosity can be expressed by
 
 \begin{equation}
 \label{viTb}
 \eta = \eta_0 \, \mathrm{exp} \left( \frac {E} {\mathcal{R} T} \right) \mathrm{exp} \left( -B \phi \right) \, ,
 \end{equation}
 
 \noindent where $\phi$ is the melt fraction which increases linearly with the temperature between $T_\mathrm{s}$ and $T_\mathrm{l}$ ($0 \leq \phi \leq 1$) and $B$ is the melt fraction coefficient ($10 \leq B \leq 40$) \citep{moore03}.
 
 At $T_\mathrm{b}$ the grains disaggregate, leading to a sudden drop in both the shear modulus and the viscosity. Above this temperature the shear modulus is set to a constant value: $\mu = 10^{-7} \mathrm{Pa}$. The viscosity follows the Roscoe-Einstein relationship so long as it reaches the liquidus temperature (where $\phi=1$) \citep{moore03}:
 
 \begin{equation}
 \label{viTl}
 \eta = 10^{-7} \mathrm{exp} \left( \frac {40000 \mathrm{K}} {T} \right) \left( 1.35 \phi - 0.35 \right)^{-5/2} \mathrm{Pa \, s} \, .
 \end{equation}
 
 Above $T_\mathrm{l}$ the shear modulus stays at $10^{-7} \mathrm{Pa}$, and the viscosity is described by \citep{moore03}
 
 \begin{equation}
 \label{viTll}
 \eta = 10^{-7} \mathrm{exp} \left( \frac {40000 \mathrm{K}} {T} \right) \mathrm{Pa \, s} \, .
 \end{equation}
 
 In our calculations rocky bodies are considered as satellites, and for this reason we follow the melting temperatures of \citet{henning09}, namely: $T_s = 1600$~K, $T_l = 2000$~K. We assume that disaggregation occurs at 50\% melt fraction, hence the breakdown temperature will be $T_b = 1800$~K.

 \subsection{Internal structure and convection}
 
 The structure of the moon in the model is the following: the body consists of an inner, homogeneous part, which is convective, and an outer, conductive layer. If the tidal forces are weak, the induced temperature will be low, resulting in a smaller convective region and a deeper conductive layer. But in case of strong tidal forces, the temperature will be higher, hence the convective zone will be larger with a thinner conductive layer.
 
 For calculating the convective heat loss, we use the iterative method described by \citet{henning09}. The convective heat flux can be obtained from
 
 \begin{equation}
 \label{qBL}
 q_\mathrm{BL} = k_\mathrm{therm} \frac {T_\mathrm{mantle} - T_\mathrm{surf}} {\delta(T)} \, ,
 \end{equation}
 
 \noindent where $k_\mathrm{therm}$ is the thermal conductivity ($\sim 2 \mathrm{W/mK}$), $T_\mathrm{mantle}$ and $T_\mathrm{surf}$ are the temperature in the mantle and on the surface, respectively, and $\delta(T)$ is the thickness of the conductive layer. We use $\delta(T)=30 \, \mathrm{km}$ as a first approximation, and then for the iteration
 
 \begin{equation}
 \label{delta}
 \delta(T) = \frac {d} {2 a_2} \left( \frac {Ra} {Ra_\mathrm{c}} \right)^{-1/4}
 \end{equation}
 
 \noindent is used, where $d$ is the mantle thickness ($\sim 3000$~km), $a_2$ is the flow geometry constant ($\sim 1$), $Ra_\mathrm{c}$ is the critical Rayleigh number ($\sim 1100$) and $Ra$ is the Rayleigh number which can be expressed by
 
 \begin{equation}
 \label{Ra}
 Ra = \frac { \alpha \, g \, \rho \, d^4 \, q_\mathrm{BL} } { \eta(T) \, \kappa \, k_\mathrm{therm} } \, .
 \end{equation}
 
 \noindent Here $\alpha$ is the thermal expansivity ($\sim 10^{-4}$) and $\kappa$ is the thermal diffusivity: $\kappa = k_\mathrm{therm} / ( \rho \, C_\mathrm{p} )$ with $C_\mathrm{p} = 1260 \, \mathrm{J/(kg \, K)}$. For detailed description see the clear explanation of \citet{henning09}.
 
 Because of the viscosity of the material the thickness of the boundary layer and convection in the underlying zone changes strongly with temperature. The weaker temperature dependencies of density and thermal expansivity are neglected in the calculations. The iteration of the convective heat flux lasts until the difference of the last two values is higher than $10^{-10} \mathrm{W/m^2}$.
 
 Calculations of tidal heat flux and convection are made for a fixed radius, density, eccentricity and orbital period of the moon. We assume that with time, the moon reaches the equilibrium state. \citet{henning09} showed that planets with significant tidal heating reach equilibrium with convection in a few million years. However, change in the eccentricity can shift, or destroy stable equilibria. After finding the stable equilibrium temperature, the tidal heat flux is calculated, from which the surface temperature can be obtained using the Stefan-Boltzmann law:
 
 \begin{equation}
 \label{Tsurf}
 T_\mathrm{surf} = \left( \frac{ \dot E_\mathrm{tidal} } { 4 \pi R_\mathrm{m}^2 \sigma } \right)^{1/4}\, ,
 \end{equation}
 
 \noindent where $\sigma$ is the Stefan-Boltzmann constant. This is the first time of using a viscoelastic model for obtaining the tidal heat induced surface temperature on exomoons.

  \subsection{Climate models} \label{climatemodel}

We adopt initial conditions essentially identical to those of \citet{forgan13} and \citet{forgan14}. The stellar mass is $M_\star = 1 M_\odot$, the mass of the host planet is $M_p = 1 M_\mathrm{Jup}$, and the mass of the moon is $M_\mathrm{s} = 1 M_\oplus$. This system has been demonstrated to be dynamically stable on time-scales comparable to the Solar System lifetime \citep{barnes02}.

The planet's orbit is given by its semi-major axis, $a_\mathrm{p}$, and eccentricity, $e_\mathrm{p}$ , and the moon's orbit by $a$ and $e$, respectively. We assume
that the planet resides at the barycentre of the moon-planet system,
which is satisfactory given the relatively large planet-to-moon mass
ratio. The inclination of the planet relative to the stellar equator is
$i_\mathrm{p} = 0$ (i.e. the planet revolves in the $x$ -- $y$ plane). The inclination of
the moon relative to the planet's equator, i.e. the inclination of the
moon relative to the $x$ -- $y$ plane, $i_\mathrm{m}$, is zero unless stated otherwise.
The orbital longitudes of the planet and moon are defined such that
they equal to zero corresponding to the $x$~axis. We also assume that the
moon's obliquity has been efficiently damped by tidal evolution,
and we therefore set it to zero.

The 1D latitude energy balance model employed in this work solves the following diffusion
equation:
 \begin{equation}
 \label{lebm}
 C \frac {\partial T} {\partial t} - \frac {\partial} {\partial x} \left( D \left( 1 - x^2 \right) \frac{\partial T} {\partial x} \right) = \left( S + S_\mathrm{p} \right) \left[ 1 - A(T) \right] + \zeta - I(T) \, ,
 \end{equation}

\noindent where $T = T(x, t)$ is the temperature at time $t, x \equiv \sin \lambda$, and $\lambda$ is
the latitude (between $-90^\circ$ and $90^\circ$). This equation is evolved with the boundary condition $\frac{\mathrm{d}T}{\mathrm{d}x} = 0$ at the poles. The ($1 - x^2$) term is a geometric factor, arising from solving the diffusion equation in spherical geometry.

$C$ is the atmospheric heat capacity, the diffusion coefficient $D$
controls latitudinal heat redistribution, $S$ and $S_\mathrm{p}$ are the stellar and
planetary insolation, respectively, $\zeta$ is the surface heating generated
by tides in the moon's interior, $I$ is the atmospheric infrared cooling
and $A$ is the albedo. $D$ and $I$ are calculated as described in \citet{forgan16}, 
and the surface heat flow can be calculated as \citep{scharf06}
 \begin{equation}
 \label{zeta}
\zeta = \frac{21} {38} \frac{\rho^2 R^5 e^2} {Q \mu} \left( \frac {G M_\mathrm{p}} {a^3} \right)^{5/2} \, .
 \end{equation}

The diffusion equation is solved using a simple explicit forward
time, centre space finite difference algorithm. A global time-step
was adopted, with constraint
 \begin{equation}
 \label{deltat}
 \delta t < \frac {\left( \Delta x \right)^2 C} {2 D \left( 1 - x^2 \right)} \, .
 \end{equation}

\noindent This time-step constraint ensures that the first term on the left-hand
side of equation \ref{lebm} is always larger than the second term, preventing
the diffusion term from setting up unphysical temperature gradients.
The parameters are diurnally averaged, i.e. a key assumption of the
model is that the moons rotate sufficiently quickly relative to their
orbital period around the primary insolation source. This is generally
true, as the star is the principal insolation source, and the moon
rotates relative to the star on time-scales of a few days.

The albedo function is
 \begin{equation}
 \label{AT}
 A(T) = 0.525 - 0.245 \tanh \left[ \frac {T - 268 \mathrm{K}} {5 \mathrm{K}} \right] \, .
 \end{equation}

\noindent This produces a rapid shift from low albedo ($\approx 0.3$) to high albedo
($\approx 0.75$) as the temperature drops below the freezing point of water,
producing highly reflective ice sheets. This shift in albedo affects the potential for global
energy balance, and for planets in circular orbits, two stable
climate solutions arise, one ice-free, and one ice-covered. \citet{spiegel08} show that such a function is sufficient to reproduce the
annual mean latitudinal temperature distribution on the Earth.
Note that we do not consider clouds in this model, which could
modify both the albedo and optical depth of the system significantly.
Also, we assume that both stellar and planetary flux are governed
by the same albedo, which in truth is not likely to be the case.

The stellar insolation flux $S$ is a function of both season and
latitude. The total diurnal insolation can be obtained from
 \begin{equation}
 \label{S}
 S = \frac {q_0} {\pi} \left( H \sin \lambda \sin \delta + \cos \lambda \cos \delta \sin H \right) \, ,
 \end{equation}

\noindent where $q_0$ is the bolometric flux received from the star at a distance of 1 AU, $H$ is the radian half-day length at a given $\lambda$ latitude and $\delta$ is the solar declination. The radian half-day length can be calculated from
 \begin{equation}
 \label{H}
 \cos H = - \tan \lambda \tan \delta \, ,
 \end{equation}

\noindent and the bolometric flux is given by
 \begin{equation}
 \label{q0}
 q_0 = 1.36 \cdot 10^6 \left( \frac {M_\star} {M_\odot} \right)^4 \mathrm{erg \, s^{-1} cm^{-2}} \, ,
 \end{equation}

We implement planetary illumination and eclipses of the moon in
the same manner as \citet{forgan14} \citep[see also][]{heller13}. While \citet{heller13} allow for the planet to be in synchronous
rotation and have a significant temperature difference between the
dayside and nightside (expressed in the free parameter $\mathrm{d}T_\mathrm{planet}$), we
assume the planetary orbit is not synchronous, and we fix $\mathrm{d}T_\mathrm{planet} = 0$.
The planetary albedo is fixed at 0.3.

The habitability function $\xi$ is
 \begin{equation}
 \label{kszi}
 \xi (t) = \left\{
        \begin{array}{ll}
            1 & \quad 273 \, \mathrm{K} < T(\lambda, t) < 373 \, \mathrm{K} \\
            0 & \quad \mathrm{otherwise.}
        \end{array}
    \right.
 \end{equation}

\noindent We then average this over latitude to calculate the fraction of habitable surface at any time-step:
\noindent and the bolometric flux is given by
 \begin{equation}
 \label{kszi-int}
 \xi (t) = \frac {1} {2} \int_{- \pi / 2}^{\pi / 2} \xi(\lambda, t) \cos \lambda \: \mathrm{d} \lambda \, .
 \end{equation}

Each simulation is allowed to evolve until it reaches a steady or
quasi-steady state, and the final ten years of climate data are used
to produce a time-averaged value of $\xi(t)$, $\bar\xi$, and the sample standard deviation, $\sigma_\xi$. We use these two parameters to classify each
simulations as follows.

\begin{enumerate}
 \item Habitable moons -- these moons possess a time-averaged $\bar\xi >
0.1$, and $\sigma_\xi < 0.1 \, \bar\xi$, i.e. the fluctuation in habitable surface is less
than 10 percent of the mean.
 \item Hot moons -- these moons have average temperatures above
373 K during all seasons, and are therefore conventionally uninhabitable, and $\bar\xi < 0.1$.
 \item Snowball moons -- these moons have undergone a snowball
transition to a state where the entire moon is frozen, and are therefore conventionally not habitable. As with hot moons, we require
$\bar\xi < 0.1$ for the moon to be classified as a snowball, but given the
nature of the snowball transition as it is modelled here, these worlds
typically have $\bar\xi = 0$.
 \item Transient moons -- these moons possess a time-averaged
$\bar\xi > 0.1$, and $\sigma_\xi > 0.1 \, \bar\xi$, i.e. the fluctuation in habitable surface is
larger than 10 percent of the mean.
\end{enumerate}

 \section{Albedo estimation} \label{albedoest}

In this section the method of albedo calculations from occultation light curve is presented. Because of the wide range of possible conditions (distance of the moon from its host planet and host star, orbital elements, size, albedo, etc.) only general and simple cases are used here. With this approach we get insight into the most important parameters without analysing all possibilities. The potential significance of parameters not used in this model are discussed in section \ref{diffpar}.

Three different configurations are possible just before and after any occultation:
\begin{enumerate}
\item the exomoon is behind the planet (both the planet and the moon are in the same line of sight),
\item the exomoon is in front of the planet (both the planet and the moon are in \ the same line of sight),
\item the exomoon and the planet are separately `visible' (although spatially not resolved).
\end{enumerate}

It can be generally assumed that the configuration does not change much in most cases during the occultation. The reason is that the time duration of the occultation varies in a couple of hours time scale, while the orbital period of the exomoon is assumed to be longer \citep{cabrera07}. A reverse scenario is unlikely, but also discussed by \citet{simon10} and \citet{sato10}.

\begin{figure*}
	\centering
	\includegraphics[width=14cm]{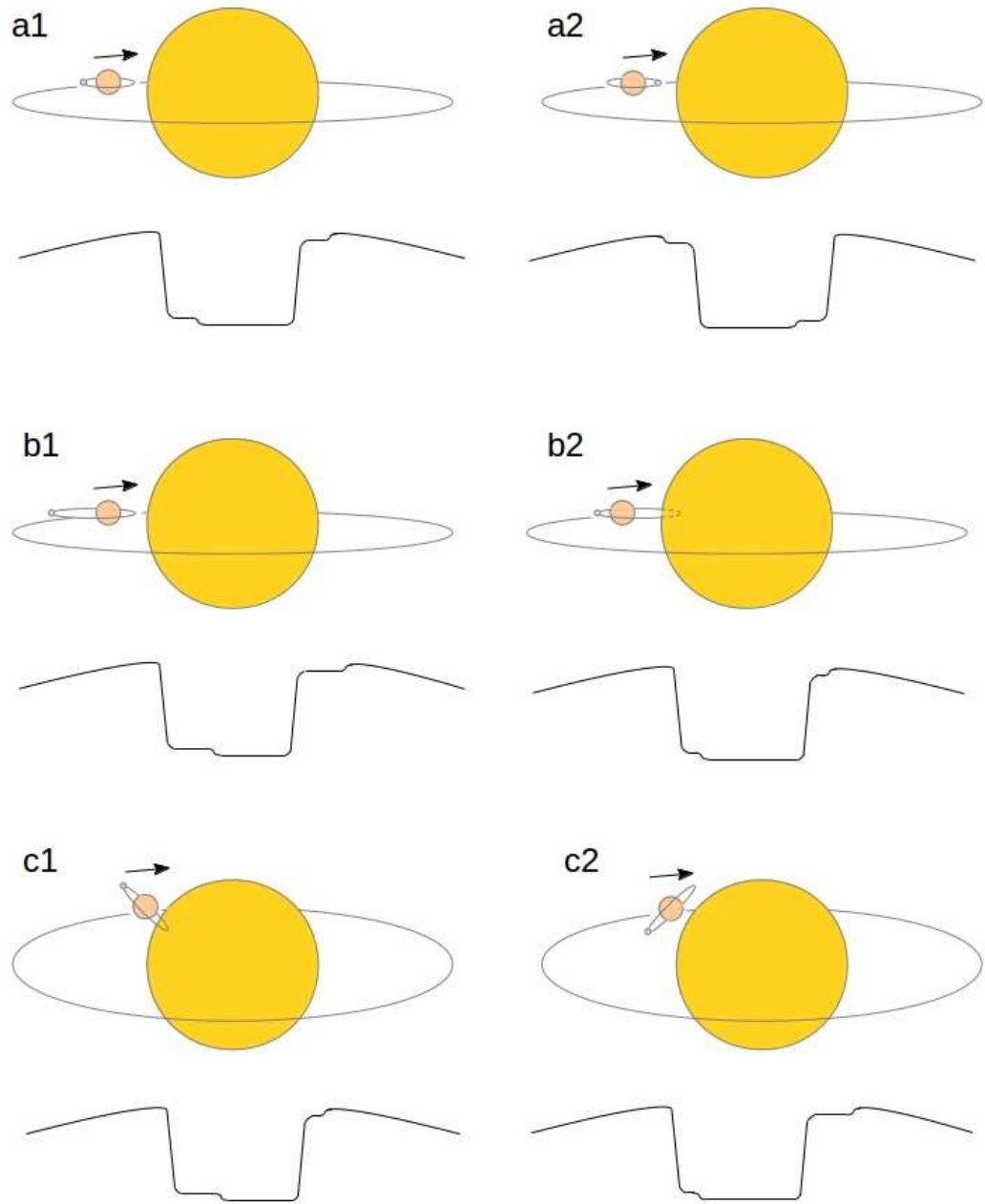}
	\caption{Schematic light curve during an occultation with an exomoon orbiting its exoplanet with some different basic conditions: a1 -- exomoon occults after the exoplanet, a2 -- exomoon occults before the exoplanet, b1 -- observationally favourable case of large eccentricity increasing the plateau duration, b2 -- unfavourable case of large eccentricity decreases the plateau duration, c1 -- favourable case of large inclination (relative to the plane perpendicular to the line of sight), c2 -- unfavourable case of large inclination. The sizes of the objects are not to scale.}
	\label{transit_types}
\end{figure*}

From the three possible configurations, the third one, which favours for the identification of the exomoon, since in the other cases the moon's presence does not change the shape of the light curve (although the depth of the flux minimum can be different). The third configuration is also the most probable of all, based on geometric considerations \citep{heller14a}. Apart from a few special cases, the moon spends just a small fraction of its orbital time in front of, or behind the planet \citep{heller12}. In this (third) case two sub-configurations are possible: the planet occults before the moon (a1 in Fig. \ref{transit_types}), or the moon occults before the planet (a2 in Fig. \ref{transit_types}). These cases can be seen in Figure \ref{transit_types} along with schematic light curves. Note that unlike at transits, stellar limb darkening does not affect the light curve at occultations.

\begin{figure}
	\centering
	\includegraphics[width=80mm]{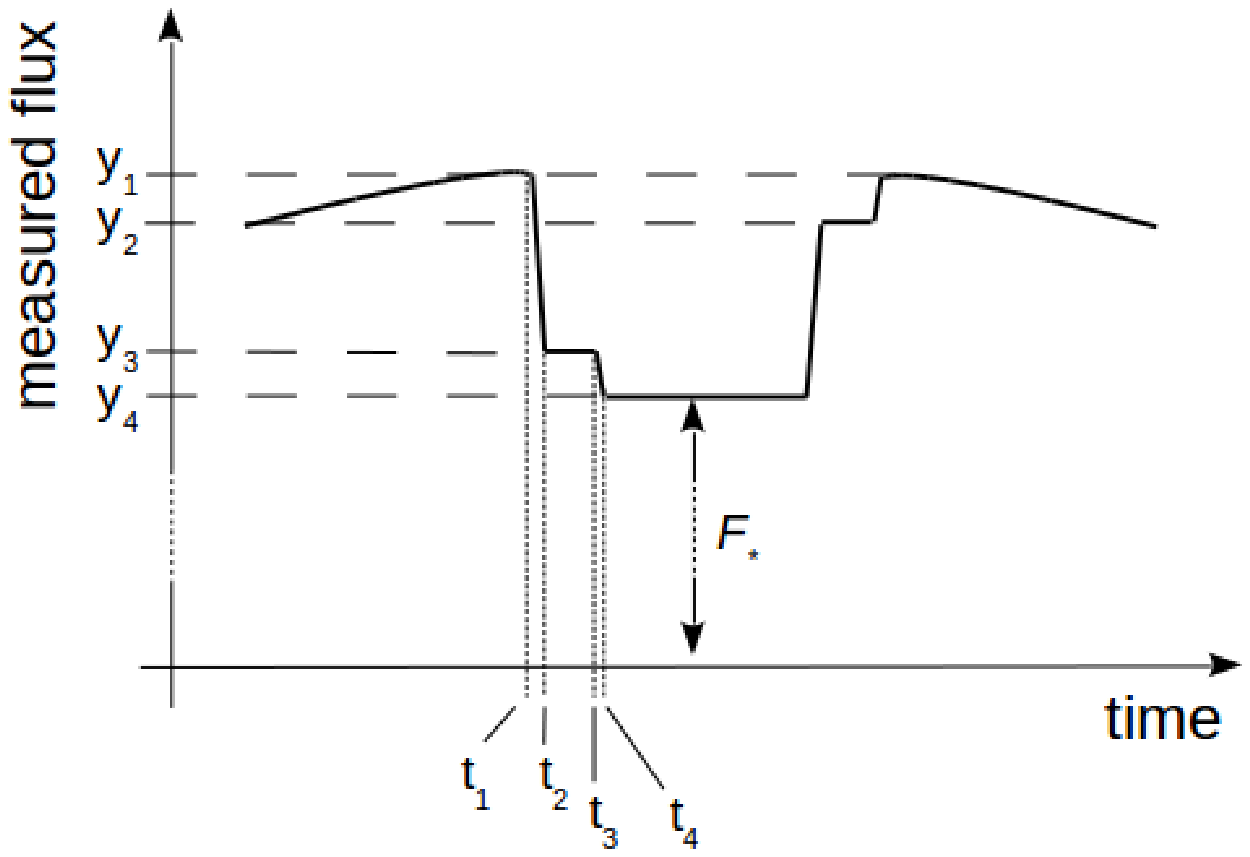}
	\caption{Schematic representation of the light curve \textup{(case a1 in Fig. \ref{transit_types})} during the occultation of an exoplanet with an exomoon.}
	\label{light_curve}
\end{figure}

Simple formulae are used for the calculations, because this is a basic phenomenon of exomoons that has not been analysed before. Simplified cases are considered because of the large number of possibilities and also to see the big picture without being lost in the details.

For the albedo calculation the following formulae and configuration are  considered. Fig. \ref{light_curve} shows a schematic light curve of an occultation, assuming that first the planet moves behind the star, followed by the moon (case a1 in Fig. \ref{transit_types}). The light curve has its maximum at $y_1$ and the minimum of the occultation is at $y_4$. The latter corresponds to the stellar flux (the planet and the moon are behind the star), denoted by $F_\star$ which is normalized so that $F_\star = 1$. Reflected stellar light from the planet and from the moon give contributions to the observed flux. Values of $y_3$ and $y_2$ refer to the measured flux before the occultation of the moon and after the occultation of the planet (the moon is still occulting), respectively. Measuring both $y_1 - y_2$ and $y_3 - y_4$ can be used to refine their value (which should be equal in theory), assuming that the apparent configuration of the system does not significantly change during the occultation. The accuracy of this value is very important, because it plays a key role in calculating the moon's albedo.

For the difference of the measured fluxes:
\begin{equation}
\label{fluxdiff}
	y_3 - y_4 = \frac{F_\mathrm{m}} {F_\star} \, ,
\end{equation}

\noindent where $F_\mathrm{m}/F_\star$ is the relative flux contribution of the moon normalized to the stellar flux. The flux contribution of the planet is irrelevant for our study. We call the $y_3 - y_4$ difference moon occultation (MO) depth. Adapting Eq. (2) from \citet{rowe08} to this case, the ratio of the fluxes can also be written as 
\begin{equation}
\label{fluxratio}
	\frac{F_\mathrm{m}} {F_\star} = A_\mathrm{g} \left( \frac{R_\mathrm{m}} {a_\mathrm{p}} \right)^2 \, ,
\end{equation}

\noindent where $A_\mathrm{g}$ is the geometric albedo, $R_\mathrm{m}$ is the radius and $a_\mathrm{p}$ is the distance of the exomoon from its host star, which is approximated by the semi-major axis of the planet. The radius of the moon is assumed to be already known from previous transit observations. The geometric albedo is the ratio of brightness at zero phase angle compared to an idealized, flat, fully reflecting, diffusely scattering disk. The Bond albedo on the other hand, is the reflectivity at all phase angles. The value of the Bond albedo is restricted between 0 and 1. The geometric albedo can occasionally be higher than 1, especially for atmosphereless bodies with icy surface (e. g. Enceladus, Tethys) because of the strong backscattering and opposition effect \citep{verbiscer07}. Because of the coherent backscatter opposition effect (CBOE), the phase difference of the light waves in the exact backscatter direction will be zero and the amplitudes add coherently as they interfere with each other, resulting in higher intensity up to a factor of two \citep{hapke12, verbiscer13}.

From Eqs. \ref{fluxdiff} and \ref{fluxratio} one can get:
\begin{equation}
\label{Ag}
	A_\mathrm{g} = \left( y_3 - y_4 \right) \left( \frac {a_\mathrm{p}} {R_\mathrm{m}} \right)^2 \, .
\end{equation}

The geometric albedo of an exomoon can be calculated by obtaining the MO depth from observations during occultation, and by using Eq. \ref{Ag}. However, very precise measurements are required, hence it is reasonable to consider the MO depth as an upper limit \citep{rowe06}. Several observations are needed for a phase-folded light curve that show the presence of the exomoon because of the photometric orbital sampling effect \citep{heller14a}.

 Using Eq. \ref{Ag} we calculated the MO depth values for different bodies, in order to investigate the required precision of future observations. These photometric measurements may lead to obtain the albedo of an exomoon with the use of the described method.

  \subsection{Calculation of `plateau duration' and `MO duration'}

In order to achieve the detection of small flux differences, sufficiently long detector integration time is required. An important limiting factor for detection is the duration while the exomoon is 'visible' alone and the exoplanet is behind the star (the time interval $t_3 - t_2$ in Fig. \ref{light_curve}), or when only the exoplanet is visible without the moon. We call this phase `plateau duration', because it causes a plateau in the light curve.

The plateau duration is calculated for ideal cases, when the exomoon is located at its maximum elongation from the exoplanet (where it has only radial velocity component). Circular orbits were assumed both for the planet and for the moon. This is a crude simplification, but in the case of captured or impact-ejected moons (like Triton around Neptune or the Moon around the Earth) tidal interactions can strongly reduce the satellite's eccentricity, almost circularizing the orbit. With this simplified model the plateau duration can easily be calculated by using the speed of the planet-moon pair around the central star:
\begin{equation}
	v_\mathrm{p} = \sqrt{ \frac{G \left( M_\star + M_\mathrm{p} \right) } {a_\mathrm{p}} } \, ,
\end{equation}

\noindent where $v_\mathrm{p}$ is the orbital speed of the planet, $G$ is the gravitational constant, $M_\star$ and $M_\mathrm{p}$ are the masses of the star and the planet, respectively. Since circular orbits are considered, $a_\mathrm{p}$ indicates the planet's orbital distance. In most cases the apparent location of the moon does not change significantly during the occultation. For this reason the plateau duration can be estimated from the orbital speed of the planet, if the orbital distance of the moon ($a$) is known (this is the distance that the moon needs to travel after the planet disappears behind the star). If the moon was not in its maximal elongation, then the distance (and hence the plateau duration) would be shorter.

However, the moon can be fast enough on its own orbit around the planet to influence the plateau duration. In case the moon passes 25\% of its orbit (starting from its maximal elongation) we terminate the plateau duration, because the moon must be in front of or behind the planet. These cases were taken into account as well. The orbital speed of the moon ($v_\mathrm{m}$) was calculated by
\begin{equation}
	v_\mathrm{m} = \sqrt{ \frac{G M_\mathrm{p}} {a} } \, .
\end{equation}

\noindent Both time intervals were calculated: $t_\mathrm{a} = a / v_\mathrm{p}$ and $t_{25} = 0.25 \left(2 a \pi \right) / v_\mathrm{m}$, and the shorter was considered as the plateau duration ($T_\mathrm{event}$) in each case.

The plateau duration was calculated for different mass host planets. It can be shown that having 1, 5, or 10 $M_\mathrm{J}$ mass planets ($M_\mathrm{J}$ denotes the mass of Jupiter) does not change the order of magnitude of the results. Table \ref{MJ} shows such cases for two moons. The stellar distance is 3 AU in each case, and a Sun-like star was used in the calculations. For the calculation, the radius of the planet was calculated from the mean density (1326 kg/m\textsuperscript{3}) and original radius (71,492 km) of Jupiter in each case. For the orbital distance of the two moons the real semi-major axes were used.

\begin{table*}
	\caption{Plateau duration for different planetary masses at 3 AU distance from a Sun-like star.}
    \label{MJ}
	\centering
	\begin{tabular}{c c c c c}
		\hline\hline
		 & & plateau duration & plateau duration & plateau duration \\
		Name & $a$ [km] & with a 1 $M_\mathrm{J}$ mass & with a 5 $M_\mathrm{J}$ mass & with a 10 $M_\mathrm{J}$ mass \\
		 & & planet [h] & planet [h] & planet [h] \\
		\hline
		Enceladus & 237,378 & 3.83 & 2.00 & 1.42 \\
		Europa & 670,900 & 10.83 & 9.52 & 6.74 \\
		\hline\hline
	\end{tabular}
\end{table*}

The plateau duration depends on many factors, including the exomoon's apparent distance from its host planet, its orbital velocity and the size of the host star. Larger stellar disk, lower velocity and larger planetary distance of the moon increase the length of the plateau duration. If the apparent planet-moon distance is larger than the diameter of the star, then the `plateau' will not appear in the light curve, because the planet and the moon will occult separately, making two separate drops in the light curve (i.e. the planet's occultation ends before the moon's occultation starts or the moon's occultation ends before the planet's one starts). In other words the occultations of the moon and the planet will not overlap in time. In this case the plateau duration is replaced by the duration of the moon's occultation. We call it `moon occultation duration', or MO duration for short.

In these cases, when the size of the star limits the MO duration, the following simple calculation was used. The velocity of the moon is already known from previous calculations, and the distance that the moon travels during the MO duration approximately equals to the diameter of the star. Since it is assumed that the impact parameter of the moon and the inclination of both the planet and the moon are zeros, the related time can be calculated from $t_\mathrm{MO} = 2 R_\star / v_\mathrm{p}$. If $t_\mathrm{MO} < t_\mathrm{a}$, then $t_\mathrm{MO}$ will be considered as the MO duration ($T_\mathrm{event}$).

\begin{figure}
	\centering
	\includegraphics[width=12cm]{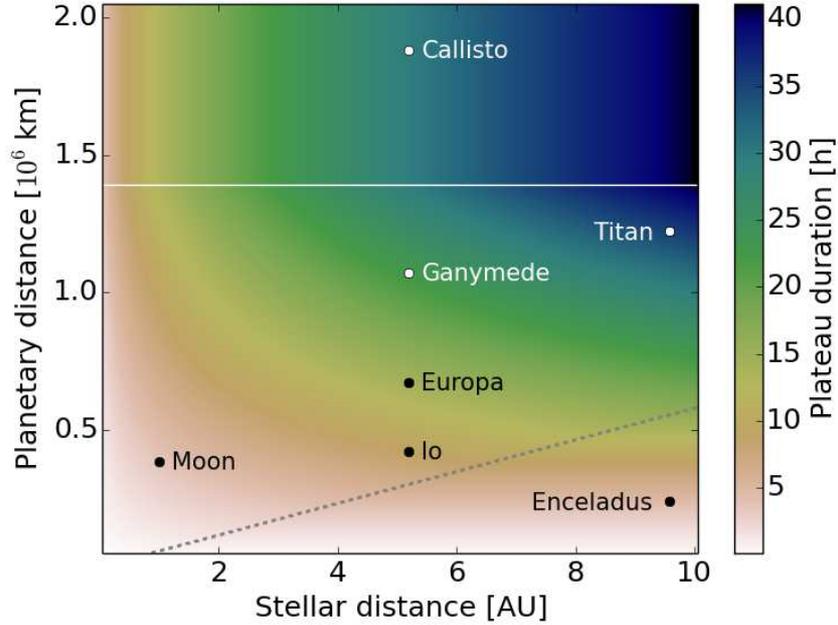}
	\caption{The so-called `plateau duration' for different stellar and planetary distances. Examples from the Solar System are plotted for comparison.}
	\label{plateau}
\end{figure}

The phase space of different stellar and planetary distances of a hypothetical moon was mapped. The plateau duration for different stellar and planetary distances can be seen in Fig. \ref{plateau}. The masses of the star and the planet are 1 solar mass and 1$M_\mathrm{J}$, respectively, for all cases. Black and white colours indicate the longest (41 hours) and the shortest (6 minutes) durations, respectively. Below the dashed grey line the plateau duration is terminated, because the moon reaches 25\% of its orbit. It mainly occurs for higher stellar distances, which produces lower orbital velocity around the star. Because of this lower velocity $t_{25}$ will be smaller than $t_\mathrm{a}$. Above the solid horizontal white line the plateau duration is constant for each stellar distance. The growing semi-major axis of the moon results in larger $t_\mathrm{a}$, but at about 1,393,000~km ($2 R_\star$) $t_\mathrm{a} = t_\mathrm{MO}$, or in other words, above this planetary distance the planet and the moon occults separately (the MO duration takes the place of the plateau duration). A few examples from the Solar System are plotted for comparison.

  \subsection{Calculating the photon noise level of different instruments} \label{calcphotonnoise}

It is assumed that the star emits perfect black-body radiation. This is 
a good assumption if we intend to estimate the photon noise level 
for a given instrument. The star is characterized by its radius $R_\star$ and 
$T_\star$, the effective  temperature of the stellar surface.
According to Planck's law, the power density on unit surface and 
unit solid angle as the function of the wavelength $\lambda$ can be written as
\begin{equation}
B_\lambda(\lambda){\rm d}\lambda=\frac{2hc^2}{\lambda^5}\frac{1}{e^{\lambda_\star/\lambda}-1}{\rm d}\lambda\label{eq:blambda}
\end{equation}

\noindent where
\begin{equation}
\lambda_\star=\frac{hc}{k_{\rm B}T_\star},
\end{equation}

\noindent which is proportional to the location of the peak of the spectral 
energy distribution. The constants are the following: $k_{\rm B}$ is the Boltzmann constant,
$h$ is the Planck constant and $c$ is the speed of light.
The integral over the stellar surface $(4\pi R_\star^2)$ as well as on the 
irradiated hemisphere according to Lambert's cosine law yields a total power 
output at a certain wavelength which can be computed as
\begin{equation}
P_\lambda(\lambda){\rm d}\lambda=4\pi^2 R_\star^2B_\lambda(\lambda){\rm d}\lambda.
\end{equation}

The number of photons emitted at a certain wavelength can then be computed
by dividing the power $P_\lambda(\lambda)$ by the photon energy $hc/\lambda$.
The total number of photons detected by a detector having a gross
quantum efficiency of $Q(\lambda)$ at unit time is then
\begin{equation}
\frac{{\rm d}n}{{\rm d}t}=
\frac{A}{4\pi d^2}
\int\limits_{\lambda=0}^\infty
Q(\lambda)P_\lambda(\lambda)\left(\frac{hc}{\lambda}\right)^{-1}{\rm d}\lambda
\end{equation}

\noindent Here $A$ is the area of the entrance aperture on the telescope 
(on which the detector is mounted) and $d$ is the distance to the star. 
In practical applications, $Q(\lambda)$ is the product of 
the transparencies of various filters, lens and reflective elements 
forming the telescope optics and the quantum efficiency of the 
detector itself. In order to simplify our calculations, $Q(\lambda)$ can be 
written as a characteristic function having a constant value of $Q$ within an interval
of $[\lambda_{\rm min},\lambda_{\rm max}]$ (and it is zero outside this interval).
Let us define $\Delta\lambda$ as the effective bandwidth of the filter as 
$\Delta\lambda=\lambda_{\rm max}-\lambda_{\rm min}$ and the central
wavelength as $\lambda_{\rm f}=(\lambda_{\rm min}+\lambda_{\rm max})/2$.

Using this assumption above, the total number of photons received at unit time
can be written as
\begin{equation}
\frac{{\rm d}n}{{\rm d}t}=
2\pi AQc\left(\frac{R_\star}{d}\right)^2
\int\limits_{\lambda=\lambda_{\rm min}}^{\lambda_{\rm max}}
\frac{1}{\lambda^4}\frac{1}{e^{\lambda_\star/\lambda}-1}{\rm d}\lambda
\end{equation}

\noindent By introducing the variable $x=\lambda_\star/\lambda$ and applying 
the respective measure of ${\rm d}\lambda=(\lambda_\star/x^2){\rm d}x$,
we can write
\begin{equation}
\frac{{\rm d}n}{{\rm d}t}=
2\pi A\left(\frac{R_\star}{d}\right)^2
c\left(\frac{k_{\rm B}T_\star}{hc}\right)^3
Q\int\limits_{x=\lambda_\star/\lambda_{\rm max}}^{\lambda_\star/\lambda_{\rm min}}
\frac{x^2}{e^x-1}{\rm d}x\label{eq:photoncountintegral}
\end{equation}

\noindent If the bandwidth of the filter is narrow, i.e. $\Delta\lambda\ll\lambda_{\rm f}$,
the integral above can further be simplified by assuming the
expression $x^2(e^x-1)^{-1}$ to be constant in the 
interval of $[\lambda_\star/\lambda_{\rm max},\lambda_\star/\lambda_{\rm min}]$.
In this case, $\Delta x=x_{\rm max}-x_{\rm min}$ can be approximated
as $\Delta x\approx\lambda_\star\Delta\lambda/\lambda_{\rm f}^2$ and hence
the photon flow is
\begin{equation}
\frac{{\rm d}n}{{\rm d}t}=
2\pi A\left(\frac{R_\star}{d}\right)^2
Qc\left(\frac{\Delta\lambda}{\lambda_{\rm f}}\right)\left(\frac{1}{\lambda_{\rm f}^3}\right)\frac{1}{e^{\lambda_\star/\lambda_{\rm f}}-1}\label{eq:pnoise}
\end{equation}

\noindent Once the photon flow, i.e. the number of photons received by the detector
is known, the photon noise of the instrument can be computed for any 
exposure time ($T$) as
\begin{equation}
\frac{\Delta n}{n}=\left(T\cdot\frac{{\rm d}n}{{\rm d}t}\right)^{-1/2}.
\end{equation}

If the filter bandwidth is comparable to the central wavelength,
the integral in Eq.~(\ref{eq:photoncountintegral}) can be computed numerically.
Since the integrand is a well-behaved function, even a few steps using
Simpson's rule can be an efficient numerical method for this computation.
We note that in the case of Kepler, TESS or PLATO 2.0, where
CCDs are used without any filters, the bandwidth of the quantum efficiency
curve is comparable to its central wavelength, so this numerical 
computation is more accurate. However, the error in the accuracy introduced
by this simplification is comparable to the discretisation of the 
$Q(\lambda)$ function (i.e. it is in the range of $10-20\%$ in total,
depending on the actual values of $\lambda_\star$, $\lambda_{\rm f}$ 
and $\Delta\lambda$).

In order to apply Eq.~(\ref{eq:pnoise}) in practice, it is worth converting
$R_\star$, $d$, $\lambda_{\rm f}$ and $\lambda_\star$ in astronomically relevant units 
(such as solar radius, parsec and microns) instead of SI units. After these
substitutions, we can write the approximation
\begin{equation}
\frac{{\rm d}n}{{\rm d}t}\cong
10^{12}\,{\rm Hz}
\left(\frac{A}{\rm m^2}\right)
\left(\frac{R_\star/R_\odot}{d/{\rm pc}}\right)^2
Q
\left(\frac{\Delta\lambda}{\lambda_{\rm f}}\right)
\frac{1}{(\lambda_{\rm f}/{\rm\mu m})^3}
\frac{1}{e^{\lambda_\star/\lambda_{\rm f}}-1}.\label{eq:pnoise2}
\end{equation}

\noindent This approximation is accurate within a few percents w.r.t.
Eq.~(\ref{eq:pnoise}). In the constant $10^{12}\,{\rm Hz}$ we included
both the dimension conversions (solar radius, parsec, microns) as well 
as the other constants such as $2\pi$ and the speed of light.

If we intend to detect a signal $S$ over the (cumulative) time $T$, 
the corresponding signal-to-noise ratio is going to be
\begin{equation}
  S/N = S\cdot\left(T\cdot\frac{{\rm d}n}{{\rm d}t}\right)^{1/2}.
\end{equation} 

\noindent The integration time is calculated as the product of the MO duration (or plateau duration) and the number of events (occultations): $T=T_\mathrm{event} \cdot N_\mathrm{event}$. For observatories, 30 events are assumed for the measurements, and for survey missions the number of events is calculated from the length of the mission campaign and the orbital period of the planet: $N_\mathrm{event}=T_\mathrm{campaign}/P_p$.

For all the calculations presented in this work, $d=50$ pc and $S/N=5$ is used, the latter corresponds to $5\,\sigma$ detection. In some cases (see Table \ref{instruments} for specific cases), instead of the $Q$ quantum efficiency, the system throughput was used, which contains $Q$ and other optical attenuations as well. In order to make these different cases comparable, either the system throughputs, or 80 percent of the quantum efficiency ($0.8Q$) was used in the calculations.


  \subsection{Orbital distances around M dwarfs}

Beside solar-like stars, I have applied the calculations to a range of small M dwarfs. The orbital distances of the planet and the moon are calculated as functions of the stellar mass. In the case of M dwarf host stars, the planet-moon pair is set to the snowline. Icy exomoons are expected to form beyond or around the so-called snowline, which is the distance where water ice condenses in the protoplanetary disk, supporting the formation of large mass planetesimals that finally evolve toward gravitationally collected gas giants \citep{ida08, qi13}.

The location of the snowline is calculated from the equilibrium temperature at the planet's sub-stellar point ($T_0$)  \citep{cowan11}: 
\begin{equation}
	a_\mathrm{p} = T_\star^2 \frac{R_\star} {T_0^2} \, .
	\label{snowline}
\end{equation}

\noindent $T_0$ is assumed to be approximately 230 K at the ice condensation boundary. To obtain $R_\star$ and $T_\star$ from the stellar mass ($M_\star$), a parabolic equation was fitted to the effective temperature and radius values given by \citet[Table 1]{kaltenegger09}. Only M4 -- M9 stars were used for the fitting. The obtained equations are:
\begin{equation}
	R_{\star} = -0.12 + 3.31 M_\star - 7.12 M_\star^2 \, ,
    \label{R_star}
\end{equation}
\begin{equation}
	T_{\star} = 1496 + 13301 M_\star - 26603 M_\star^2 \, ,
    \label{T_star}
\end{equation}

\noindent that can be used if $0.075 < M_\star < 0.2$.

The moon's orbital distance is set to 0.45 Hill radius ($R_\mathrm{H}$), hence it also depends on the stellar mass. It was calculated from the following formula:
\begin{equation}
	a = 0.45 \, R_\mathrm{H} = 0.45 \, a_\mathrm{p} \left( \frac{M_\mathrm{p}} {3 M_\star} \right)^{1/3} \, .
    \label{M_star}
\end{equation}

\noindent We chose this distance in order to avoid orbital instability, since \citet{domingos06} and \citet{donnison10} showed that beyond approximately half of the Hill radius, the orbit of a direct orbiting exomoon becomes unstable.

\chapter{Calculating the Liquid Water Habitable Zone} \label{calcHZ}
  
  In this work the liquid water habitable zone is investigated as a function of the
  mass of the main sequence star. Section \ref{habzone} gave an overview of different
  calculation methods for the boundaries of the habitable zone and for the
  effective stellar temperature and luminosity. These methods are compared to
  each other, discussed in this section, and new empirical formulae are proposed. The following description is based on the work of \citet{dobos13}.

  \section{Effective temperature -- mass relation} \label{T-M_rel}
    
\begin{figure}
    \centering
    \includegraphics[width=16cm]{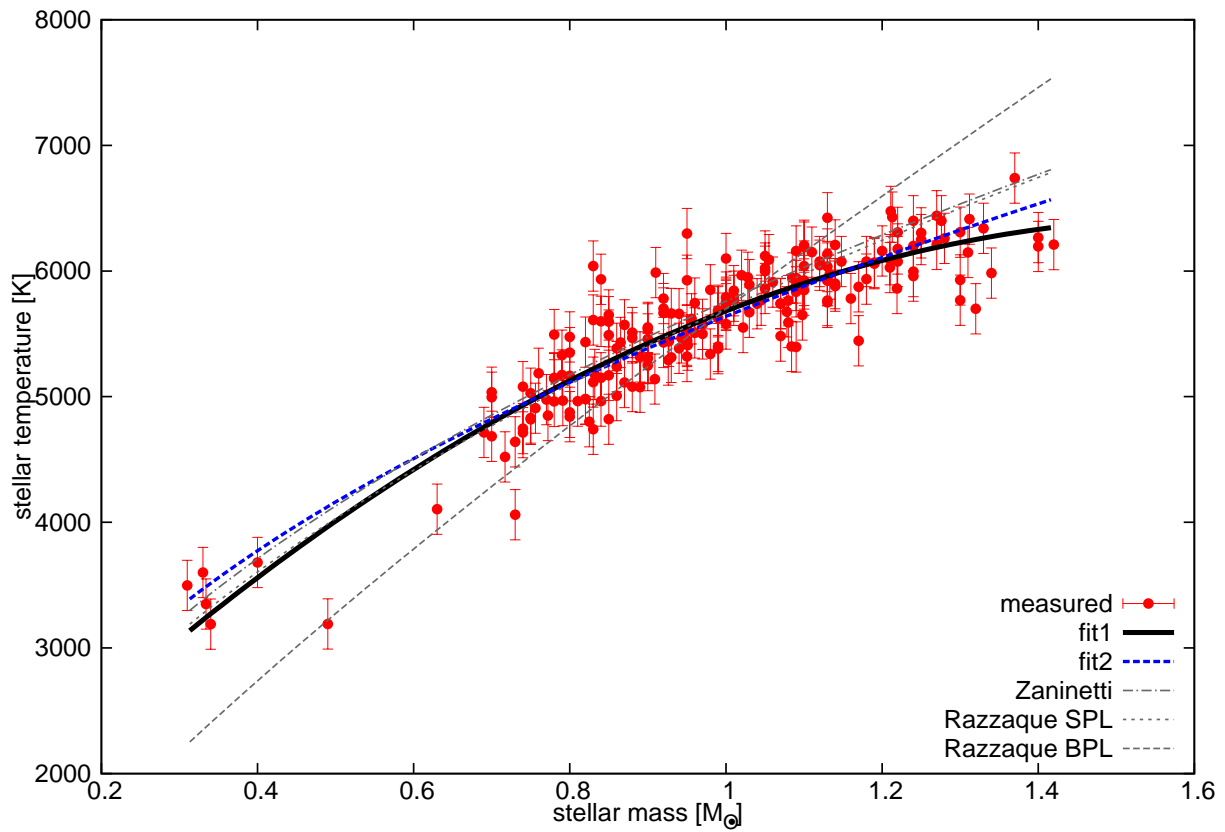}
    \caption{Effective temperature -- mass relation. Red dots with error bars indicate measured effective temperature values as a function of the stellar mass. The uncertainty in the measurements is considered to be $200$~K in each case. The curves represent calculated values: three previously defined models (\textit{Zaninetti, Razzaque SPL} and \textit{Razzaque BPL}) and two fitted equations (\textit{fit1} and \textit{fit2}) are plotted.}
    \label{T}
    \end{figure}

    The boundaries of the HZ strongly depend on the stellar temperature, but
    different calculation methods give different results.
    In Fig. \ref{T} the effective temperature can be seen as a function of the stellar
    mass. The red dots indicate measured parameters of main sequence stars with
    known planets \citep[data from exoplanet.eu, 2012 November,][]{schneider12}. The curves labelled as
    \textit{Zaninetti}, \textit{Razzaque SPL}, and \textit{Razzaque BPL} show the
    values calculated from Eqs.\,\ref{T_Zaninetti}, \ref{T_SPL}, and \ref{T_BPL},
    respectively. It can be clearly seen that the broken power
    law gives the worst estimation compared to the measured data points, because the
    calculated values are too low for lower mass stars (0.3 -- 0.7 $M_\odot$) and
    too high for higher mass stars (1.1 -- 1.4 $M_\odot$). The other two equations
    give much better results comparing to the data points, but still give a bit
    higher temperature for stars with masses of 1.2 -- 1.4 $M_\odot$.
    
    For this reason, two equations were fitted to the data points. The first equation
    (labelled as \textit{fit1} on Fig. \ref{T}) is a second order polynomial, and
    the other equation (\textit{fit2}) has a similar expression than the ones from
    the previous models (Eqs.\,\ref{T_Zaninetti}, \ref{T_SPL}, and
    \ref{T_BPL}). The fitted equations for \textit{fit1} and \textit{fit2} are 
    \begin{subequations}
      \begin{align}
        \label{fit1}
          f_1 = a + b \cdot M + c \cdot M^2 \, , \\
        \label{fit2}
          f_2 = p \cdot M^{q} \, ,
      \end{align}
    \end{subequations}
    \noindent respectively, where $f_i$ is $T_i(M)$ or $L_i(M)$, i.e. these equations are used for calculating both the stellar temperature and the luminosity. The fitted values
    can be seen in Table \ref{fit}. These equations can be used in the 0.3 -- 1.4 $M_\odot$ 
    interval. For higher mass values $T_1(M)$, and for smaller masses $T_2(M)$ have large 
    deviations.
    
    \begin{table}
     \centering
      \caption{Constants of the fitted equations for calculating the stellar 
    temperature ($T$) and the luminosity ($L$).}
       \label{fit}
        \begin{tabular}{lccccc}\hline\hline
          & \multicolumn{3}{c}{fit1} & \multicolumn{2}{c}{fit2}\\
          & $a$ & $b$ & $c$  & $p$ & $q$ \\ 
         \hline
         $T$ & $1379 \pm 232$ & $6219 \pm 494$ & $-1915 \pm 262$ & $5639 \pm 19$ & $0.438 \pm 0.016$\\
         $L$ & $0.465 \pm 0.102$ & $-2.345 \pm 0.409$ & $2.950 \pm 0.330$ & $1.050 \pm 0.045$ & $4.266 \pm 0.284$\\
         \hline\hline
        \end{tabular}
    \end{table}
    
    \begin{table}
     \centering
      \caption{Reduced chi-square ($\chi^2$) values for the five different models.
    The first column describes the name of the used method; the second, third, fourth and fifth columns show the $\chi^2$ values for the stellar temperature, luminosity, inner and outer boundaries of the HZ, respectively.}
       \label{RSS}
        \begin{tabular}{lcccc}\hline\hline
         Method & $\chi^2_T$ & $\chi^2_L$ & $\chi^2_{\mathrm{inner}}$ & $\chi^2_{\mathrm{outer}}$\\ 
         \hline
         Zaninetti & $2.11$ & $8.13$ & $9.33$ & $9.03$\\
         Razzaque SPL & $2.26$ & $7.58$ & $10.94$ & $10.40$\\
         Razzaque BPL & $6.48$ & $7.35$ & $10.53$ & $9.93$\\
         fit1 & $1.53$ & $8.01$ & $9.20$ & $8.85$\\
         fit2 & $1.66$ & $7.14$ & $9.20$ & $8.84$\\
         \hline\hline
        \end{tabular}
    \end{table}

	For calculating $\chi^2$, $200$~K was considered as uncertainty in the temperature measurements.
    The two fitted curves reproduce the measured parameters more precisely, than the curves of the
    other models. For numeric comparison, the reduced chi-square values were calculated for all five cases and the results can be seen in the
    second column of Table \ref{RSS}. From the three previously known models the
    \textit{Zaninetti} method, and from all cases the \textit{fit1} have the lowest $\chi^2$
    values. The \textit{fit2} model has nearly as low $\chi^2$ value as the \textit{fit1} model. The fitted $p$ and 
    $q$ parameters are in a good agreement with the parameters in Eqs.\,\ref{T_Zaninetti}, 
    \ref{T_SPL}, and \ref{T_BPL}, but have slightly lower values.

   \section{Luminosity -- mass relation} \label{L-M_rel}
    
    \begin{figure}
    \centering
    \includegraphics[width=16cm]{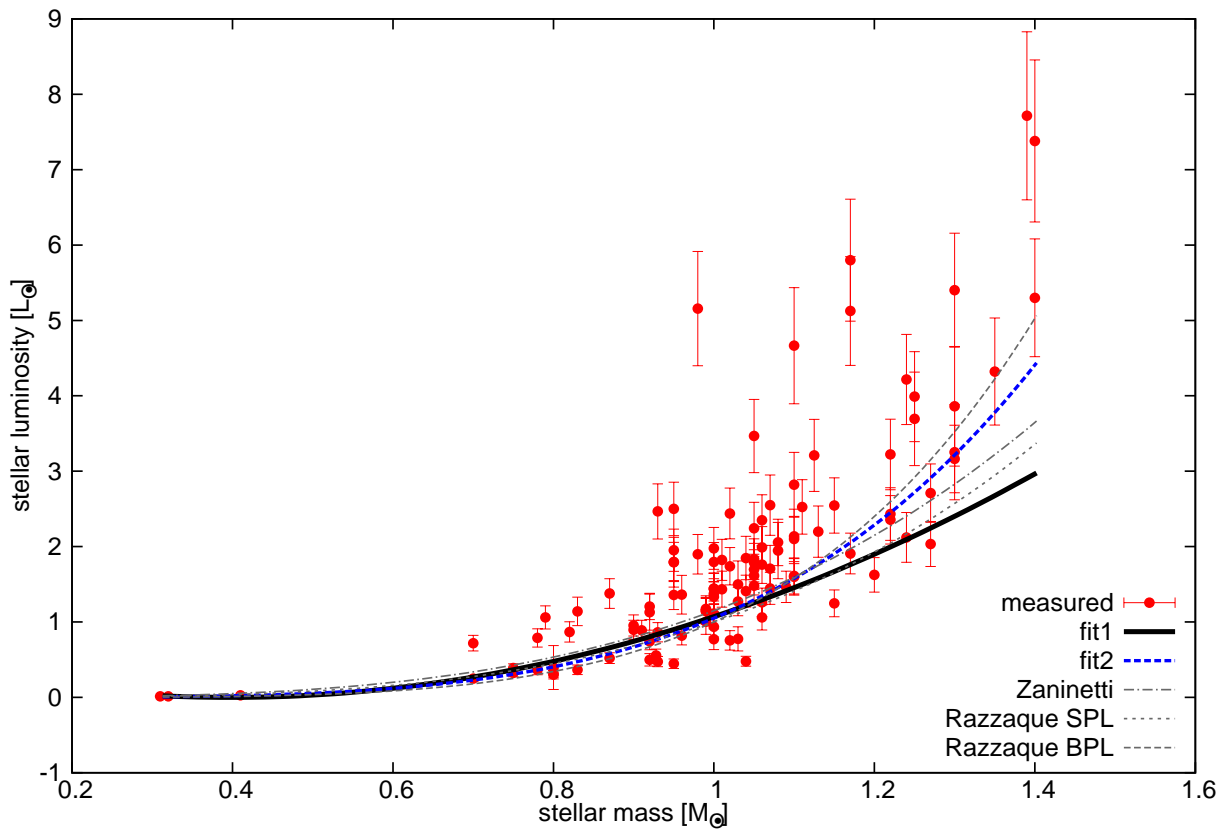}
    \caption{Luminosity -- mass relation. Red dots with error bars indicate luminosity values that were calculated from measured stellar parameters given by \citet[][Table 1]{jones06}. The curves represent the three previously defined models (\textit{Zaninetti, Razzaque SPL} and \textit{Razzaque BPL}) and the two fitted equations (\textit{fit1} and \textit{fit2}).}
    \label{L}
    \end{figure}

    Similar calculations were made for determining the stellar luminosity. All five 
    methods were used as described in Eqs.\,\ref{L_Zaninetti}, \ref{L_SPL}, 
    \ref{L_BPL}, \ref{fit1} and \ref{fit2}. The results are shown in Fig. 
    \ref{L}. Here the values represented by red dots are calculated from measured stellar parameters given by 
    \citet{jones06}. These parameters are the distance ($d$), the apparent visual brightness ($V$) and the bolometric correction ($BC$). The luminosity is calculated from the following expression \citep{jones06}
\begin{equation}
	L = 0.787 d^2 \cdot 10^{-0.4(V+BC)} \, .
    \label{L_Jones}
\end{equation}
    
When calculating the stellar luminosity values, the following uncertainties were considered: $d \pm 1 \mathrm{pc}$, $V \pm 0.1^\mathrm{m}$, $BC \pm 0.1^\mathrm{m}$. These uncertainties propagate into the luminosity and are shown in Fig.~\ref{L} as error bars. The same uncertentaintes were taken into account when fitting the luminosity functions with Eqs.\,\ref{fit1} and \ref{fit2}. If the uncertainties were not considered in the fitting, then the parameters would significantly differ from the ones that are shown in the second row of Table~\ref{fit}.
    
    The $\chi^2$ values were calculated again for all cases, and they show that 
    the formula of \textit{fit2} reproduces the stellar luminosity more precisely than the other formulae
    (see the $\chi^2_L$ column of Table \ref{RSS}). The value of $q$ in the 
    $L_2(M)$ equation is between the corresponding parameters of the \textit{SPL} and \textit{BPL} cases 
    (3.6 and 4.8, respectively), and like in the \textit{Zaninetti} 
    model, a multiplier ($p$) is also used, which improves the fitting. The \textit{fit1} formula, however, gives negative value for the stellar luminosity between approximately $0.38$ and $0.41\,M_\odot$, which obviously cannot be a good estimation for any star.

   \section{Habitable zone}
    
    For calculating the boundaries of the habitable zone, Eqs.\,\ref{r_inner} 
    and \ref{r_outer} were used. As the effective temperature and the luminosity 
    can be calculated from different equations, a comparison was made between the 
    different methods. The used equations for each model can be seen in 
    Table \ref{HZ}. The first column contains the names of the models, and the 
    second, third and fourth columns show the number of the used equations for 
    calculating the effective temperature, the stellar luminosity and the boundaries of 
    the HZ, respectively. As the $L_1(M)$ equation gave negative values for the luminosity for some stellar masses, $L_2(M)$ (Eq.\,\ref{fit2}) is 
    used instead.
    
    \begin{table}
     \centering
      \caption{Number of used equations for calculating the effective temperature ($T$), the
    luminosity ($L$) and the boundaries of the HZ.}
       \label{HZ}
        \begin{tabular}{lccc}\hline\hline
         Method & $T$ & $L$ & HZ\\ 
         \hline
         Zaninetti & \ref{T_Zaninetti} & \ref{L_Zaninetti} &
    \ref{r_inner}, \ref{r_outer}\\
         Razzaque SPL &\ref{T_SPL} & \ref{L_SPL} & \ref{r_inner},
    \ref{r_outer}\\
         Razzaque BPL & \ref{T_BPL} & \ref{L_BPL} & \ref{r_inner},
    \ref{r_outer}\\
         fit1 & \ref{fit1} & \ref{fit2} & \ref{r_inner},
    \ref{r_outer}\\
         fit2 & \ref{fit2} & \ref{fit2} & \ref{r_inner},
    \ref{r_outer}\\
         \hline\hline
        \end{tabular}
    \end{table}
    
\begin{figure}
    \centering
    \includegraphics[width=14.5cm]{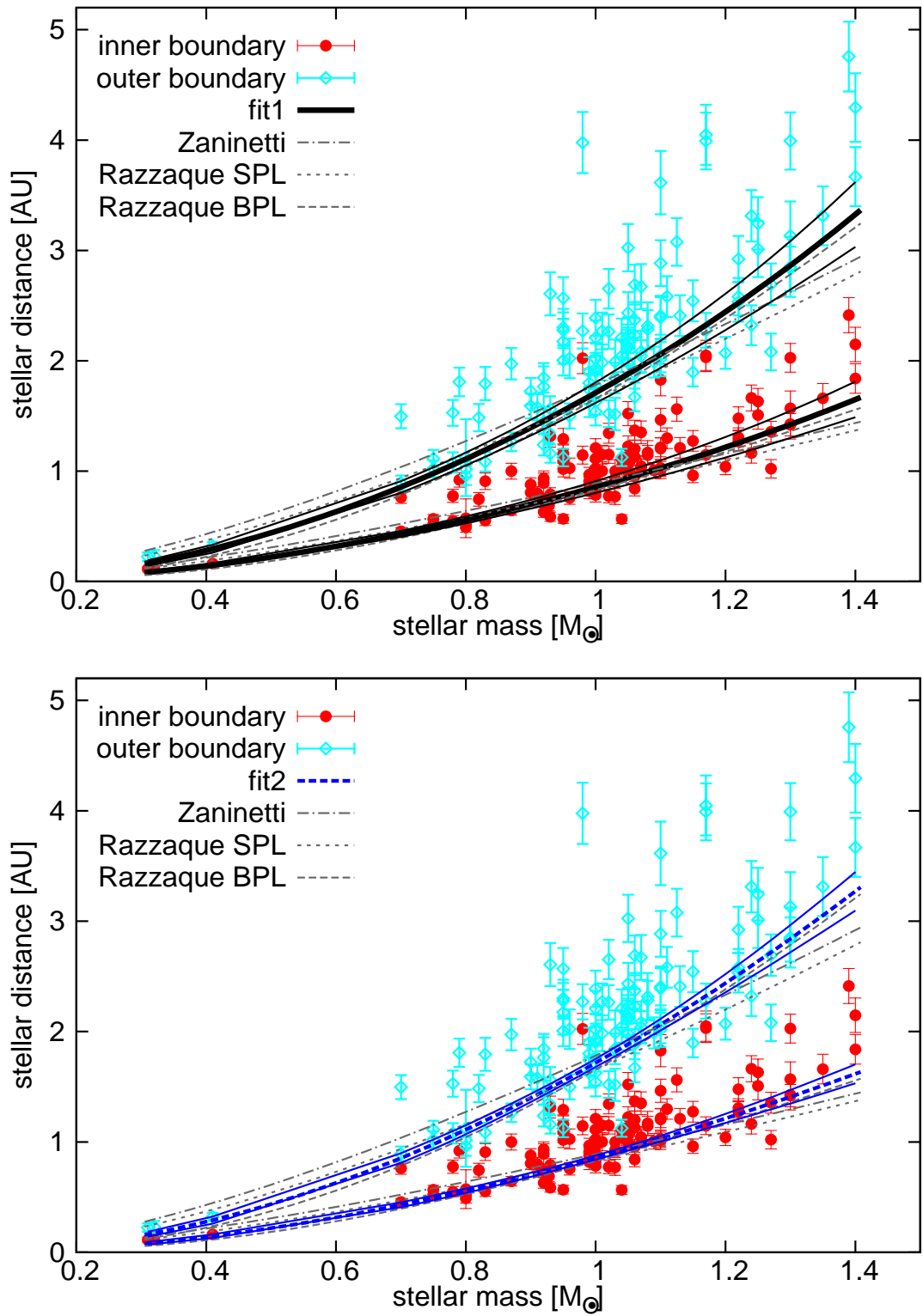}
    \caption{Boundaries of the habitable zone as a function of the stellar mass.
    Red dots and light blue diamonds indicate the inner and outer boundaries, respectively, as given by \citet{jones06}. The curves show the calculated values using the equations given in Table \ref{HZ}. The top panel shows the \textit{fit1}, and the bottom panel shows the \textit{fit2} models along with the other three models.}
    \label{LWHZ}
    \end{figure}

    The results of the calculations are shown in Fig.~\ref{LWHZ} and in the last two 
    columns of Table~\ref{RSS}. In Fig.~\ref{LWHZ} the red dots
    indicate the location of the inner, and the light blue diamonds indicate the outer
    boundary of the HZ, calculated by \citet{jones06} using 
    Eq.\,\ref{S_J_inner} and \ref{S_J_outer}. These values
    were calculated from measured stellar parameters, hence we consider them as good
    references for different calculation models. The error bars shown are the propagated uncertainties originating from the measurements of the distance, stellar temperature, visual brightness and bolometric correction. The curves in Fig.~\ref{LWHZ} indicate the different models as described 
    in Table~\ref{HZ}. (The names of the methods in Fig.~\ref{LWHZ} are the same as 
    in the first column of the table.) With these calculation methods both the inner and 
    the outer boundaries were calculated for the same stellar masses that are 
    presented by the data points (red dots and light blue diamonds). The \textit{fit1} and \textit{fit2} curves shown in the top and bottom panels, respectively, are nearly identical.

The solid curves surrounding the \textit{fit1} and the \textit{fit2} curves indicate the uncertainties (variance) of the fitted models. They were estimated for each stellar mass from the following expression:
\begin{equation}
	\Delta r_i \approx \left[ \left( \sqrt{ \frac{L_i + \Delta L_i} {S_i} } - r_i \right)^2 + \left( \sqrt{ \frac{L_i} {S_i + \Delta S_i} } - r_i \right)^2 \right]^{1/2} \, ,
    \label{uncert}
\end{equation}

\noindent where $\Delta L$ and $\Delta S$ are the estimated uncertainties of the luminosity and the stellar flux, respectively. The values of $\Delta L$ and $\Delta S$ were calculated with the same method (but with their own equations, as shown in Table~\ref{HZ}), using the uncertainties of the parameters presented in Table~\ref{fit}.

It can be clearly seen in Fig.~\ref{LWHZ} that the uncertainty of the fitted models is larger for more massive stars, than for smaller stars. It is probably caused by the larger deviation and uncertainty of the luminosity observed for more massive stars. The uncertainty region at 1.4 stellar masses is wider in the case of the \textit{fit1} model (about 0.32 and 0.58~AU at the inner and the outer boundaries, respectively), than for the \textit{fit2} model (about 0.17 and 0.35~AU, respectively).
    
    From the reduced chi-square values (see the last two columns of Table \ref{RSS}) it can be seen that the \textit{fit1} and \textit{fit2} models provide nearly the same precision 
    and they have the lowest $\chi^2$ values among the different models. It means that these two models give the closest results to the calculation made by \citet{jones06} in the 0.3 -- 1.4 $M_\odot$ interval.

    \section{Conclusion}
    
    Different calculation methods of the liquid water habitable zone were compared
    to each other. 
    Three previously defined and two new fitted models were used for the calculations. The first 
    fitted equation (\textit{fit1}) is a parabolic function, and the other (\textit{fit2}) 
    has a similar form than the other models. The 
    equations of the \textit{fit2} model are:
    
    \begin{subequations}
      \begin{align}
        \label{f2T}
          T = 5640 \cdot M^{0.44} \, , \\
        \label{f2L}
          L = 1.05 \cdot M^{4.27} \, .
      \end{align}
    \end{subequations}
    
    These models were used for calculating the effective temperature and stellar 
    luminosity. From the three already known models the one called \textit{Zaninetti} gave the 
    best results for the effective temperature, and for the luminosity the 
    \textit{Razzaque BPL} model gave the best reproduction to measured stellar parametes. The \textit{fit2} model gave even better results for both the stellar temperature and luminosity values. Using the \textit{fit2} method, better 
    estimations can be obtained for the boundaries of the habitable zone of F, G, K and M class (0.3 -- 1.4 
    $M_\odot$) main sequence stars.

Since the observed deviation of stellar parameters is large for similar stellar masses, it is possible, that even more precise models could be obtained, if another parameter than the stellar mass (for example the effective temperature) would be considered in the fitting, as well.

\chapter{Viscoelastic Mod\-els of Tid\-ally Heat\-ed Exomoons} \label{viscoelastictidal}

 The idea of a circumplanetary, tidally-heated habitable zone has emerged and was investigated by several authors \citep[e.g.][]{reynolds87, scharf06, heller13}. For the first time, a viscoelastic model was applied for studying tidal heat in exomoons. This work aims to give a detailed study of the circumplanetary Tidal Temperate Zone (TTZ), and discusses the differences to other models. For a complex investigation of habitability on exomoons, the viscoelastic model was also coupled to a climate model. The work described in sections \ref{viscmodel} and \ref{comparison} are based on the paper of \citet{dobos15}, section \ref{climatemod} is based on the work of \citet{forgan16} and section \ref{tidalconclusion} presents the conclusions of both works.

 \section{Viscoelastic model} \label{viscmodel}
  
 The satellite's surface temperature is calculated as described in section \ref{tidalheat}, for different orbital periods and radii, at a fixed density and eccentricity. Stellar radiation and other heat sources are not considered, and have been neglected. The orbital period and the radius of the moon vary between 2 and 20 days, and between 250~km and 6550~km, respectively. It is common to consider Earth-mass moons in extra-Solar Systems when speaking of habitability, however, their existence is not proven. In the Solar System the largest moon, Ganymede has only 0.025 Earth mass. But the mass of satellite systems is proportional to the mass of their host planet. \citet{canup06} showed that this might be the case for extra-solar satellite systems as well, giving an upper limit for the mass ratio at around $10^{-4}$. This means, that 10 Jupiter-mass planets may have 0.3 Earth-mass satellites. Besides accretion, large moons can also form from collisions, as in the case of the Earth's Moon, and in such cases the mass ratio of the moon and the planet can be even larger than $10^{-4}$. Other possibility is the capturing of terrestrial-sized bodies through a close planetary encounter, as described by \citet{williams13}. For these reasons, we also take Earth-like moons into account.
 
\begin{figure}
 	\centering   
 	\includegraphics[width=13cm]{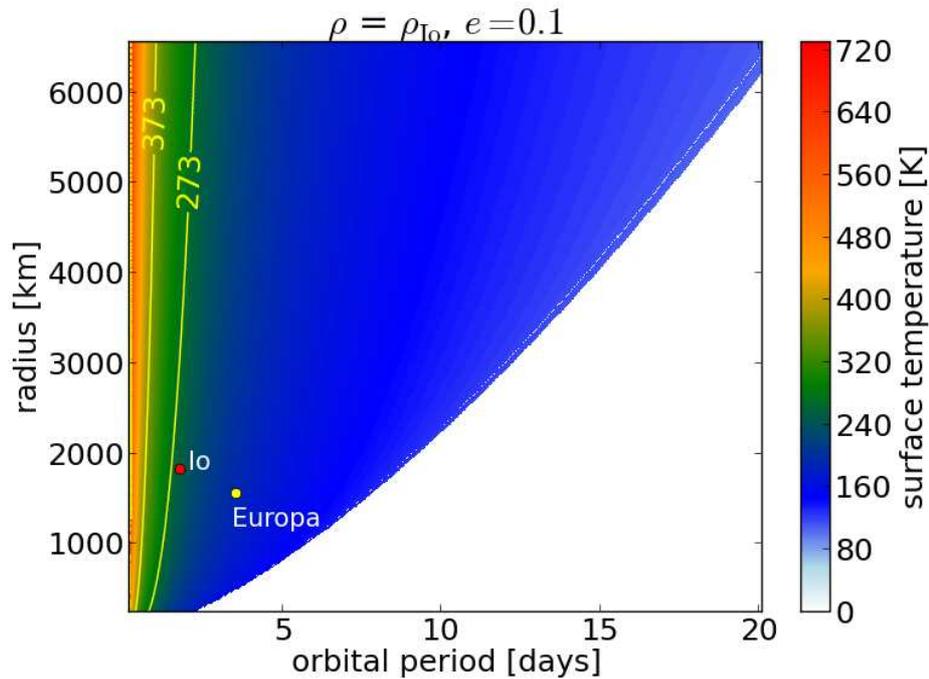}     
 	\caption{\label{e01}Tidal heat induced surface temperature for moons with similar density to Io at 0.1 orbital eccentricity. The yellow curves at 273 and 373 K indicate the boundaries of the TTZ. The orbital period and radius of Io and Europa are plotted for comparison (however their surface temperature differs from the values shown here, since their orbital eccentricites are 0.004 and 0.009, respectively).}
 \end{figure}

 The results of the calculations can be seen in Fig. \ref{e01}, where the density of the moon is that of Io, and its eccentricity is set to 0.1. Different colours indicate different surface temperatures. In the white region there is no stable equilibrium between tidal heat and convective cooling. In other words, tidal heat is not strong enough to induce convection. For comparison, a few Solar System moons are plotted that have similar densities to Io's. Yellow contour curves denote 0 and $100\,^\circ$C. The green area between these curves indicates that water may be liquid on the surface of the moon (atmospheric considerations were not applied). We define this territory as the Tidal Temperate Zone.
 
 Interestingly, the location of the TTZ strongly depends on the orbital period, and less on the radius of the moon. Low radii are less relevant, since smaller bodies are less capable of maintaining significant atmospheres.
 
\begin{figure}
 	\centering   
 	\includegraphics[width=13cm]{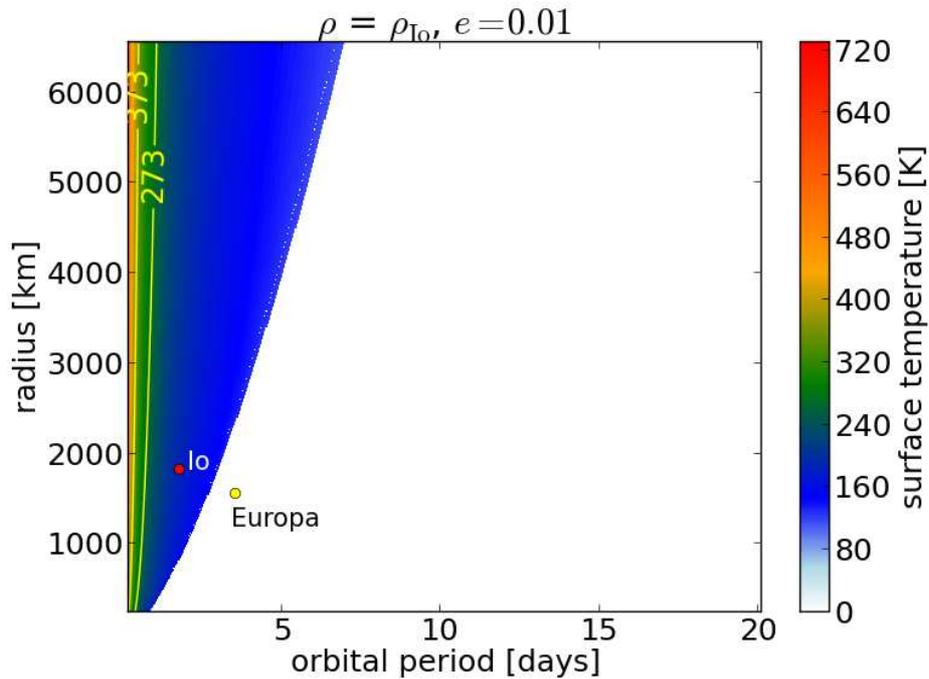}     
 	\caption{\label{e001}Tidal heat induced surface temperature for moons with similar density to Io at 0.01 orbital eccentricity. The yellow curves at 273 and 373 K indicate the boundaries of the TTZ. The orbital period and radius of Io and Europa are plotted for comparison (however their surface temperature differs from the values shown here, since their orbital eccentricites are 0.004 and 0.009, respectively).}
 \end{figure}

 The dependency on the eccentricity can be seen by comparing Figs. \ref{e01} and \ref{e001}. In the case of the latter figure the moon's eccentricity is 0.01. For most of the {\it orbital period--radius} pairs there is no solution (white area). Due to this drastic difference, Europa analogues get out of equilibrium for smaller eccentricities, and the TTZ becomes narrower and shifts to shorter orbital periods. Note, that radiogenic heat is not considered in the model, which could push back the moon into equilibrium state, and would result in higher surface temperature.
 
\begin{figure}
 	\centering $
 	\begin{array}{c}
 	\includegraphics[width=13cm]{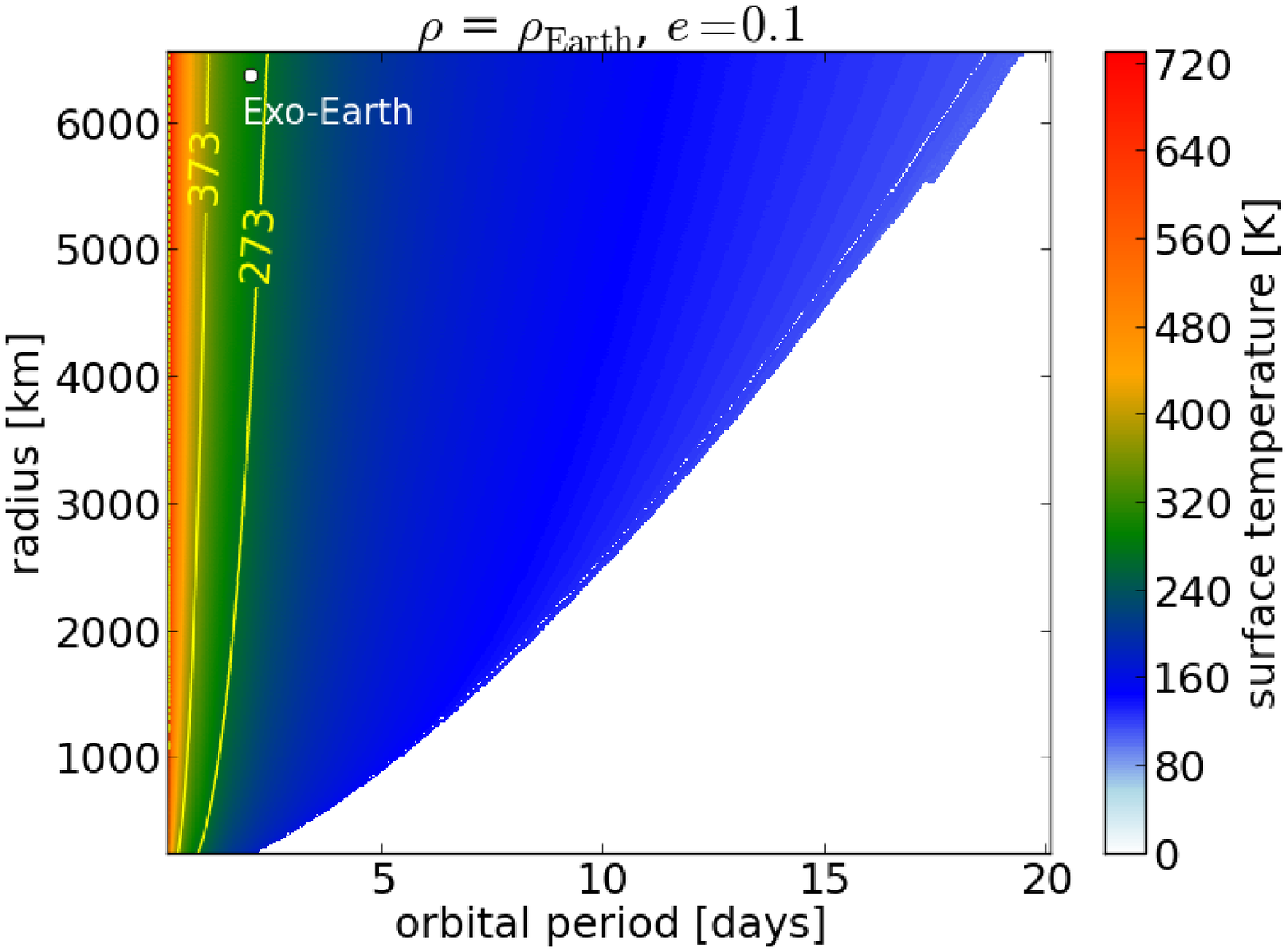} \\
    \includegraphics[width=13cm]{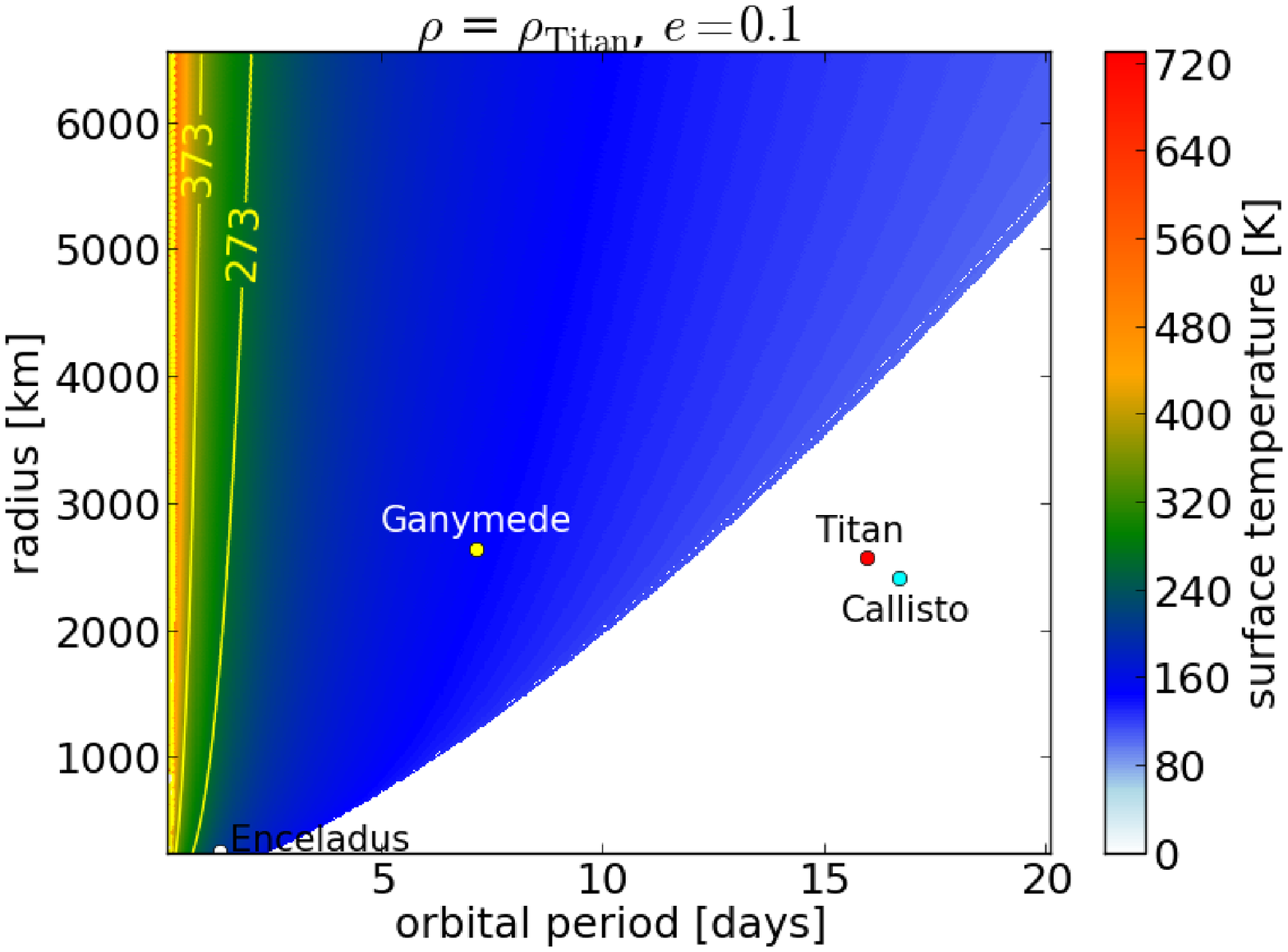} 
 	\end{array}$
 	\caption{\label{EarthTitan}Tidal heat induced surface temperature for moons with similar density to the Earth (top panel) and Titan (bottom panel). The yellow curves at 273 and 373 K indicate the boundaries of the TTZ. Exo-Earth indicates an imaginary satellite with the radius and mean surface temperature of the Earth. The orbital period and radius of Enceladus, Ganymede, Titan and Callisto are plotted for comparison (however their surface temperature differs from the values shown here).}
 \end{figure}

 Similar calculations were made for the density of the Earth and Titan (top and bottom panels of Fig. \ref{EarthTitan}, respectively). The densities do not have high influence on the tidally induced surface temperature, however, the TTZ slightly shifts to lower orbital parameters for higher densities. (The density of Earth, Io and Titan are $5515 \, \mathrm{kg/m^3}$, $3528 \, \mathrm{kg/m^3}$ and $1880 \, \mathrm{kg/m^3}$, respectively.)
 
 In the left panel of Fig. \ref{EarthTitan} an example Earth-like moon is plotted inside the TTZ. This hypothetical body has the same mean surface temperature (288~K), radius (6370~km) and density as the Earth, hence its orbital period is 2.06~days. In the right panel a few Solar System satellites are plotted that have similar densities to that of Titan.
 
 The stellar flux for moons with ambient temperatures of $\sim$100~K (which is similar to the case of the Galilean and Saturnian moons in the Solar System) is about one percent of the tidal flux in the TTZ. For this reason, stellar insolation may be safely ignored if the planet-moon system orbits the star at a far distance, or if they are free-floating. In systems in which the stellar irradiation alone is sufficient to heat the surface to levels of order the melting temperature or higher, the models presented here would need to be replaced by more complex hybrid ones to take into account both sources of heat and their very different spatial distributions on and within the moon.

 \section{Comparison to the fixed \textit{Q} model} \label{comparison}

 \subsection{Method}
 
 It is clear from the results, that the viscoelastic model does not give solution in the case of small tidal forces. In other words, the amount of heat that is produced by tidal interactions is insufficient to induce convective movements inside the body, and for this reason there is no equilibrium between them. In contrast, the fixed $Q$ model provides solution in these cases, as well. However, the viscoelastic model describes the tidal heating of the body more realistically than the fixed $Q$ model, due to the temperature dependency of the $Q$ and $\mu$ parameters. How are the results of the two models related to each other?
 
 For comparing the results of the two kinds of models, we use the expression of Eq.\,7 from \citet{peters13} for the fixed $Q$ calculation:
 
 \begin{equation}
 \label{temp}
 T_\mathrm{surf} = \left( \left( \frac {392 \pi^5 G^5} {9747 \sigma^2} \right)^{1/2} \left( \frac {R_\mathrm{m}^5 \rho^{9/2}} {\mu Q} \right) \left( \frac {e^2} {\beta^{15/2}} \right) \right)^{1/4} \, ,
 \end{equation}
 
 \noindent where $T_\mathrm{surf}$ is the surface temperature of the moon induced by tidal heating, $G$ is the gravitational constant, $\sigma$ is the Stefan-Boltzmann constant, $R_\mathrm{m}$ is the radius, $\rho$ is the density, $\mu$ is the elastic rigidity, and $Q$ is the dissipation function of the moon, $e$ is the eccentricity of the moon's orbit, and $\beta$ is expressed with the semi-major axis ($a$) and the mass of the planet ($M_\mathrm{p}$):
 \begin{equation}
 \label{a}
 a = \beta a_R = \beta \left( \frac {3 M_\mathrm{p}} {2 \pi \rho} \right)^{1/3} \, ,
 \end{equation}
 
 \noindent where $a_R$ is the Roche radius of the host planet. These equations can be used for calculating the surface temperature of the moon heated solely by tidal forces.
 
 The satellite's mean motion can be expressed from $\beta$ by
 
 \begin{equation}
 \label{n}
 n = \sqrt {\frac {2 \pi G} {3} \frac {\rho} {\beta^3} } \, ,
 \end{equation}
 
 \noindent which makes the comparison of the two models easier.

 \subsection{Surface temperature}
 
 \begin{figure}
 	\centering   
 	\includegraphics[width=12cm]{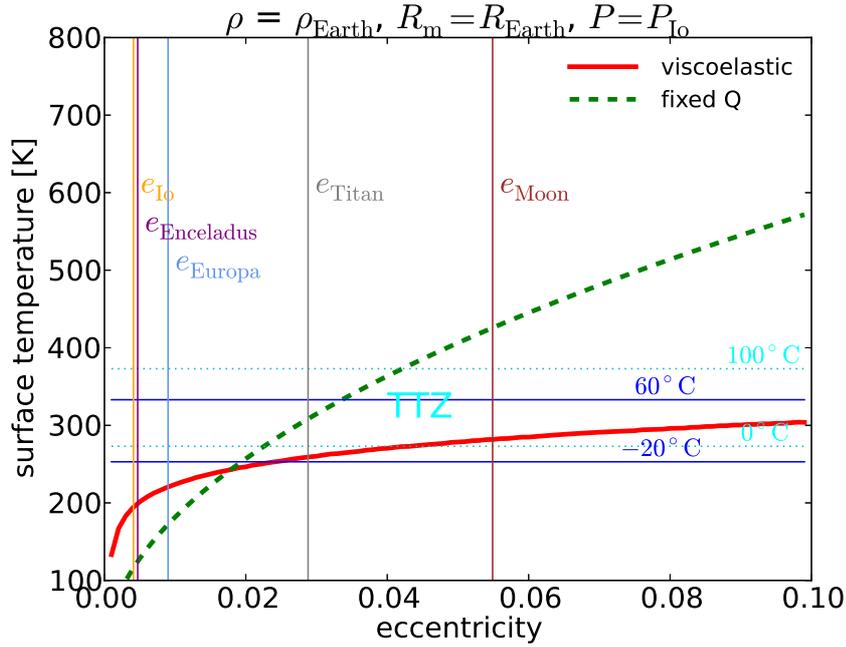}     
 	\caption{\label{T-e}Tidally induced surface temperature of the satellite as a function of its eccentricity.}
 \end{figure}
 
\begin{figure}
 	\centering   
 	\includegraphics[width=12cm]{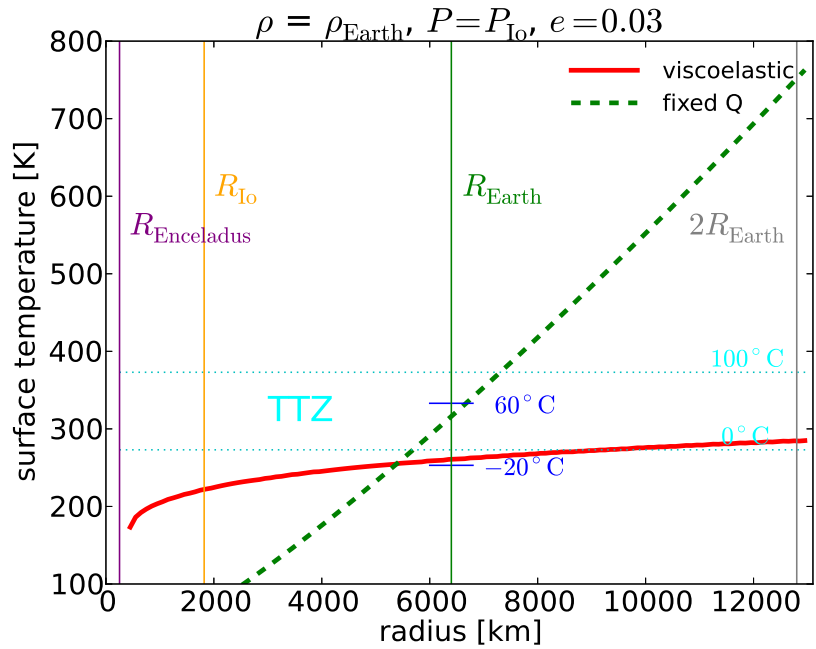}     
 	\caption{\label{T-R}Tidally induced surface temperature of the satellite as a function of its radius.}
 \end{figure}

\begin{figure}
 	\centering   
 	\includegraphics[width=12cm]{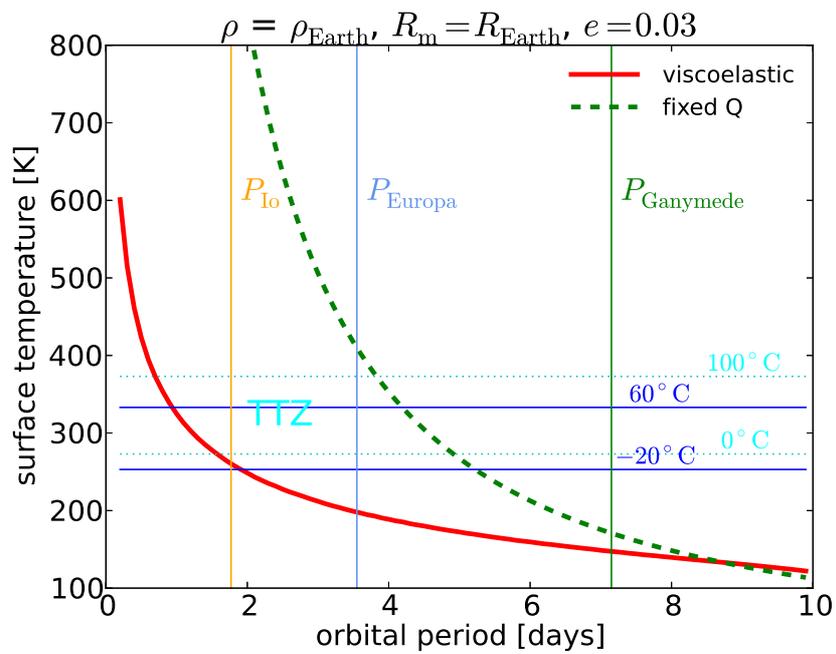}     
 	\caption{\label{T-P}Tidally induced surface temperature of the satellite as a function of its orbital period.}
 \end{figure}

 For comparison of the fixed $Q$ and the viscoelastic model, see Figs. \ref{T-e}, \ref{T-R} and \ref{T-P}, which show the surface temperature of a moon, calculated with both the viscoelastic (red solid curve) and the fixed $Q$ model (green dashed curve) as functions of the eccentricity, radius and orbital period of the satellite. For the density of the moon we used 5515~$\mathrm{kg/m^3}$, which is the density of the Earth, and for the fixed $Q$ model we used $Q=280$ and $\mu = 12 \cdot 10^{10} \, \mathrm{kg / ( m \, s^2 ) }$ in each case \citep[Table 1]{peters13}. The radius, orbital period and eccentricity of the satellite are set to that of the Earth, Io and 0.03, respectively, except that one of these parameters is varied in each figure (horizontal axes). The horizontal light blue, dashed lines indicate 0 and 100~$^\circ$C (making the boundaries of the TTZ), and the solid blue lines denote the minimum and maximum temperatures ($-20$ and $60\,^\circ$C) that are probable limits of habitability on an Earth-like body \citep[][Chapter 4]{sullivan07}. In salty solutions the lower limit for microbial activity is around $-20\,^\circ$C, and the upper limit for complex eukaryotic life is $60\,^\circ$C. The latter temperature is also about the runaway greenhouse limit for Earth. These limits are only used for Earth-like bodies ($\rho = \rho_\mathrm{Earth}$ and $R_\mathrm{m} \approx R_\mathrm{Earth}$). Vertical lines show a few examples from the Solar System for different eccentricities, radii and orbital periods.
 
 It is noticeable that the red curve is less steep than the green one, and larger portion of it is located inside the TTZ, especially in Figs.~\ref{T-e} and \ref{T-R}. It shows that the viscoelastic model stabilizes the surface temperature comparing to the fixed $Q$ model. These are just a few examples indicating that the viscoelastic model is less sensitive to these parameters, and that there are huge differences in the results of the models. In the next section the volume of the TTZ is investigated more thoroughly.

 \subsection{Occurrence rate of `habitable' moons}
 
 \begin{figure}
 	\centering   
 	\includegraphics[width=12cm]{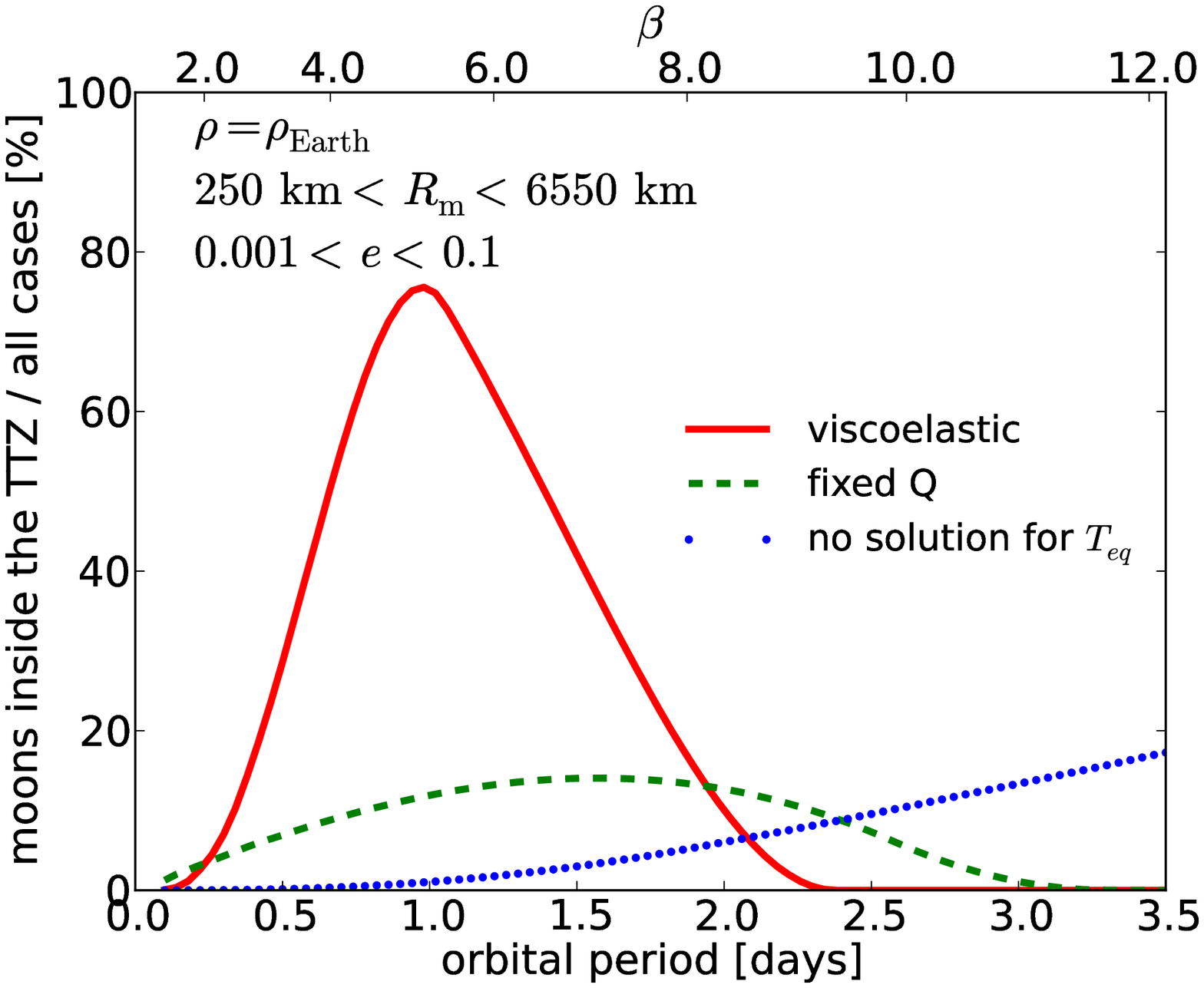}     
 	\caption{\label{beta}Percentage of cases that give surface temperatures between 0 and 100~$^\circ$C. Solid red and dashed green curves represent the results of the viscoelastic and the fixed $Q$ model, respectively. The density of the moon is that of the Earth in each case. The dotted blue curve shows the ratio in percentage of those cases where the viscoelastic model did not have solution for the equilibrium temperature.}
 \end{figure}
 
 Habitability on extraterrestrial bodies is an exciting, but complex question. Here we consider solely the tidally induced surface temperature of a hypothetical moon. We were curious about the occurrence rate of moons with suitable surface temperature for life. For this reason we mapped the phase space evenly with hypothetical moons that have different radii (between 250 and 6550~km) and eccentricities (between 0.001 and 0.1), and their densities are that of the Earth. We used both the viscoelastic and the fixed $Q$ model for calculating the surface temperature of these bodies, and then calculated the percentage of those that have suitable surface temperature, i.e. that are located inside the TTZ ($0 \leq T_\mathrm{surf} \leq 100~^\circ$C). The calculation was made for different orbital periods between 0.1 to 3.5~days, and for each value there were 63100 hypothetical moons distributed in the radius-eccentricity phase space. The result can be seen in Fig. \ref{beta}. Red solid and green dashed curves indicate the percentage of being inside the TTZ for the viscoelastic and for the fixed $Q$ model, respectively. The blue dotted curve shows the percentage of those cases that do not give result for the viscoelastic model. The top axis shows the $\beta$ parameter, which is the ratio of the moon's semi-major axis and the planet's Roche radius. It can be clearly seen that the red and green curves have a peak, which means that the probability of having suitable surface temperature has a maximum at a certain orbital period. The viscoelastic model predicts a much more efficient heating than the fixed $Q$ model, i.e. a much larger fraction of the hypothetical moons have their surface temperature between 0 and 100~$^\circ$C. The ratio of the integral under the red to that under the green curve is 2.8, meaning that 2.8 times more exomoons are predicted in the TTZ with the viscoelastic model. For the viscoelastic model the maximum percentage appears around 1~day orbital period, and here the probability for the moon of being inside the TTZ is almost 80\%. For higher orbital periods, this probability rapidly falls down, which is in contrast with the fixed $Q$ model. The latter has its peak around 1.5~days, and it has less than 20\% chance for satellites being in the TTZ. Despite the high probabilities achieved by the viscoelastic model for small orbital periods, the fixed $Q$ model give more promising results for those moons that have their orbital periods at 2~days or more.
 
 \begin{figure}
 	\centering   
 	\includegraphics[width=12cm]{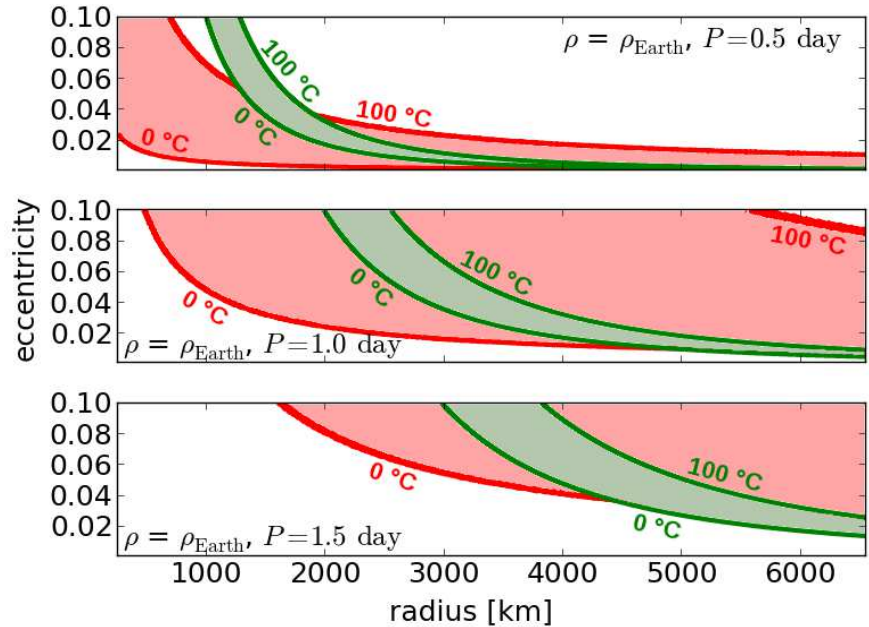}     
 	\caption{\label{slice}Temperature contours for the two kinds of models. Red colour: viscoelastic model, green: fixed $Q$ model. Top panel: orbital period $P = 0.5$~day, middle: $P = 1$~day, bottom: $P = 1.5$~days.}
 \end{figure}
 
 For detailed study, the 0 and 100~$^\circ$C temperature contours were plotted in the radius-eccentricity plane for a few, specific orbital periods, namely $P = 0.5$~day (top panel), $P = 1$~day (middle panel) and $P = 1.5$~days (bottom panel) (see Fig. \ref{slice}). Again, red and green colours represent the viscoelastic and the fixed $Q$ model, respectively. Between the contour curves the region of the TTZ is filled with light red and light green colours. The result shows that the viscoelastic model mostly favours the small moons, especially at high eccentricities, but also some large moons at small eccentricities over the fixed $Q$ model. This suggests that the viscoelastic model is less sensitive to the parameters of the moon, and holds the temperature more steady than the fixed $Q$ model. This is due to the melting of the inner material of the moon that leads to a less elevated temperature, as discussed by \citet{peters13}. On the other hand, the lower temperature implies that the total irradiated flux of the moon will be also lower, hence making the detection of the moon more difficult.
 
 \begin{figure}
 	\centering   
 	\includegraphics[width=12cm]{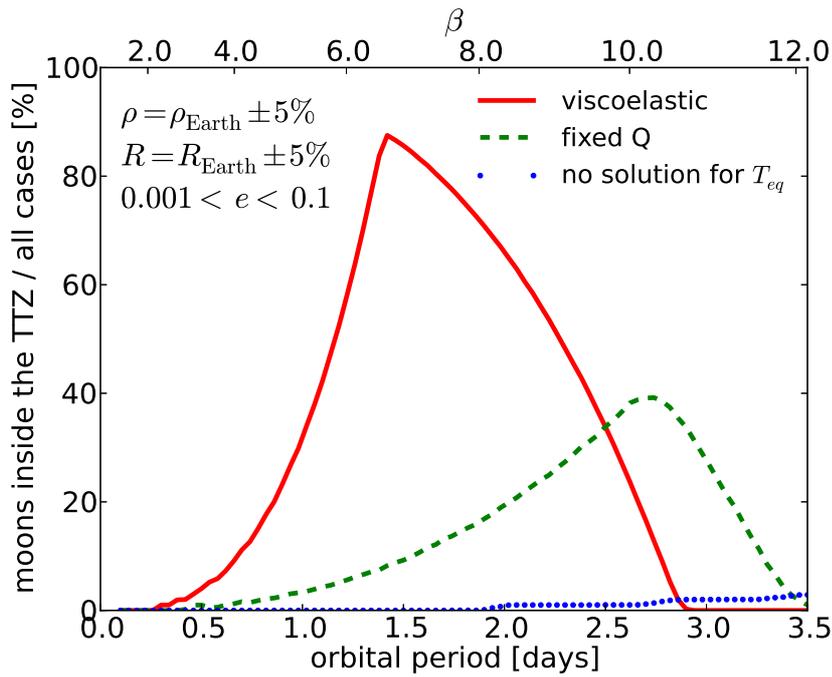}     
 	\caption{\label{betaEarth}Percentage of cases that give surface temperatures between $-20$ and 60~$^\circ$C. Solid red and dashed green curves represent the results of the viscoelastic and the fixed $Q$ model, respectively. The density of the moon is that of the Earth in each case. The dotted blue curve shows the ratio in percentage of those cases where the viscoelastic model did not have solution for the equilibrium temperature.}
 \end{figure}
 
 We were also interested in `Earth-twins' as satellites and in the probability of their `habitability'. For this reason we made similar calculations, but the radius and the density of the hypothetical moons were set to be close to that of the Earth: $R_\mathrm{m} = 6378\, \mathrm{km} \, (\pm 5\%)$ and  $\rho = 5515\, \mathrm{kg/m^3} \, (\pm 5\%)$. The radius and density values were chosen randomly from these intervals. The eccentricity was altered similarly as in the previous case (uniformly between 0.001 and 0.1). Altogether 200000 cases were considered for each orbital period. The temperature limits were set to $-20$ and 60~$^\circ$C, which are the probable limits for life on Earth. Fig. \ref{betaEarth} shows the results of this calculation. Note, that the peaks of the red solid (viscoelastic model) and the green dashed (fixed $Q$ model) curves are shifted to higher orbital periods, comparing to Fig. \ref{beta}. This is in part caused by the changed temperature limits, and in part by the much shorter radius range. The maximum probabilities are also higher, that is especially visible in the case of the fixed $Q$ model, which reaches more than 40\% at the curve's peak (in the previous case it was less that 20\%). As one would expect, it suggests that larger moons are more probable of maintaining warm surfaces. The ratio of the areas under the red and the green curves is 2.3.
 
 In general, it can be concluded that the viscoelastic model is not just more realistic than the fixed $Q$ model, but also gives more promising results for exomoons, since much larger fraction of the hypothetical satellites have been found in the TTZ. In those cases when the viscoelastic model does not give solution for the equilibrium temperature, one can use the fixed $Q$ model instead, however, the values of $Q$ and $\mu$ are highly uncertain.

 \subsection{The value of \textit{Q\textmu}} \label{qmuvalue}
 
 With the product of $Q$ and $\mu$, one can easily calculate the tidally induced surface temperature of a moon without using a complex viscoelastic model. Using Eq.\,\ref{temp} is a fast way to obtain $T_\mathrm{surf}$, but a good approximation is needed for the $Q \mu$ value. For such calculations, in the following, we give the $Q \mu$ values for hypothetical moons. Because of the large number of possible variations in the physical and orbital parameters of the moons, only a few, Solar System-like bodies were considered. Since the $Q \mu$ varies several orders of magnitudes for different rocky bodies, a good estimation can serve almost as well as the exact value. The following examples can be used as a guideline for making such estimations. Note, that the used model can be applied to rocky bodies, but for icy satellites, such as Enceladus or Europa, the results may be misleading, because of the more complex structure and different behaviour of icy material.
 
 \begin{table*}
    \centering
 	\begin{tabular}{lcccccc}\hline\hline
 		label & $R_\mathrm{m} \, \mathrm{[km]}$ & $\rho \, \mathrm{[kg/m^3]}$ & $e$ & $P \, \mathrm{[days]}$ & $T_\mathrm{surf} \, \mathrm{[K]}$ & $\mathrm{log_{10}}(Q \mu)$ [Pa] \\
 		\hline
 		\multirow{6}{*}{Earth-like} & \multirow{6}{*}{6378} & \multirow{6}{*}{5515} & \multirow{3}{*}{0.01} & 1 & 281 & 14.0 \\
 		& & & & 2 & 213 & 13.0 \\
 		& & & & 3 & 180 & 12.4 \\
 		\cline{4-7}
 		& & & \multirow{3}{*}{0.1} & 1 & 378 & 15.5 \\
 		& & & & 2 & 291 & 14.4 \\
 		& & & & 3 & 249 & 13.8 \\
 		\hline
 		\multirow{6}{*}{Mars-like} & \multirow{6}{*}{3394} & \multirow{6}{*}{3933} & \multirow{3}{*}{0.01} & 1 & 256 & 12.5 \\
 		& & & & 2 & 194 & 11.5 \\
 		& & & & 3 & 163 & 10.9 \\
 		\cline{4-7}
 		& & & \multirow{3}{*}{0.1} & 1 & 342 & 14.0 \\
 		& & & & 2 & 263 & 13.0 \\
 		& & & & 3 & 225 & 12.4 \\
 		\hline
 		\multirow{6}{*}{Moon-like} & \multirow{6}{*}{1738} & \multirow{6}{*}{3342} & \multirow{3}{*}{0.01} & 1 & 232 & 11.1 \\
 		& & & & 2 & 176 & 10.1 \\
 		& & & & 3 & n.a. & n.a. \\
 		\cline{4-7}
 		& & & \multirow{3}{*}{0.1} & 1 & 313 & 12.6 \\
 		& & & & 2 & 240 & 11.5 \\
 		& & & & 3 & 205 & 10.9 \\
 		\hline
 		\multirow{6}{*}{Io-like} & \multirow{6}{*}{1821} & \multirow{6}{*}{3532} & \multirow{3}{*}{0.01} & 1 & 235 & 11.2 \\
 		& & & & 2 & 177 & 10.2 \\
 		& & & & 3 & n.a. & n.a. \\
 		\cline{4-7}
 		& & & \multirow{3}{*}{0.1} & 1 & 315 & 12.7 \\
 		& & & & 2 & 242 & 11.7 \\
 		& & & & 3 & 207 & 11.1 \\
        \hline\hline
 	\end{tabular}
 	\caption{\label{Qmu}$Q \mu$ values for Solar System-like, rocky moons. The radii and densities are those of the corresponding Solar System bodies. The reference for these values is \citet[Appendix A]{murray99}.}
 \end{table*}
 
 From the surface temperature of the moon, that was calculated from the viscoelastic model, the $Q \mu$ value was determined using Eq.\,\ref{temp} for six orbital period--eccentricity pairs. In Table \ref{Qmu} the tidally induced surface temperatures and the logarithm of the $Q \mu$ values can be seen. The radius and density of the satellites are those of the corresponding Solar System bodies (see the first column), and the values are from \citet{murray99}. The eccentricities are set to 0.01 and 0.1, and the orbital periods to 1, 2 and 3 days. `N.a.' indicates that there was no solution (weak tidal forces).

 \subsection{Scaling the Galilean satellite system}
 
 Since no satellite has been discovered so far outside the Solar System, we used the Galilean system as a prototype for realistic calculations. Io, Europa and Ganymede are orbiting in a 1:2:4 mean motion resonance, that maintains their eccentricities, which play an essential role in forcing continuously their tidal heating. \citet{ogihara12} investigated satellite formation in the circumplanetary disc of giant planets using $N$-body simulations including gravitational interactions with the circumplanetary gas disc. They have found that 2:1 mean motion resonances are almost inevitable in Galilean-like satellite systems, and based on their results they predict that mean motion resonances may be common in exoplanetary systems. For these reasons the Galilean satellite system seems to be a representative example for realistic calculations, since the moons are in resonance, and their scaled-up versions will probably stay in resonance, too.
 
 The test systems consist of a planet (Jupiter) and the four Galilean moons. 91 cases are considered, one is the real Galilean system, and the others are the scaled-up versions: the masses of the planet and the moons were multiplied by the $scale$ factor ($scale = 1.0, 1.1, 1.2, ... \, 10.0$), and the semi-major axes of the moons were altered with constant orbital periods for each $scale$ value.
 
 \begin{equation}
 \label{P}
 P = 2 \pi \sqrt { \frac {a^3} {scale \cdot G (M_\mathrm{p} + M_\mathrm{s})} } \, ,
 \end{equation}
 
 \noindent where $a$ is the semi-major axis of the moon, $M_\mathrm{p}$ and $M_\mathrm{s}$ are the masses of the planet and the moon, respectively. The fixed orbital periods guarantee that the satellites approximately stay in resonances. This calculation resulted in constant $\beta$ values for all $scale$ parameters.
 
 \begin{figure}
 	\centering   
 	\includegraphics[width=12cm]{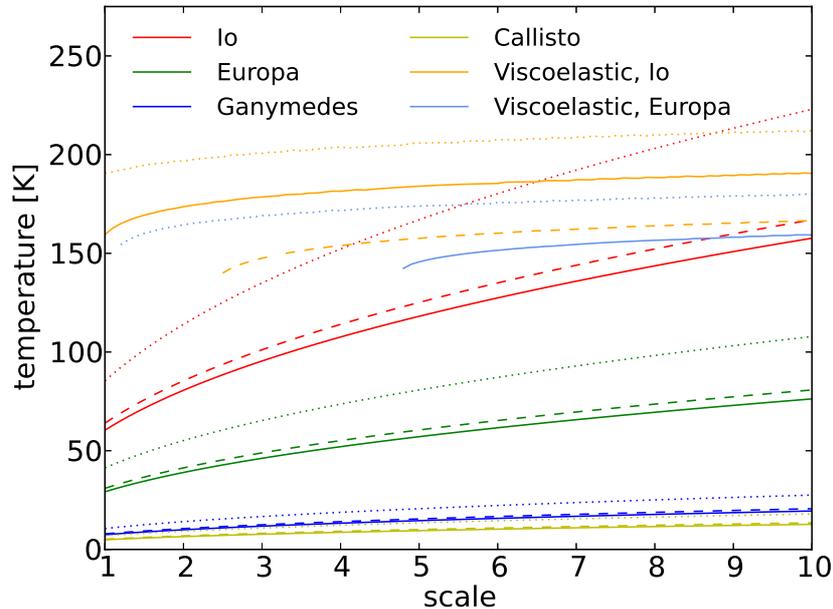}
 	\caption{\label{scaleJup}Surface temperature of the Galilean moons as function of the $scale$ parameter, calculated by both the fixed $Q$ and the viscoelastic model. Dashed and dotted curves indicate that the density or the eccentricity of the moon is doubled compared to the Solar System case, respectively.}
 \end{figure}
 
 Using both the fixed $Q$ and the viscoelastic models, the warmth of tidal heat was investigated in each case. The tidal heat induced surface temperature can be seen in Fig. \ref{scaleJup}, where the 91 cases are connected with solid curves for each satellite. 182 other cases were calculated, too, which are shown with dashed and dotted curves in the figure. These curves indicate that the densities (dashed curve) and the eccentricities (dotted curves) of the satellites are doubled compared to their original values in the Solar System. In the calculations $\mu$ and $Q$ were set for all satellites to that of Io, namely $10^{10} \mathrm{\,kg / (m \, s^2} )$ and 36, respectively, except for Europa, which has the following parameters: $\mu = 4 \cdot 10^{9} \mathrm{\,kg / (m \, s^2} )$ and $Q = 100$ \citep{peters13}. Densities of the moons are from \citet{lodders98}, and the reference for the semi-major axis, eccentricity and mass values is \citet[Appendix A]{murray99}.
 
 For Io, in the $scale = 1$ (Solar System) case the fixed $Q$ and the viscoelastic models give 60~K and 160~K, respectively. The observed surface heat flux induced by tidal heat on Io is around $2 \mathrm{\,W / m^2}$, which is a lower limit \citep{spencer00}. In other words, tidal forces produce at least 77~K heat on the surface of Io. The fixed $Q$ model resulted in a lower value than this limit, but note that $Q$ and $\mu$ are very difficult to estimate. The viscoelastic model gave much higher temperature than the observation, but keep in mind, that the heat is concentrated in hotspots, and is not evenly distributed on the surface of Io. The temperature of the warmest volcano, Loki is higher than 300~K \citep{spencer00}.
 
 The viscoelastic model gives solution only for Io (orange curves) and Europa (light blue curves), but not for all $scale$ values, as shown in Fig.~\ref{scaleJup}. In those cases when the densities are twice than those in the Solar System (dashed curves), the surface temperatures are just slightly higher. In fact, in the viscoelastic model, higher densities result in less tidal heat because of the imaginary part of the second order Love number. Doubling the eccentricity instead of the density (dotted curves) makes the surface temperature higher in each case.

  \section{Tidal heating with climate models} \label{climatemod}
 
For a complete habitability investigation of exomoons other energy sources are also needed to be considered besides tidal heating. Although it is possible that stellar radiation can be neglected (either because there is no central star, or because the planet-moon system orbits the star in a far distance), a general description must include the irradiated energy, as well. For this reason a complex climate model was used as described in section \ref{climatemodel}, and the viscoelastic tidal heating model was applied as a part of the whole calculation. From the viscoelastic model the $Q\mu$ value is obtained as described in section \ref{qmuvalue}. From this value the surface heat flow ($\zeta$) was calculated and then inserted into Eq. \ref{lebm}. The stellar mass in the simulations is $1 M_\odot$, the mass of the planet and the moon are $1 M_\mathrm{Jup}$ and $1 M_\oplus$, respectively. The planet revolves the star on a circular orbit at $a_\mathrm{p} = 1$ AU distance.

\begin{figure}
 	\centering   
 	\includegraphics[width=16cm]{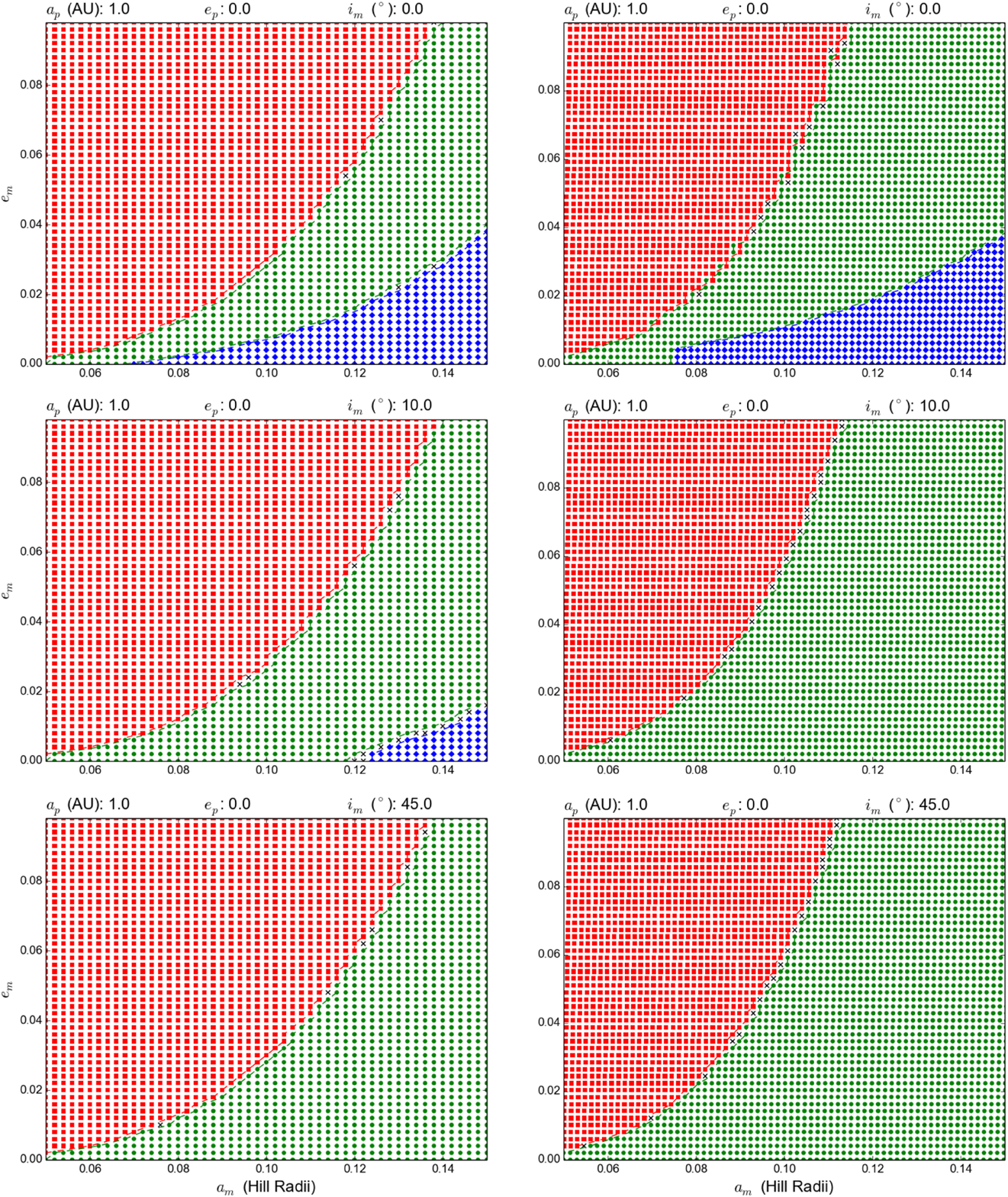}
 	\caption{\label{climate}The circumplanetary habitable zone as a function of the semi-major axis ($a_m$) and eccentricity ($e_m$) of the satellite's orbit. The colours indicate the following cases: \textit{red squares} -- `hot moons' with global mean temperatures above 373 K; \textit{blue diamonds} -- `snowball moons' with global mean temperatures below 273 K; \textit{green circles} -- `habitable moons' with mean temperatures between 273 and 373 K and low fluctuations; \textit{black crosses} -- `transient moons' with mean temperatures between 273 and 373 K, but with strong fluctuations. For the left-hand column the fixed $Q$, and for the right-hand column the viscoelastic tidal heating models were used. For the three rows different moon orbital inclinations (relative to the planet's equator) were used: $i = 0^\circ$ (top), $i = 10^\circ$ (middle) and $i = 45^\circ$ (bottom).}
 \end{figure}

Fig. \ref{climate} shows the results of the simulation data as functions of the semi-major axis and eccentricity of the moon's orbit. For the data in the left-hand column the fixed $Q$ tidal heating was applied (calculated with Eq. \ref{zeta}), compared to
viscoelastic tidal heating model in the right-hand column. Different rows show different orbital inclinations of the moon relative to the planet's equator. The four colours correspond to the four possible states of the satellite as described in section \ref{climatemodel} and also summarised in the caption of the figure.

The circumplanetary HZ is considerably wider when the viscoelastic tidal heating is used (right-hand column of Fig. \ref{climate}) compared to the fixed $Q$ model. While the weaker tidal heating at low $e$ tends to pinch the HZ slightly, at high $e$ the HZ is much wider, since tidal heating prevents catastrophic heating of the moon. For zero inclination the HZ is at least as wide as 0.06 Hill radius, and likely wider than that is measured in this parameter space. Perhaps most notably, for $a \lesssim 0.07$ Hill radius, the outer edge is not present at any eccentricity. These moons experience rapid, brief eclipses which are compensated by thermal inertia and the carbonate-silicate cycle.

If the inclination of the moon is increased (middle and bottom rows) than the outer edge begins to disappear, even at relatively large moon semi-major axis. For the outer edge to become
apparent, eclipses must be sufficiently frequent for their effects to
be cumulative and induce a snowball state. Increasing inclination
reduces the eclipse rate \citep[see e.g.][]{heller12} until it is close to
zero. (For polar lunar orbits ($i = 90^\circ$), the eclipse rate is not exactly zero --
eclipses can still occur twice per year, when the lunar nodes align with the
planet's orbital position vector.) With both implementations of tidal heating, relatively large
inclinations ($45^\circ$) erase the outer habitable edge completely in the investigated parameter space. At
intermediate inclinations ($10^\circ$), we see a slightly mixed picture.
The eclipse rate should be effectively zero, yet the fixed $Q$ tidal
heating run shows that an edge still exists (albeit much further from
the planet). This is an indication that here the ice-albedo feedback
mechanism is activating due to the moon's distance from the star
rather than due to eclipses, which in turn may suggest that the albedo
prescription we use may in fact be too simple for this scenario.

 \section{Conclusions}  \label{tidalconclusion}
 
 We have used, for the first time, a viscoelastic model for calculating the surface temperature of tidally heated exomoons. The viscoelastic model gives more reliable results than the widely used fixed $Q$ model, because it takes into account that the tidal dissipation factor ($Q$) and rigidity ($\mu$) strongly depend on the temperature. Besides, these values are poorly known even for planets and satellites in the Solar System. Using the viscoelastic model for exomoons helps to get a more realistic estimation of their surface temperature, and to determine a circumplanetary region, where liquid water may exist on them. It may help future missions in selecting targets for exomoon detections.
 
 We have defined the Tidal Temperate Zone, which is the region around a planet where the surface temperature of the satellite is between $0~^\circ$C and $100~^\circ$C. No sources of heat were considered other than tidal forces. Assuming, that the planet-moon system orbits the star at a far distance, or the stellar radiation is low due to the spectral type, tidal heat can be the dominant heat source affecting the satellite. We have investigated such systems, and found that the TTZ strongly depends on the orbital period, and less on the radius of the moon. For higher densities or eccentricities of the moon, the location of the TTZ is slightly closer to the planet.
 
 Comparing this model to the traditionally used fixed $Q$ model revealed that there are huge differences in the results. Generally, the viscoelastic model is less sensitive to moon radius than the uniform $Q$ model, keeping the surface temperature of the body more steady. The reason is that higher tidal forces induce higher melt fraction which results in a lower temperature than the fixed $Q$ model. The viscoelastic model demonstrates the way in which partially melting of a moon can act as a thermostat and tend to fix its temperature somewhere near its melting point over a wide range of physical and orbital parameters. As a consequence, the statistic volume of the TTZ is much larger in the viscoelastic case, which is favourable for life. But this lower temperature also means that the detectability of such moons is lower in the infrared. In addition, for low tidal forces there is no equilibrium with the convective cooling; hence, only the fixed $Q$ model provides solution. In these cases the challenge is to determine the values of $Q$ and $\mu$.
 
 For a few characteristic cases the product of the tidal dissipation factor and rigidity was calculated from the viscoelastic model, in order to help in quick estimations of tidally heated exomoon surface temperatures. Since the viscoelastic model is more realistic because of the inner melting and the temperature dependency of the parameters, but the fixed $Q$ model is easier to use, these $Q \mu$ values (along with the surface temperature) are provided in Table \ref{Qmu}. By inserting $Q \mu$ into Eq.\,\ref{temp}, one can get the estimation of the tidally induced surface temperature of a moon. Connection between the quality factor ($Q$) and the viscoelastic parameters (viscosity and shear modulus) was given for the Maxwell model, too, by \citet{remus12}.
 
 Earth-like bodies were also investigated as satellites, and in these cases the $-20$ and $60~^\circ$C temperatures were used as limits of habitability. The results are similar, but the volume of this habitable zone is larger than that of the TTZ for wide range of satellite radii. This habitable zone includes atmospheric considerations of the moon, but stellar radiation was neglected in the calculations. In case of significant radiation from other sources, the surface temperature of the moon will be higher. Additional heat sources (such as stellar radiation, radiogenic processes, reflected stellar and emitted thermal radiation from the planet), and the effects of eclipses, or the obliquity of the satellite are thoroughly discussed by \citet{heller13}. 
 
 For simulating realistic systems, the Galilean moons were used as a prototype. Their surface temperature was calculated with both models for different, scaled up masses. The mean motion resonance between the satellites helps to maintain their eccentricity, and consequently to maintain the tidal forces. By raising their masses, the temperatures of Io and Europa elevates less drastically in the viscoelastic model, than in the fixed $Q$ model (see Fig. \ref{scaleJup}). At $scale = 5$ (masses are five times as in the Solar System case) the surface temperature of Europa is $\sim150$~K calculated from the viscoelastic model. Assuming that its density does not change, its radius will be approximately 0.25 Earth radii. In case of an additional 100--120~K heat (e. g. from stellar radiation), the ice would melt, and this super-Europa would become an `ocean moon', covered entirely by global water ocean. The used viscoelastic model might not be adequate, and can be oversimplified for such bodies that consist of rocky and icy layers, as well. Salty ice mixtures may also modify the results. The applied model ignores the structure, pressure and other effects, and applies melting for the whole body. However, it provides a global picture of the tidally heated moon. Even with a more detailed viscoelastic model, that describes Enceladus as a three layered body (rocky core, ocean and ice shell), \citet{barr08} have found that tidal heat is $\sim10$ times lower than that was observed by the Cassini Composite Infrared Spectrometer. Similarly, \citet{moore03} concluded that observed heat flux on Io is about an order of magnitude higher than that can be explained with a multilayered, viscoelastic model. These results suggest that tidal heat can be much more relevant than what is predicted by models.
 
We have also used a 1D energy balance model of Earth-like exomoon climates, which contain stellar and planetary insolation, atmospheric circulation, infrared cooling, eclipses and tidal heating as the principal contributors to the moon's radiative energy budget. The model also takes into account the carbonate-silicate cycle, a negative feedback mechanism to regulate planetary temperatures and the positive ice-albedo feedback system. We calculated tidal heating as part of this whole climate model with both the fixed $Q$ and the viscoelastic models.

We have found that the results show many more potentially habitable configurations when the viscoelastic model is used, even in our somewhat limited parameter space. The circumplanetary HZ is significantly wider with viscoelastic tidal heating, and extends much further from the planet (provided the moon's orbital eccentricity is larger than about 0.02, which is slightly less than that of Titan). We have found that in case of zero inclination of the moon's orbit there is a well-defined outer edge to the circumplanetary HZ, inside the orbital stability limit, due to a combination of eclipses and ice-albedo feedback. The outer edge requires the orbits of the planet and the moon to be quite closely aligned, so that eclipses are sufficiently frequent. We have shown that if the moon's orbit is inclined so that eclipses are unlikely, the outer edge completely disappears for any eccentricity of the moon's orbit, even for relatively large semi-major axes of the moon.

\chapter{The possibility for albedo estimation of exomoons} \label{possibalbedoest}
 
The aim of this work is to propose a new method to identify elevated albedo and thus the possibility of water ice on the surface of an exomoon using photometric measurements during occultations. We will also discuss the role of different orbital and physical parameters that may influence the observed data. The following work is based on the article by \citet{dobos16}.

 Surface ice cover might be more frequent for exomoons than for exoplanets, based on examples of the Solar System and theoretical argumentation. Although the photometric signal would be small, it is worth considering the possibility and potential results of such measurements, because of the following reasons: the findings might orient the instrumental development, and search programs to focus on given stellar type, and also give a hint on surface properties of exomoons, or provide information on their habitability. Analysing different configurations helps to see the best possibility to estimate exomoons' albedo. We also aim to compare the scale and possible role of such configurations of exomoons that are realistic but not present in the Solar System. Just like in the case of exoplanets, unusual situations could be present and occasionally dominate because of observational selection effects (see for example the case of many hot Jupiters during the first decade of exoplanet discoveries). In this work we analyse these possibilities, pointing to the difficulties, possible ways and best conditions for such estimation.

Although such precise measurements are difficult to make, the method could be used to differentiate among exomoon surface properties. Despite no exomoon has been discovered yet, probably those are numerous in other planetary systems, and will play an important role in astrobiology research in the future \citep{kaltenegger10, heller13} giving rationality to elucidate the potential method outlined here. Because of the large number of possible configurations and different parameters of exomoon systems, specific cases are considered in this paper with the aim to give general insights about the factors that are important for albedo estimation. We also consider such possible conditions that are not present in the Solar System but could be observed at certain exomoons, e.g. large exomoons, objects on eccentric, inclined or retrograde orbits (section \ref{diffpar}).

The basic idea behind this work is that the exomoons' albedo could be estimated from the flux change during occultation. High albedo values might suggest water ice on the surface \citep{verbiscer13}. With the analysis of periodic changes in the transit time and its duration \citep{kipping09a, kipping09b}, the mass of the moon can be calculated, from which the bulk density could also be estimated. The density of the moon helps in determining whether or not the moon contains significant amount of ice. For the albedo calculations Solar System moons and hypothetical bodies were considered as exomoons, in order to determine their observability during the occultation. The possible usage, problems and uncertainties of this method are described in section \ref{albedodiscuss}.

  \section{Solar-type stars}

First, the flux change caused by the presence of different hypothetical exomoons was calculated using Eq. \ref{Ag}. As there might be a great variety of sizes, orbital distances and stellar luminosities, we restricted our analysis to certain number of example cases. For stellar distance 3 AU was used, which is the approximate location of the snowline in the Solar System. In the following calculations the Sun was used as central object, because future surveys and monitoring programs will probably focus on Sun-like stars to search for Earth-like exomoons \citep{kaltenegger10, peters13}. Beside solar type stars we also calculated some example cases for M dwarfs (see section \ref{nextgen}), as they came into the focus of exoplanet related research recently \citep[see the MEarth project,][]{berta13}.

A few parameters of interesting examples are shown in Table \ref{MOdepth}. Beside Solar System bodies (Earth, Europa, Enceladus, Io), `icy Earth' is introduced as a hypothetical moon. Icy Earth is considered to reflect as much light as Enceladus (almost 100\% of the incident light) and its radius equals to that of the Earth. It is important to note that Europa and Enceladus have no relevant atmospheres and thus we also consider the hypothetical `icy Earth' to satisfy this criterion as well. Io does not contain water ice on the surface, but still has moderately elevated albedo because of fresh sulphur containing material. Although there is no Earth-sized moon in the Solar System, a natural satellite with this size might be considered realistic, and can serve as a useful for comparison to other objects \citep{ogihara12}. As it can be seen from Table \ref{MOdepth}, the required photometric precision is many orders of magnitude beyond the capability of current technology and such measurements will not be feasible in the foreseeable future.

\begin{table}
	\caption{Calculated values of the MO depth for different bodies around a Sun-like star. $A_\mathrm{g}$: geometric albedo, $R_\mathrm{m}$: radius of the exomoon. The distance from the central star is 3 AU in each case. The geometric albedo values are obtained from \citet{pang81, verbiscer07, verbiscer13}.}
	\centering
    \label{MOdepth}
	\begin{tabular}{c c c c}
		\hline\hline
		Name & $A_\mathrm{g}$ & $R_\mathrm{m}$ [km] & $y_3 - y_4$ [ppm] \\
		\hline
		Enceladus & 1.38 & 252 & $4.3 \cdot 10^{-7}$ \\
		Europa & 1.02 & 1561 & $1.2 \cdot 10^{-5}$ \\
		Io & 0.74 & 1821 & $1.2 \cdot 10^{-5}$ \\
		Earth & 0.37 & 6371 & $7.5 \cdot 10^{-5}$ \\
		icy Earth & 1.38 & 6371 & $2.8 \cdot 10^{-4}$ \\
		\hline\hline
	\end{tabular}
\end{table}

  \section{Moons in systems of M dwarfs}

   \subsection{Plausibility of large icy moons}

Moons are more likely to form by ejection around Neptune than around Jupiter-like giants because. Jupiter-like gas giants are composed of mostly liquid and gaseous hydrogen at their outer layers, thus the ejected satellite-forming material might be $\mathrm{H_2 O}$ poor, composed mostly of gaseous $\mathrm{H_2}$ and could easily escape. But if this event happens before the dispersion of the circumplanetary disk, the co-accreted ice in the disk could provide icy surface for such satellite. Neptune-like planets seem to be better candidates. They are mainly composed of $\mathrm{H_2 O}$, and the ejected satellite, too. The occurrence of impacts at giant planets are favoured by certain theories, as a contributor to the continuous gas accretion \citep{broeg12}, increasing the probability of impact-ejected icy satellites.

Besides ejection, capture could also produce a large moon. In the so-called \textit{binary-exchange} captures, a terrestrial sized moon can be captured by a 5 Jupiter-radii planet around an M dwarf, if the planetary encounter is close enough \citep{williams13}. For Neptune-mass planets the capture is easier than for Jupiter-mass planets, because generally the encounter speeds are lower. The capturing is much more difficult at small stellar distances, but if it occurs before or during the planetary migration, then it is possible that such a planet-moon pair will orbit an M dwarf at a close distance (around the snowline). In the Solar System, Triton may be an example for such binary-exchange capture \citep{agnor06}. For these reasons we consider the existence of large exomoons (up to a few Earth-masses) of Neptune-mass planets around M dwarfs.

   \subsection{Different wavelengths} \label{wavelength}

It is well known that M stars are much brighter in the infrared than in the visible spectrum. For this reason it seems obvious to measure the flux at longer wavelengths, however, water ice is most reflective in the visible. In this section we discuss which wavelengths are optimal to use for successful measurements of icy moons.

\begin{figure*}
	\centering
	\includegraphics[width=14cm]{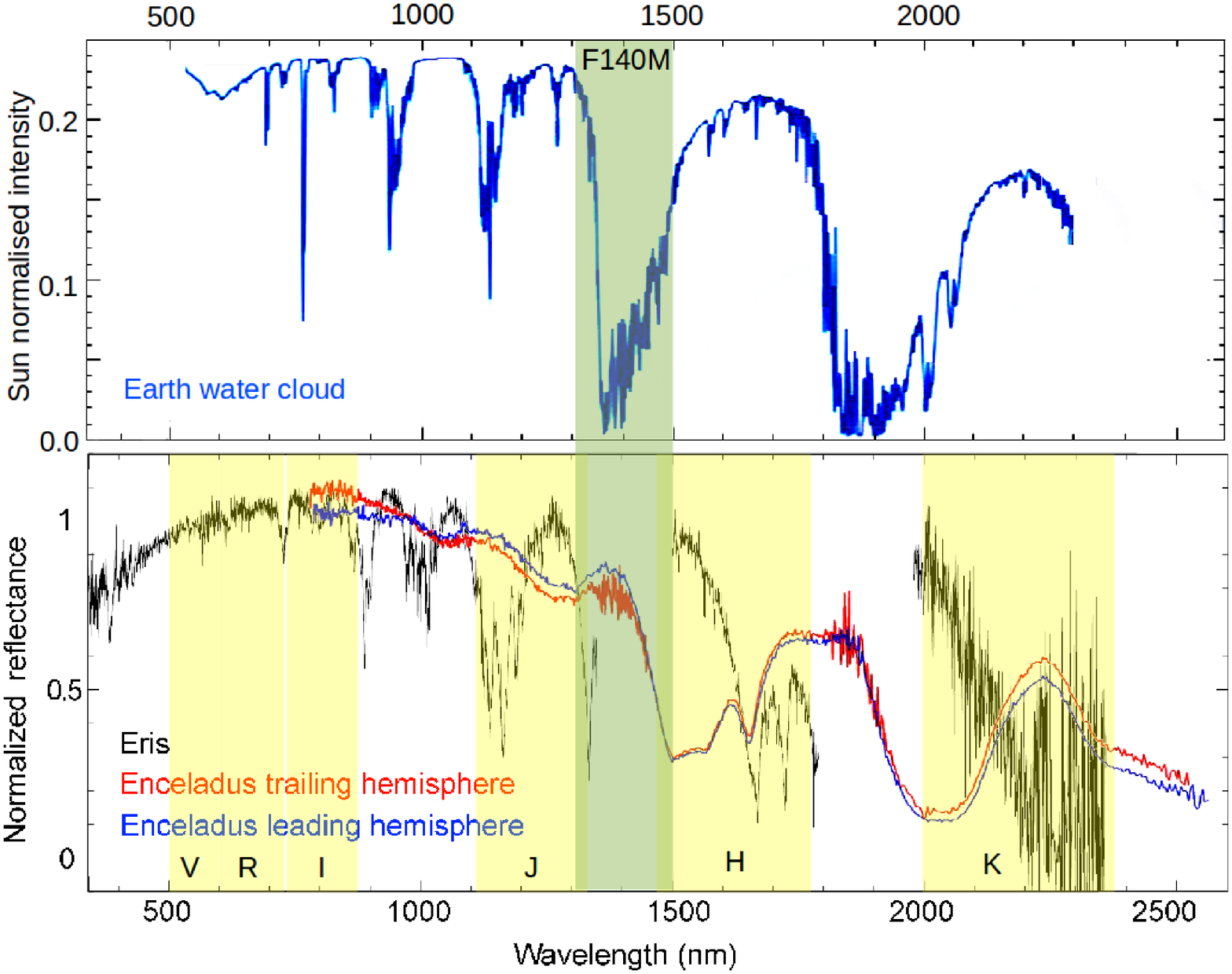}
	\caption{Top panel: Spectrum of water clouds on Earth. Bottom panel: Spectra of Enceladus (both the leading and the trailing hemispheres are indicated with blue and red colours, respectively) and Eris (black) from \citet{verbiscer06} and \citet{alvarez11}, respectively. The visible spectrum of Enceladus is not shown, but the geometric albedo is known to be 1.375 \citep{verbiscer07}. Yellow vertical bands indicate the full widths at half maximum of the V, R, I, J, H and K photometric bands. The green vertical band indicates the F140M filter band where the cloud spectrum is low but the water ice spectrum is relatively high.}
	\label{spectra}
\end{figure*}

Beside water ice, methane ice is also very reflective and such bodies are also known in the Solar System that are covered at least partially with methane ice. For this reason, if the estimated albedo is high, it may be challenging to decide whether water ice or methane ice is present on the surface. To make the distinction easier, see the differences in the spectra of Enceladus and Eris in the bottom panel of Fig. \ref{spectra} (the spectra are from \citet{verbiscer06} and \citet{alvarez11}). Eris is a good example for a small body covered by methane ice, and Enceladus is used as a representation of a moon with water ice on the surface. The spectra are normalized, but these normalized values must be in good correspondence with the geometrical albedo for both bodies: the spectra of Enceladus are normalized at 0.889 $\mu$m where the geometric albedo (at phase angle ${\approx} 3^{\circ}$) of the leading and the trailing hemispheres are 1.02 and 1.06, respectively \citep{buratti98}, and the reflectance of Eris is normalized at 0.6 $\mu$m \citep{alvarez11}, and its geometric albedo in the V band is 0.96 \citep{verbiscer13} which is very close to one. From their spectra, the albedo values at different photometric bands are estimated and presented in Table \ref{bands}. All values in the table are rough estimations, which can differ for various exomoons. The albedo of Enceladus (water ice case in the table) in the V band is from \citet{verbiscer07}.

\begin{table}
	\caption{Estimated albedo values at different photometric bands, based on the spectrum of Enceladus (water ice surface, \citet{verbiscer06, verbiscer07}) and Eris (methane ice surface, \citet{alvarez11}).}
    \label{bands}
	\centering
	\begin{tabular}{c c c c c c c}
		\hline\hline
		Surface & V & R & I & J & H & K \\
		\hline
		Water ice & 1.375 & n.a. & 1.00 & 0.84 & 0.52 & 0.43 \\
		Methane ice & 0.97 & 1.02 & 0.98 & 0.65 & 0.45 & 0.42 \\
		\hline\hline
	\end{tabular}
\end{table}

To see the albedo variations respective to wavelength, it is best to measure the flux in all photometric bands indicated in Table \ref{bands}, but it seems that the V and the J bands are the most relevant. Measuring in these two bands provide useful data, partly because of the apparent difference in the albedo for the two kinds of ice, which helps determining whether water or methane ice is present on the surface. Another reason for the V band is the maximum reflectance of water ice, and for the J band is the maximum spectral radiance of low-mass M dwarfs. The radiance of an M6 -- M7 main-sequence star in the J band is about 5 times stronger than in the V band. Because of the larger number of photons that arrive to the detector, it is easier to observe the occultation in the J band, where the albedo of water ice is still relatively high. However, the shape of the light curve will not change in different bands.

Clouds are also very reflective and for this reason it may be challenging to determine whether we see an icy surface or clouds in the atmosphere. Since water is very abundant is the universe, we assume that clouds form from H\textsubscript{2}O. The upper panel of Fig. \ref{spectra} shows the intensity of water cloud on Earth \citep[Chapter~7]{gottwald06}. In the F140M filter band (indicated by a green vertical band in the figure) the intensity of the clouds are very low while the reflectance of water ice is moderately high (see the Enceladus' spectrum in the bottom panel). This photometric band gives opportunity to separate clouds from ice.

   \subsection{Next generation telescopes and the occultation of exomoons} \label{nextgen}

It seems easier to estimate the albedo of satellites around M dwarfs than around larger stars. This is because the fact that the lower mass the star has, the closer the snowline is located. It means that the planet-moon system can be closer to the central star without the sublimation and/or melting of the ice on the surface. If the moon is closer to the star, then the light drop caused by its presence will be more significant in the light curve. This effect favours detection. In other words, it may be easier to observe icy moons around lower mass stars in occultation, than around more massive stars. It also means, that smaller moons may also be detected in occultation around late type stars.

The capabilities of next generation telescopes are of fundamental importance in successful albedo estimations. Beside Kepler, the most relevant missions for exomoon detection are the CHaracterising ExOPlanets Satellite \citep[CHEOPS, ][]{broeg13}, the Transiting Exoplanet Survey Satellite \citep[TESS, ][]{ricker10, ricker14}, the James Webb Space Telescope \citep[JWST, ][]{clanton12}, the PLAnetary Transit and Oscillations of stars \citep[PLATO 2.0, ][]{rauer11, rauer14} and the European Extremely Large Telescope \citep[E-ELT, ][]{ramsay14}. The photon noise levels of these instruments were calculated as described in section \ref{calcphotonnoise}. The data used for the calculations for each instrument are shown in Table \ref{instruments}. All these numbers are subject to change, since most of these missions are in the planning and testing phase. The numbers in the last column indicate references as explained in the footnote\footnote{1: \citet{ricker14}, 2: \citet{auvergne09}, 3: \citet{broeg13}, 4: \citet{benz13}, 5: \citet{rauer14}, 6: \citet{vancleve09}, 7: \citet{avila16}, 8: \citet{dressel16}, 9: JWST webpage: http://www.stsci.edu/jwst/instruments/nircam/instrumentdesign/filters/, 10: \citet{davies10}}.

\begin{sidewaystable}
	\caption{Instrument parameters for calculating the photon noise level. $A$: aperture, $Q$: quantum efficiency, $T_\mathrm{campaign}$: observation period per target field for survey instruments. In most cases rough estimations were used. Reference numbers shown in the last column are explained in the footnote. In the cases of TESS and CHEOPS the quantum efficiencies are estimated based on the $Q(\lambda)$ curves provided by the manufacturers. Asterisk indicates that instead of the quantum efficiency, the system throughput is presented.}
	\centering
	\begin{tabular}{c c c c c c c c c}
		\hline\hline
		Instrument & $A$ [m] & Filter & $\lambda_\mathrm{f}$ [nm] & $\lambda_\mathrm{min}$ [nm] & $\lambda_\mathrm{max}$ [nm] & $Q$ [\%] & $T_\mathrm{campaign}$ & Ref. \\
		\hline
		TESS & 0.1 & -- & 800 & 600 & 1000 & 80 & 80 days & 1\\
		CoRoT & 0.27 & -- & 650 & 400 & 900 & 63 & 150 days & 2\\
		CHEOPS & 0.3 & -- & 750 & 400 & 1100 & 67 & -- & 3, 4\\
		PLATO 2.0 & 0.68 & -- & 750 & 500 & 1000 & 50* & 2 years & 5\\
		Kepler & 0.95 & -- & 660 & 420 & 900 & 50* & 4 years & 6\\
		HST ACS WFC & 2.4 & F555W & 541 & 458 & 621 & 35* & -- & 7\\
		HST WFC3 IR & 2.4 & F125W & 1250 & 1100 & 1400 & 52* & -- & 8\\
		JWST NIRCam & 6.5 & F115W & 1200 & 1000 & 1400 & 40* & -- & 9\\
		JWST NIRCam & 6.5 & F140M & 1400 & 1300 & 1500 & 45* & -- & 9\\
		E-ELT MICADO & 39 & J & 1258 & 1181 & 1335 & 60* & -- & 10\\
		\hline\hline
	\end{tabular}
	\label{instruments}
\end{sidewaystable}

\begin{figure}
	\centering
	\includegraphics[width=14cm]{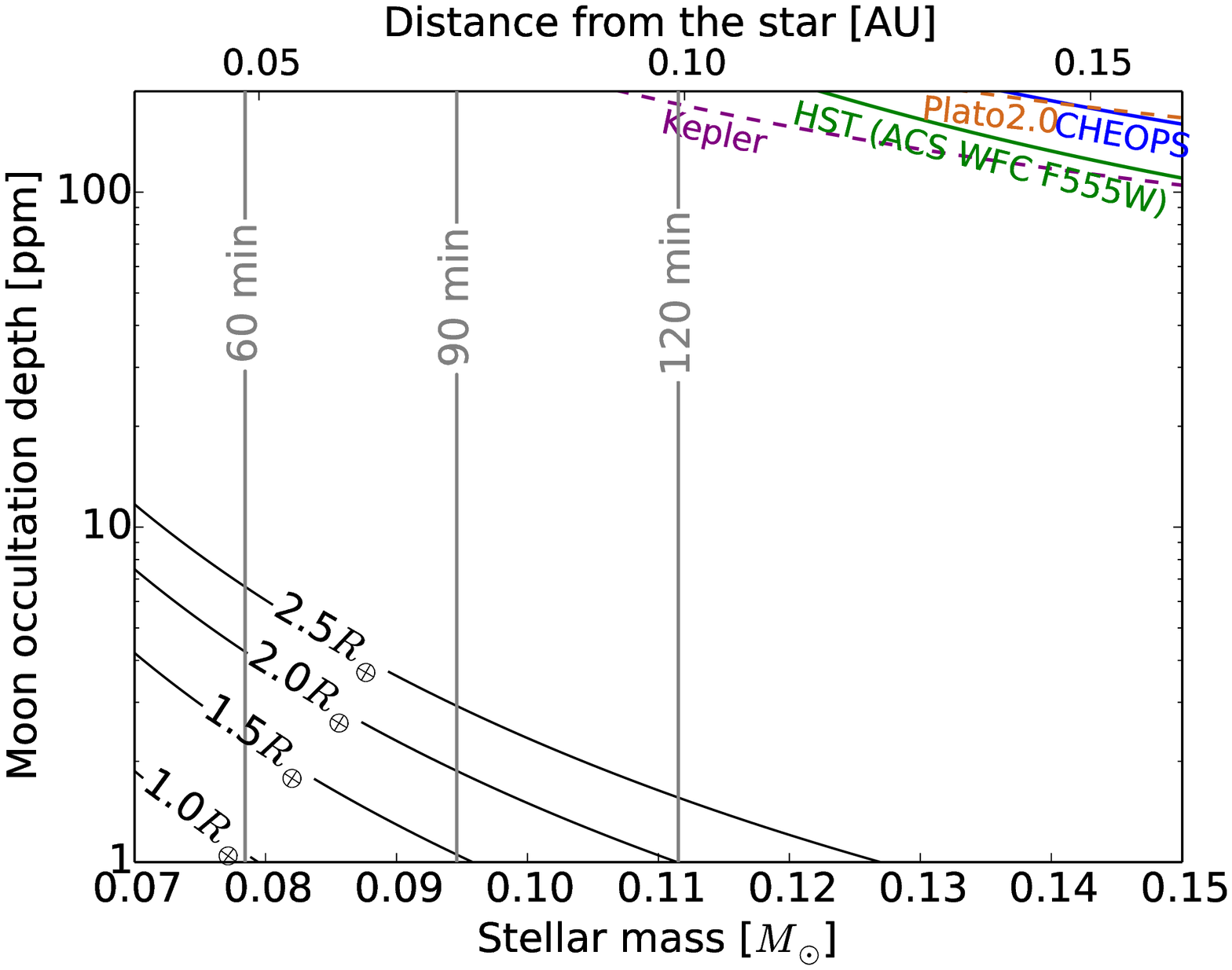}
	\caption{Radius of icy moons around different late type stars with a given MO depth. The moon's geometric albedo was set to $A_\mathrm{g}$ = 1.38 in each case (according to the reflectance of ice in the V photometric band), and the distance from the star is at the snowline, that is calculated from Eq.~\ref{snowline}, and indicated in the top axis, too. Brown, blue, green and purple curves indicate the estimated photon noise levels of PLATO 2.0, CHEOPS, HST and Kepler, respectively. The dashed curves for PLATO 2.0 and Kepler indicate that these are survey missions in contrast with the HST and CHEOPS observatories. Grey vertical lines indicate the 60, 90 and 120 minute MO duration contours for a moon in circular orbit around a Neptune-mass planet.}
	\label{telescopes}
\end{figure}

\begin{figure}
	\centering
	\includegraphics[width=14cm]{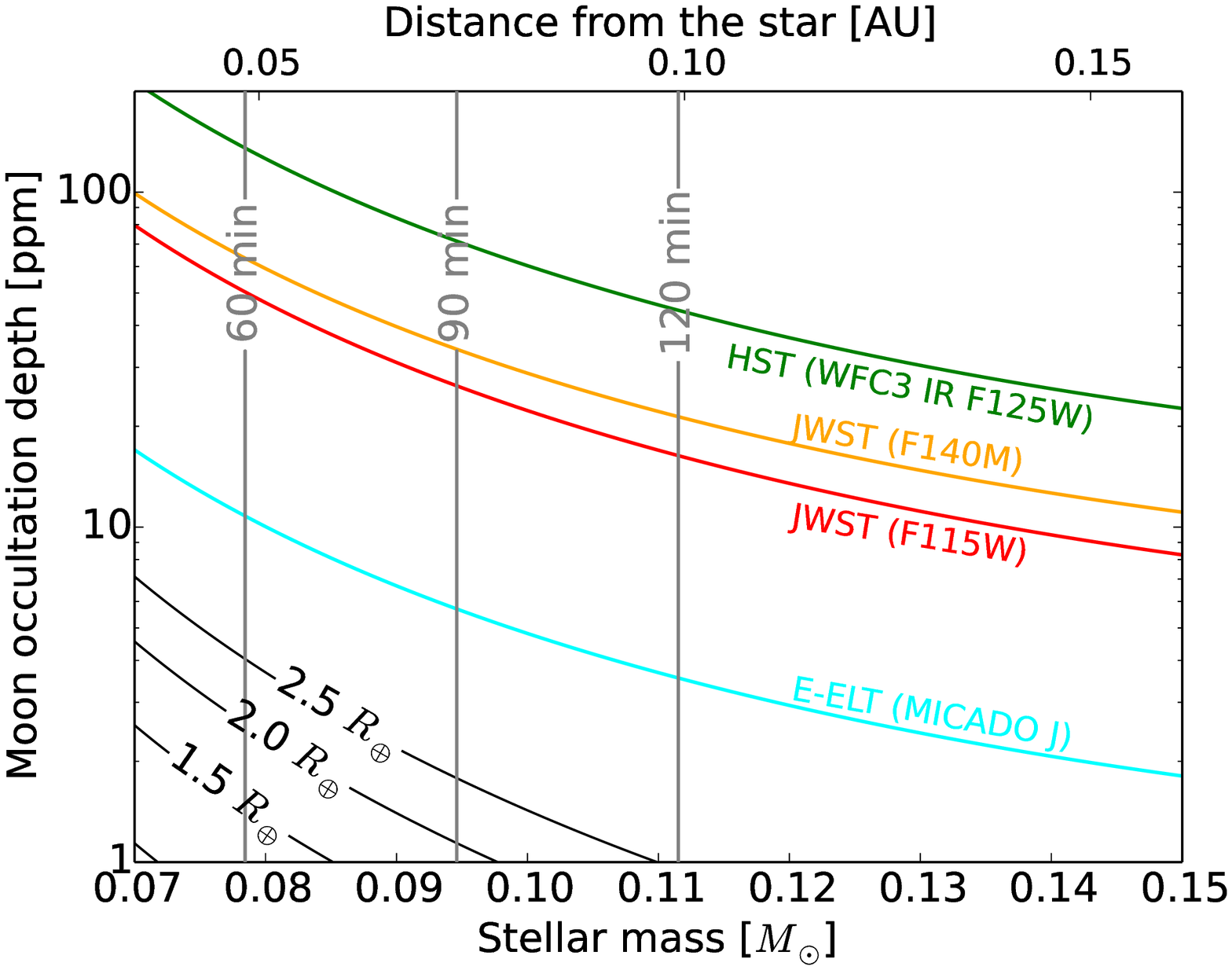}
	\caption{Radius of icy moons around different late type stars with a given MO depth. The moon's geometric albedo was set to $A_\mathrm{g}$ = 0.84 in each case (according to the reflectance of ice in the J photometric band), and the distance from the star is at the snowline, that is calculated from Eq.~\ref{snowline}, and indicated in the top axis, too. Green and light blue coloured curves indicate the estimated photon noise levels of HST and E-ELT, respectively. Orange and red curves depict the same for JWST but with two different filters (F140N and F115W, respectively). Grey vertical lines indicate the 60, 90 and 120 minute MO duration contours for a moon in circular orbit around a Neptune-mass planet.}
	\label{telescopes_J}
\end{figure}

Figs. \ref{telescopes} and \ref{telescopes_J} show the relation between the MO depth of icy moons and the mass of the host star in the V and J photometric bands, respectively. The objects covered in this figure are low-mass M dwarfs: M9 -- M5 according to \citet{kaltenegger09}. The MO depth is in logarithmic scale. The radius of the moon indicated by black curves was calculated for all stellar mass and MO depth pairs. For this calculation Eq. \ref{Ag} was used and the albedo was fixed to that of Enceladus in each case, i.e. 1.38 for Fig. \ref{telescopes} and 0.84 for Fig. \ref{telescopes_J}. The distance from the star was set to the snowline, which was calculated from Eq.~\ref{snowline}, and is indicated in the upper axis. This is the minimum stellar distance that is required in order to prevent the melting of the surface ice. It also means that our calculation is an optimistic approach: the minimum distance and the approximately minimum moon radius are determined for a successful detection. The vertical grey lines indicate the 60, 90 and 120 minute MO durations. In all investigated cases, the moon and the planet occults separately. For larger stellar masses the MO duration is longer, because of the larger stellar distance. The expected noise levels for the telescopes are indicated by coloured curves. Those instruments that measure in or near the J photometric band (HST WFC3 IR, JWST NIRCam and E-ELT MICADO) are shown in Fig. \ref{telescopes_J}, all the others are in Fig. \ref{telescopes}, however, because of their limited capabilities, TESS, and CoRoT do not appear in Fig. \ref{telescopes}.

The highest radius that is shown in Figs. \ref{telescopes} and \ref{telescopes_J} is 2.5 Earth-radii (for comparison, the radius of Neptune is 3.9 Earth-radii). We choose this 2.5 Earth-radii as an upper limit for a rocky moon with surface ice to guarantee that its mass is not larger than 5 Earth-masses. (The radius of a body with the density of Enceladus and with a mass of 5 Earth-masses is 2.58 Earth-radii.) Such large body around a Neptune-mass planet could also be called a binary planet, which would probably be a different system than the Saturn--Enceladus planet--moon pair. For simplicity, and also because we use this 2.5 Earth-radii as an upper limit, we still call them moons in the rest of the paper.

In the visible spectrum, the photon noise of the telescopes is much higher than the required precision for detecting large, rocky moons in occultation. However, in the J photometric band (see Fig. \ref{telescopes_J}) the E-ELT's photon noise is just a few ppm greater than the MO depth of large icy exomoons. (Note that the MO depth is shown in a logarithmic scale.) For example, for a 2.5 Earth-radii moon around an 0.11 stellar mass star, the difference is about 3-4 ppms, and for a 1.5 Earth-radii moon at an 0.085 stellar mass star, it's about 7 ppms. The accuracy of the measurements and additional parameters that can influence the success of the observations are discussed thoroughly in the next section.

 \section{Discussion} \label{albedodiscuss}

The findings on the observability of different exomoons, the estimation of their surface albedo, and other related considerations are summarised below.

\begin{itemize}
	\item \textit{Smaller stellar distance:} The depth of the flux drop caused by the moon in occultation as well as its observability increases with decreasing stellar distance. In the case of smaller stars, the surface of the moon can keep its icy composition even for small stellar distances, because the star's luminosity is low. Small stellar distances (such as 0.01--0.1 AU in our calculations) are important regarding the long term stability of a planet-moon system, too. According to migration theories and perturbation effects, numerous criteria have to be satisfied in order to support the presence of an exomoon that survived a certain time period after it got close to the central star \citep{barnes02}. During this process they might be able to keep their satellite system protected against perturbations by the central star and other planetary objects \citep{mosqueira03}. However, the migration is not relevant, if the moon is captured or impact-ejected (which seems more likely for large satellites) after the planet's migration phase.
	\item \textit{Larger size:} Among the analysed cases, the best possibility to determine the albedo is for Earth-sized, or even larger exomoons, because the reflecting surface is larger.
	\item \textit{Higher albedo:} The higher the albedo that an exomoon has, the more stellar light it reflects, and the flux difference in the light curve is larger during the occultation. Extremely high geometric albedos (for example 1.375 in the case of Enceladus in the Solar System) are indicators for the presence of relatively young and fresh water ice since this is the only known material that could reflect such high percentage of light. The presence of water ice on the surface may imply astrobiological importance. Considering strongly reflecting atmospheric clouds, their presence could also cause high albedo. The separation of cloud and ice reflection with optical methods could be difficult, but in case of successful measurements with an F140M filter, it may be feasible, as discussed in section \ref{wavelength}. However, our results show that such measurements will not be achievable with the instruments presented in section \ref{nextgen}, meaning that next generation telescopes will not yet be able to separate atmospheric clouds from surface ice on exoplanetary bodies.
\end{itemize}

Beside the challenge in observation of the small flux contribution of the exomoon, analysing the measured data will also be difficult, as to fit a realistic model, because many possible configurations are needed to consider. To see the possible roles of these factors, and in order to help to plan and target future observations, in the following we evaluate how several basic parameters affect the observation and analysis.

  \subsection{Accuracy}

For an adequate result from the light curve measurements, repeated observations are needed. It is not necessary that the moon is positioned exactly at the same apparent position every time when the occultation occurs. For this reason the small flux drop caused by the moon may be shorter or longer, and may not be seen at every observation. \citet{simon12} propose using an averaged light curve of several observations to deal with the scattered position of the flux change caused by the presence of the moon. They also present a four-step detection strategy for discovering exomoons by transit method. Occultation events with the same apparent positions of the moon and the planet might be rare, but in tidally locked situations, a given configuration during subsequent occultations could be frequent or regular and repeated at every revolution of the exoplanet, and the exomoon might be at the same position viewed from our line of sight.

Variable surface features of the host star, such as granules, micro-flares, etc. could influence the result, producing small flux fluctuations. Such variability is expected especially in the late type stellar atmospheres related to convective motion \citep{ludwig06}. From the moon's transit measurements and from the occultation of the host planet, the expected time of the moon's occultation can be estimated, which may help in distinguishing the MO signal from the stellar surface fluctuations, especially because the occultation will cause a periodic change in the light curve.

Beside the uncertainties in flux measurements, the accuracy of albedo estimation also depends on the accuracy of the exomoon's size and the stellar distance. In our work both are assumed to be already known from transit detections, and determining their accuracy was already discussed by several authors \citep[see e.g.][]{pal08, carter08}. However, the errors from transit measurements propagate into the determination of the geometric albedo. The total error of the albedo can be obtained from
\begin{equation}
	\left( \frac{\Delta A_\mathrm{g}} {A_\mathrm{g}} \right)^2 = \frac {\Delta (y_3 - y_4)^2} {(y_3 - y_4)^2} + \frac {4 \Delta (a_\mathrm{p} / R_\star)^2} {(a_\mathrm{p} / R_\star)^2} + \frac {4 \Delta (R_\mathrm{m}/R_\star)^2} {(R_\mathrm{m}/R_\star)^2}  \, ,
    \label{error}
\end{equation}

\noindent where the three terms of the right side of the equation can safely be considered to be independent of each other. The first term is $(1/5)^2$ since we set the measurements to $5\,\sigma$. The second term is typically around (1--5\%$)^2$ \citep{pal08}. Assuming that $a_m/R_\star \geq 1$ (i.e. the moon and the planet occult separately, which was true in all cases described in section \ref{nextgen}), and that the same number of occultations are detected for the moon as for the planet, and that these measurements were made with the same instrument, the third term can be expressed as
\begin{equation}
	\frac {\Delta (R_\mathrm{m}/R_\star)} {(R_\mathrm{m}/R_\star)} = \frac {\Delta (R_\mathrm{p}/R_\star)} {(R_\mathrm{p}/R_\star)} \cdot \frac {(R_\mathrm{p}/R_\star)^2} {Q (R_\mathrm{m}/R_\star)^2}  \, ,
\end{equation}

\noindent where $Q$ is the quantum efficiency. This error can be calculated for given geometric configurations. In any case, the last term must be smaller than the first one in Eq. \ref{error}, because the S/N ratio is much larger for transits than for occultations, hence transit measurements are much more precise. See e.g. the case of CoRoT-1b which was measured in transit and occultation, as well \citep{snellen09}. The same effect is true for exomoons. It means that the first term of Eq. \ref{error} is dominating in the error of the albedo estimation.

In the calculations shown in section \ref{nextgen} only photon noise was considered based on the quantum efficiency or on the system throughput. However, in near infrared measurements other noise sources become significant, as well: CCD read-out, CCD dark signal, Earth's sky variations and zodiacal light. The produced noise will be comparable to the photon noise. The system throughput that was considered in the calculations (even when the quantum efficiency was used, since in these cases the quantum efficiency was multiplied by $0.8$) already contains some of these extra noise sources, but in future missions these are subject to change, since their precise values will be known only after the commissioning of the instruments. In addition, we assumed 30 occultation detections for observatories that would require at least 5 years for E-ELT because the orbital period of the planet at the snowline is approximately 31 days, and the occultation can only be measured at nights. Considering that not all noise sources were included in our estimations, and that measuring 30 occultations may be unrealistic in most cases, it can be concluded that our calculations are too optimistic, and the real noise level of the instruments will be higher than presented.

According to \citet{simon07}, measuring the photometric transit timing variations (TTV\textsubscript{p}) leads to the estimation of the size of the exomoon. By analysing transit duration variations (TDVs) as well, rough mass estimation is also possible under favourable conditions \citep{kipping09a, kipping09b}. Using the derived mass and radius together, information on bulk density could be gained, and putting together this value with the presence of surface ice gives strong hint on the possible existence and ratio of water ice in the interior.

  \subsection{Considering different parameters} \label{diffpar}

In the following we categorize parameters by their importance in influencing the results of the observations. Specific cases that might be relevant for certain exomoons, but are not usual in our Solar System are discussed below.

\begin{itemize}
\item \textit{Eccentricity:} For the analysed, relatively large exomoons we do not expect highly eccentric orbits as the large satellites in the Solar System (even the captured ones) have low eccentricity. However, perturbations might elongate the orbit, or resonances with other moons may maintain a higher eccentricity. The variable orbital velocity around the host planet influences the apparent speed of the exomoon perpendicular to our line of sight. The changing velocity influences the length of the plateau/occultation duration. Shorter plateau/MO duration means less integration time for the telescope which results in larger photon noise. The moon-star distance periodically change with the orbital phase of the moon, influencing its brightness, but this effect is very small. Eccentricity of the moon may enhance this phenomenon. It could be substantial in cases when the ratio between the stellar distance and the exomoon's semi-major axis is relatively small. \citet{sato10} described the effect of the satellite's eccentricity on the transit light curve in details, which can similarly be used to occultations with some changes.

\item \textit{Inclination:} Large tilt of the orbital plane of the exomoon relatively to the planet's orbital plane around the host star could also influence the plateau / occultation duration, especially when the exoplanet orbits at large distance from the host star. In case of large inclinations, instead of occulting, the moon may pass beside the star. The length of the plateau/occultation duration also depends on the inclination and the ratio of the exomoon's apparent distance from its host planet and the diameter of the host star. Larger stellar disk and smaller orbital distance increase the influencing effect of higher inclinations, however if the stellar disk is smaller, then the moon will not occult for high inclinations. For transits the effect of the exomoon's inclination is discussed by \citet{sato10}.

\item \textit{Reflected light from planet:} In ideal cases the exomoon's brightness might be elevated by reflected light from its host planet. The strongest reflected light flux reaches the exomoon during the `new moon' phase (watching from the exoplanet). In different configurations mutual transits and mutual shadows affect the shape of the light curve \citep{cabrera07}. In the albedo estimation this effect of the reflected light from the planet is not significant, as the exomoon shows almost full moon phase during the occultation (looking from the Earth). This reflected light flux is negligibly small comparing to the stellar irradiation \citep{hinkel13}. The illumination from the planet with other light sources is thoroughly discussed by \citet{heller13}.

\item \textit{Orbital direction:} Retrograde orbits affect the plateau/occultation duration as stellar e\-clips\-es by the host planet are more frequent, than with a prograde orbit and can be prolonged because of the large size of giant planets \citep{forgan13}. Such eclipses may occur just before or after the occultation with the star, changing the shape of the light curve. Using the phase-folded light curve of several observations may help in filtering out such cases. Beside eclipses, it does not influence the observability whether the exomoon is approaching to or receding from the host star viewing from the Earth.

\item \textit{Effect of hemispherical asymmetry:} An exomoon that faces with different sides toward the star (and the toward the Earth) might have different brightness at different occultations, if significant albedo difference exists between the two hemispheres. Calculating with a Solar System analogy, Iapetus shows huge difference in its Bond albedo with $\sim$0.31 at the trailing, and maximum 0.1 at the leading hemisphere \citep{howett10}. The difference between the observations of the two hemispheres produces such brightness difference that can not be neglected. This effect may influence the result of the albedo estimation, but using the phase-folded light curve such flux changes may be noticed, and from the average light curve the average albedo can be estimated.
\end{itemize}

\begin{table*}
	\caption{Summary of various factors that influence the successful albedo estimation for an exomoon. The most influencing ones are listed in the left column, the less important and indifferent ones are listed in the right column.}
    \label{diffparameters}
	\centering
	\begin{tabular}{l l l}
		\hline\hline
		Substantial influence & moderate influence & minimal influence \\
		\hline
		$\bullet$ stellar distance & $\bullet$ eccentricity of the exomoon & $\bullet$ direct/retrograde orbital \\
		$\bullet$ exomoon size & $\bullet$ inclination of the exomoon & direction \\
		$\bullet$ albedo & $\bullet$ hemispherical asymmetry & $\bullet$ reflected light from the \\
		$\bullet$ star--planet size ratio & of the moon & planet \\
		$\bullet$ planetary distance & $\bullet$ mass of the planet \\
		\hline\hline
	\end{tabular}
\end{table*}

Based on this subsection, different parameters are categorized by their importance in occultation observations. The results can be seen in Table \ref{diffparameters}. The most relevant factors influencing the observation are: 1. close stellar distance that is observationally favourable especially at small main sequence stars; 2. star/planet diameter ratio: the larger the ratio is the longer MO duration could be observed which helps in reaching smaller photon noise for the measurement (only relevant at small stars); 3. larger exomoons are easier targets, and are favourable to form by impact ejection or by capture; 4. larger planetary distance of the moon can increase the length of the plateau duration; 5. higher albedo means that the satellite is brighter, hence makes the observation easier.

  \subsection{Importance of ice}

Icy moons with subsurface oceans might hold large volume of liquid water for geologic time-scales and thus may potentially be habitable or at least could be considered as favourable environments for life. In the Solar System such subsurface oceans exist in Europa, Titan, Enceladus, and possibly also in Ganymede and Callisto. The liquid phase state of these oceans are maintained by tidal heat production, freezing point depressing solved ingredients, and/or internal heat due to accretional energy conservation and radiogenic heat \citep{carr98, khurana98, kargel00, zimmer00, mccord01, schenk02, collins07, roberts08, lorenz08, postberg09, iess14, dobos15}. Having information on the existence of surface ice, using mass estimation from TTV and TDV might increase the probability to identify exomoons that are probable of having subsurface oceans.

The brightness of the ice could point to relatively young surface and active resurfacing processes. In the case of Europa the ocean gives possibility for active resurfacing thus for the existence of clean and bright ice there, although locations with non-ice ingredients can also be found on Europa, while on Enceladus the subsurface liquid water contributes to the geyser-like eruptions, and to the increase of surface albedo caused by fresh water ice crystals that fall back to the surface \citep{kadel00, fagents03, porco06, spencer06, verbiscer07, howett11}. Although other reasons could also produce elevated albedo, the very high albedo is still a strong indication of the water ice on the surface, and even of a liquid subsurface ocean.

Estimation of moons' albedo is much easier at M type stars, if the planet and the moon are close to the snowline. However, strong, regular flares and coronal mass ejections are expected, that potentially make the orbiting rocky bodies uninhabitable, because of high UV irradiation and loss of the atmosphere \citep{scalo07}. Exomoons still might be habitable because the ice cover attenuate UV radiation, and serve as a physical screening mechanism \citep{cockell98, cockell00}. The loss of the atmosphere could also easily happen at small stellar distance, but does not influence directly the oceans beneath ice sheets.

 \section{Conclusions}

In this work we proposed a method for the first time that hints on the surface albedo of an exomoon and thus might indicate the existence of water ice. The argumentation presented in this work can be useful to orient future research and to identify the best candidate systems for such observations. The methods and the results of our calculations are presented firstly, then secondly those parameters are listed which substantially influence such measurements and also those that are not relevant -- as such information is useful in planning and targeting observations.

Moons of smaller stars seem to be easier to detect, if their stellar distance is smaller. In addition, the smaller the star, the closer the snowline located, meaning that rocky bodies can stay icy even if they are orbiting the star in a closer orbit (but still outside the snowline). The vicinity to the star makes the body brighter, thus an icy exomoon is easier to detect in occultation close to the snowline of an M dwarf, than at the snowline of a solar-like star. The largest flux difference is expected from large, $\sim$ 2.5 Earth-radii icy moons (about 5 Earth-mass body with similar density of Enceladus) orbiting small M dwarfs ($\sim$ 0.07--0.13 solar masses) close to the snowline. We have found that such moons cannot be observed in occultation with next generation space missions in the visual spectrum, because the instruments' photon noise is far greater than the flux drop caused by the moon. However, in the near infrared (J band), the E-ELT's photon noise is just about 5-6 ppm greater than the MO depth, despite the lower albedo of ice. This result, however, is too optimistic, because it does not take into account other noise sources, such as CCD read-out and dark signal, which become significant in the near infrared measurements. Considering that not all noise sources were included in our estimations, and that measuring 30 occultations may be unrealistic in most cases, since it would require years to achieve, it can be concluded that the real noise level of the instruments in the near infrared wavelengths will be higher than presented in section \ref{nextgen}. Flux fluctuations caused by stellar activity makes the measurements even more difficult. We conclude that occultation measurements with next generation missions are far too challenging, even in the case of large, icy moons at the snowline of small M dwarfs.

We outlined the parameters that can influence the detection. Based on assumptions, the orbital and physical parameters of exomoons might be highly diverse, thus we evaluated a range in the space of several parameters. We categorized these parameters by their influence on the success of MO detections (see Table \ref{diffparameters}).

We also discussed the possible properties of the most suitable exomoons for characterization which may be useful once the instrumentation is available. Exomoons around Neptune-sized exoplanets of small stars seem to be the best targets for occultation-based albedo estimation opposite to Jupiter-like exoplanets, if the satellites, formed by impact ejection or capturing, are large enough. In the case of such satellites the surface ice cover might be co-accreted if the ejection happened before the clearing up of the circumplanetary disk, or formed by ejection from planets composed mostly of water ice, like Neptune and not of hydrogen, like Jupiter. Binary-exchange capture is also a possibility for a giant planet to have a large icy satellite (neptunes are better candidates for this process than jupiters).

Based on our work, the first albedo estimations of exomoons from occultations are expected around low mass M dwarfs at small stellar distances. These moons will probably be large, and have high albedo which may imply the presence of ice on the surface.

 \chapter{Summary} \label{bigsummary}

In this work I investigated the habitability of exoplanets and their satellites. Three major topics were described in details: the habitable zone, tidally heated exomoons and the possibility of albedo estimation of icy satellites. These topics cover not only circumstellar and circumplanetary habitability, but also connect observation with potentially habitable moons.

The habitable zone is widely used to investigate the habitability of planets around different stars. 
This zone is determined as a region around a star in which an Earth-like planet could support liquid water on its surface for a long period of time. 
For calculating the location of habitable zones around different stars, usually complex climate models are used which take into account the atmospheric properties of the Earth and the carbonate-silicate cycle, which regulates the temperature of the planet. 
Beside this complex climate model, simple equations can be used, too. 
These simple formulae are fitted to the results of the more complex climate models. 
For this simple calculation the stellar temperature and luminosity are needed as input parameters, which can be calculated from the stellar mass. 
However, different calculation methods give different results. 

We applied three previously defined models, and we also proposed new empirical formulae for calculating the temperature and luminosity from the stellar mass. 
We numerically compared the results to measured stellar parameters, by calculating the reduced chi-square values.
The new equations result in a better reproduction of the boundaries of the HZ (calculated from measured stellar parameters) than the other models.

Beside exoplanets, their moons can also be habitable. 
Tidal heating may play a key role in their habitability, since the elevated temperature can melt the ice on the body and prevent the snowball state even without significant stellar radiation. 
For calculating the tidal heating rate in exomoons, usually the so-called fixed $Q$ models are used, which highly underestimate the tidal heat of the body.
These models use $Q$ and $\mu$, the tidal dissipation and rigidity (respectively) as constants, however these parameters are functions of the temperature. 
In addition, the correct values of $Q$ and $\mu$ are extremely difficult to estimate even for Solar System bodies. 
Their values can vary a few orders of magnitude for similar bodies, which means that their estimation for yet unknown exomoons is not possible, one can only guess or assume some value for these parameters.
For these reasons, we applied a viscoelastic tidal heating model for exomoons for the first time.
This model is more realistic than the widely used, fixed $Q$ models, because it takes into account the temperature dependence of the tidal heat flux and the melting of the inner material. 
Because of the temperature dependency of the shear modulus and viscosity that are used in the viscoelastic model, the tidal heating will also be a function of the temperature.
The tidally heated surface temperature of exomoons is calculated for the equilibrium temperature of the body, which corresponds to the stable equilibrium between tidal heating and convective cooling.
In the calculations no stellar radiation, or atmosphere were considered.

Using the viscoelastic model, we introduced the Tidal Temperate Zone, which is a circumplanetary region where the tidally induced surface temperature of the satellite is between 273 and 373 K.
We have found that the location of the TTZ strongly depends on the orbital period of the moon and less on its radius. 
We applied the calculation method to different densities and orbital eccentricities of the moon. 
In some cases there was no solution (no equilibrium temperature), because the tidal forces were so weak that they could not induce convective cooling inside the body.
We compared the results with the fixed $Q$ model and investigated the statistical volume of the TTZ using both models. 
We have found that the viscoelastic model predicts 2.8 times more exomoons in the TTZ with orbital periods between 0.1 and 3.5 days than the fixed $Q$ model for plausible distributions of physical and orbital parameters. 
The viscoelastic model provides more promising results in terms of habitability because the inner melting of the body moderates the surface temperature, acting like a thermostat.

For Earth-like bodies other habitability limits than the 0 and $100\,^\circ$C were used, as well. 
The lower limit for microbial activity in salty solutions is around $-20\,^\circ$C, and the upper limit for complex eukaryotic life is approximately $60\,^\circ$C, which also corresponds to the runaway greenhouse limit on Earth.
For most of the calculations these probable limits of habitability were also considered when applicable ($\rho = \rho_\mathrm{Earth}$ and $R_\mathrm{m} \approx R_\mathrm{Earth}$), because these are related to biological and atmospherical constraints. 
When investigating the volume of habitable orbits, we found that the results were similar to the ones of the TTZ, but the volume was larger for Earth-like moons.

We have also applied a 1D energy balance model of Earth-like exomoon climates, which contain stellar and planetary insolation, atmospheric circulation, infrared cooling, eclipses and tidal heating as the principal contributors to the moon's radiative energy budget. 
The model also takes into account the carbonate-silicate cycle and the positive ice-albedo feedback system. 
We used both the viscoelastic and the fixed $Q$ models for calculating the tidal heating inside the moons.
We have found that the circumplanetary HZ is significantly wider with viscoelastic tidal heating, and extends much further from the planet, than with the fixed $Q$ model. 
In the case of zero inclination of the moon's orbit relative to the planet's equator there is a well-defined outer edge to the circumplanetary HZ, due to a combination of eclipses and ice-albedo feedback. 
We have shown that if the moon's orbit is inclined so that eclipses are unlikely, the outer edge completely disappears for any eccentricity of the moon's orbit, even for relatively large semi-major axes of the moon.

Icy moons might have subsurface oceans, such as Europa or Enceladus. 
If the moon is tidally heated, then life might emerge at the bottom of the ocean, where the rocky seafloor interacts with water through hydrothermal vents. 
In order to separate icy satellites from other exomoons, we propose a new method for obtaining their albedo. 
The albedo may be estimated from photometric measurements during occultations, when the reflected light from the satellite is blocked out and a small flux drop occurs in the light curve.
The depth of the flux drop is proportional to the albedo and the radius of the moon.

If the albedo is high, it implies that the surface is covered with ice, however methane ice is very reflective, too. 
In order to determine whether the surface ice is of water or methane, we suggest to make measurements in several photometric bands, especially in the J band, where the two kinds of ices show different spectral features, but they are still very reflective.
Clouds can also cause high albedo on a planetary body. 
Since water is very abundant in the Universe, we assume that water clouds might be common, and we compared the spectrum of the Earth's water cloud with ice spectrum. 
We found that the cloud's reflectance significantly drops in the F140M photometric band, where the ice's albedo is still very high. 
In order to distinguish clouds from surface ice, we propose to make measurements in this band, too, upon the instrumentation is available for such precise observations.

We applied simple calculations for different stellar masses in the V and J photometric bands, and compared the flux drop caused by the moon's occultation and the estimated photon noise of next generation missions with $5\,\sigma$ confidence.
We found that albedo estimation by this method is not feasible for moons of solar-like stars, but small M dwarfs are better candidates for such measurements, because their weaker stellar radiation brings the snowline closer to the star, and if the planet-moon system orbits the star at a closer distance, then they will reflect more light.
Our calculations in the J photometric band show that E-ELT MICADO's photon noise is just about 4 ppm larger than the flux difference caused by a 2 Earth-radii icy satellite in a circular orbit at the snowline of an 0.1 stellar mass star.
Because of other noise sources that were not included in our calculations, we conclude that occultation measurements with next generation missions are far too challenging, even in the case of large, icy moons at the snowline of small M dwarfs.

We also discussed the role of different parameters (some of them were neglected in the calculations, e.g. inclination, eccentricity, orbiting direction of the moon), and categorised them by their influence on the measurements. 
The most substantial influencing factors are the stellar distance, the size of the exomoon, its albedo, the star-planet size ratio and the planetary distance.
We also predict that the first albedo estimations of exomoons will probably be made for large icy moons around the snowline of M4 -- M9 type main sequence stars.

 \chapter{Short description}

I investigated the habitability conditions of extra-solar planets and satellites.

We proposed new empirical equations for calculating the temperature and luminosity from the stellar mass.
By using numeric investigations, we have found that these formulae reproduce the stellar parameters more precisely than other models that we applied.
With the empirical equations, the boundaries of the circumstellar habitable zone fit well to the ones which were calculated from measured stellar parameters of F, G, K and M main sequence stars. 

Tidal heating of exomoons may play a key role in their habitability, since the elevated temperature can prevent the body to enter a snowball state even without significant stellar radiation. 
We used a viscoelastic model for exomoons for the first time, which is more realistic than the widely used, so-called fixed $Q$ models.
We introduced the circumplanetary Tidal Temperate Zone (TTZ), and compared the results to the fixed $Q$ model. 
We have found that the viscoelastic model predicts 2.8 times more exomoons in the TTZ, than the fixed $Q$ model, for plausible distributions of physical and orbital parameters. 
The viscoelastic model provides more promising results in terms of habitability because the inner melting of the body moderates the surface temperature, acting like a thermostat.
We have also used a 1D energy balance model of Earth-like exomoon climates, and found that the circumplanetary HZ is significantly wider with viscoelastic tidal heating, and extends much farther from the planet, than with the fixed $Q$ model. 
We have shown that if the moon's orbit is inclined, then the outer edge completely disappears for any eccentricity of the moon's orbit.

Icy moons might also be habitable, if they have tidally heated subsurface oceans.
Occultation light curves of exomoons may give information on their albedo and hence indicate the presence of ice cover on the surface.
We compared the flux drop caused by the moon's occultation with the estimated photon noise of next generation missions with $5\,\sigma$ confidence.
We found that albedo estimation by this method is not feasible for moons of solar-like stars, but small M dwarfs are better candidates for such measurements.
Our calculations in the J photometric band show that E-ELT MICADO's photon noise is just about 4 ppm greater than the flux difference caused by a 2 Earth-radii icy satellite in a circular orbit at the snowline of an 0.1 stellar mass star.
We conclude that occultation measurements with next generation missions are far too challenging.
We predict that the first albedo estimations of exomoons will probably be made for large icy moons around the snowline of M4 -- M9 type main sequence stars.

 \chapter{Kivonat}

Disszert\'aci\'omban a Naprendszeren k\'iv\"uli bolyg\'ok \'es holdak lakhat\'os\'ag\'at vizsg\'altam.

\'Uj, em\-pi\-ri\-kus formul\'ak haszn\'alatat javaslom a h\H{o}\-m\'er\-s\'ek\-le\-t \'es a luminozit\'as csillagt\"omegb\H{o}l t\"ort\'en\H{o} meghat\'aroz\'as\'ara.
Numerikus vizsg\'alatok alapj\'an arra a k\"ovetkeztet\'esre jutottam, hogy ezek a k\'epletek pon\-to\-sabban visszaadj\'ak a m\'ert csillagparam\'etereket, mint a t\"obbi vizsg\'alt mo\-dell.
Az \'uj egyenletek hasz\-n\'a\-la\-t\'a\-val a lakhat\'o z\'ona hat\'araira kapott e\-red\-m\'enyek j\'ol illeszked\-nek a m\'ert param\'eterek alapj\'an sz\'amolt \'ert\'ekekhez F, G, K \'es M t\'ipus\'u f\H{o}sorozati csillagokra.

Az exoholdak \'arap\'alyf\H{u}t\'ese kulcsszerepet j\'atszhat a holdak lakhat\'os\'ag\'aban, hiszen az \'ar\-a\-p\'aly\-e\-r\H{o}k elhanyagolhat\'o csillagsug\'arz\'as mellett is megemelik az \'egitest h\H{o}\-m\'er\-s\'ek\-le\-t\'et, ami megakad\'alyozhatja a h\'olabda \'allapot kialakul\'as\'at.
Els\H{o} alkalommal haszn\'altam visz\-ko\-e\-lasz\-ti\-kus modellt az exoholdak \'arap\'alyf\H{u}t\'es\'enek le\'ir\'as\'ara, ami realisztikusabb a sz\'e\-les k\"or\-ben hasz\-n\'alt, \'un. r\"ogz\'itett $Q$ modellekn\'el.
Bevezettem a bolyg\'o k\"or\"uli m\'ers\'ekelt \'ar\-a\-p\'aly\-f\H{u}\-t\'e\-si z\'o\-na (M\'AZ) fogalm\'at, \'es az eredm\'enyeket \"osszehasonl\'itottam a r\"ogz\'itett $Q$ mo\-del\-l\'e\-vel. 
Arra ju\-tot\-tam, hogy \'eszszer\H{u} fizikai \'es kering\'esi param\'eterek haszn\'alata mellett a visz\-ko\-e\-lasz\-ti\-kus modell 2,8-szer t\"obb exoholdat j\'osol a M\'AZ-ban, \'igy a lakhat\'os\'ag szempontj\'ab\'ol kedvez\H{o}bb eredm\'enyeket ad, mert a hold belsej\'enek megolvad\'asa m\'ers\'ekli a felsz\'ini h\H{o}m\'ers\'ekletet.
A f\"oldihez hasonl\'o l\'egk\"ort felt\'etelezve az 1D energiaegyenleg modell alkalmaz\'as\'aval arra ju\-tot\-tunk, hogy a bolyg\'o k\"or\"uli lakhat\'o z\'ona sokkal sz\'elesebb a viszkoelasztikus modell haszn\'alata eset\'en, \'es a bolyg\'ot\'ol nagyobb t\'avols\'agokig kiterjed, mint a r\"ogz\'itett $Q$ modellel.
Megmutattuk azt is, hogy ha a hold p\'alyas\'ikja d\H{o}lt, akkor a k\"uls\H{o} hat\'ar b\'armely excentricit\'as \'ert\'ekre elt\H{u}nik. 

Jeges exoholdak is lehetnek lakhat\'ok, ha a felsz\'in alatti \'oce\'an \'arap\'alyf\H{u}t\"ott.
Az exoholdak okkult\'aci\'os f\'enyg\"orb\'eje inform\'aci\'ot adhat az albed\'or\'ol, amivel a felsz\'ini j\'eg jelenl\'et\'ere k\"o\-vet\-kez\-tet\-he\-t\"unk.
\"Osszehasonl\'itottam a hold okkult\'aci\'ojakor bek\"ovetkez\H{o} fluxuscs\"okken\'est a k\"o\-vet\-ke\-z\H{o} gener\'aci\'os t\'avcs\"ovek fotonzaj\'aval $5\,\sigma$ sz\'or\'as mellett.
Meg\'allap\'itottam, hogy a Nap\-hoz hasonl\'o csillagok eset\'en az albed\'o becsl\'ese nem val\'os\'ithat\'o meg ezzel a m\'odszerrel, de a kis M t\"orp\'ek \'ig\'eretesebb c\'elpontok lehetnek.
Megmutattam, hogy a J fotometrikus s\'avban az E-ELT MICADO fotonzaja csak 4 ppm-mel haladja meg egy 2 f\"oldsugar\'u jeges hold fluxuscs\"okken\'es\'et egy 0,1 napt\"omeg\H{u} csillag h\'ohat\'ar\'an\'al h\'uz\'od\'o k\"orp\'aly\'an keringve.
Arra a k\"o\-vet\-kez\-te\-t\'es\-re jutottam, hogy az okkult\'aci\'os m\'er\'esek t\'uls\'agosan meghaladj\'ak a k\"ovetkez\H{o} gener\'aci\'os t\'avcs\"ovek k\'epess\'egeit.
Az els\H{o} exoholdas albed\'obecsl\'eseket v\'arhat\'oan a nagy, jeges holdakra fogj\'ak el\-v\'e\-gez\-ni, melyek M4 -- M9 t\'ipus\'u f\H{o}sorozati csillagok h\'ohat\'ar\'anak k\"ozel\'eben keringenek.


\bibliographystyle{apj}
\bibliography{ref}

\chapter{Acknowledgements}

I would like to thank the work and help that I got from my supervisors during the years. 
I express my grateful thanks to Prof.~L\'aszl\'o L. Kiss, who helped with useful discussions and ideas that improved my research, and funded my participation at conferences and summer schools through his projects, which led to new collaborations and research works, part of them presented in this thesis.
I am also grateful to Prof.~Edwin L. Turner for his invitation to work at Princeton University, for his hospitality, and the helpful research consultations in person and also online across continents.
I thank the support of Dr.~Judit Orgov\'anyi who sacrificed her free time to help me along the way to my PhD.
I thank Prof.~J\'ozsef Vink\'o the help he provided with the recalculation and improvement of the habitable zone boundaries.
I thank Prof.~L\'aszl\'o Szabados for reading my dissertation, providing useful comments and grammar checking.
I also thank all of the co-authors whom I worked with, without them I couldn't have written my thesis.
I am indebted to my parents and family for helping me through my education, which finally, after following a long and curvy road, led me to becoming an astronomer and hopefully to getting a doctorate degree in the near future.
I owe my husband a debt of gratitude for being patient and very supportive even during the most stressful times.
Special thanks to my friends, especially the (extended) \textit{flight~team}, for the happy moments that we spent together, which helped me to `recharge my batteries' so I could continue working.
Finally, I would like to thank the \textit{Eurest} cantine in the Ericsson building in the Infopark, the \textit{bonapp\'etit} food service, and the \textit{Genomics Caf\'e} in the Icahn building for the lunches, that saved me a lot of time, since I didn't have to cook on weekdays.

I have been supported by the Hungarian OTKA Grant K104607, the Hungarian National Research, Development and Innovation Office (NKFIH) grant K-115709, the Lend\"ulet-2009 Young Researchers Program of the Hungarian Academy of Sciences, the T\'ET-14FR-1-2015-0012 project, and the ESA PECS Contract No. 4000110889/14/NL/NDe.

\end{document}